\documentclass[10pt]{article}

\usepackage{amsmath}
\usepackage{graphicx}
\usepackage{amssymb}
\usepackage{changepage}
\usepackage{amsthm}
\usepackage{enumerate}
\usepackage{enumitem}
\usepackage{multicol}
\usepackage{bm}
\usepackage{physics}
\usepackage{hyperref}
\usepackage{float}
\usepackage{subcaption}
\usepackage{listings}
\usepackage{natbib}
\usepackage{bbm}
\usepackage[linesnumbered,ruled,vlined]{algorithm2e}
\usepackage{blkarray}
\usepackage{setspace}

\newcommand{\reffig}[2]{\hyperref[#1]{figure~\ref*{#1}#2}}
\newcommand{\refeq}[1]{\hyperref[#1]{equation~(\ref*{#1})}}
\newcommand{\refeqsystem}[1]{\hyperref[#1]{equations~(\ref*{#1})}}
\newcommand{\refeqshort}[1]{\hyperref[#1]{Eq~(\ref*{#1})}}
\newcommand{\refeqs}[2]{\hyperref[#1]{equations~(\ref*{#1})} and \hyperref[#2]{(\ref*{#2})}}
\newcommand{\RN}[1]{\textup{\uppercase\expandafter{\romannumeral#1}}}

\begin{document}

\title{A flexible, random histogram kernel for discrete-time Hawkes processes}
\author{Raiha Browning\textsuperscript{*1,2}, Judith Rousseau\textsuperscript{3,4}, Kerrie Mengersen\textsuperscript{1,2}}
\date{}

\maketitle

\textsuperscript{1} School of Mathematical Sciences, Queensland University of Technology, Brisbane, Australia 

\textsuperscript{2} Centre for Data Science, Queensland University of Technology
 
\textsuperscript{3} Department of Statistics, University of Oxford, Oxford, United Kingdom

\textsuperscript{4} Ceremade, Universit\'{e} Paris-Dauphine, Paris, France \\

* raiha.browning@icloud.com

\section*{Abstract}

Hawkes processes are a self-exciting stochastic process used to describe phenomena whereby past events increase the probability of the occurrence of future events. This work presents a flexible approach for modelling a variant of these, namely discrete-time Hawkes processes. Most standard models of Hawkes processes rely on a parametric form for the function describing the influence of past events, referred to as the triggering kernel. This is likely to be insufficient to capture the true excitation pattern, particularly for complex data. By utilising trans-dimensional Markov chain Monte Carlo inference techniques, our proposed model for the triggering kernel can take the form of any step function, affording significantly more flexibility than a parametric form. We first demonstrate the utility of the proposed model through a comprehensive simulation study. This includes univariate scenarios, and multivariate scenarios whereby there are multiple interacting Hawkes processes. We then apply the proposed model to several case studies: the interaction between two countries during the early to middle stages of the COVID-19 pandemic, taking Italy and France as an example, and the interaction of terrorist activity between two countries in close spatial proximity, Indonesia and the Philippines, and then within three regions of the Philippines.

\section{Introduction}

A Hawkes process (HP) is a self-exciting stochastic process, first introduced by \cite{Hawkes:1971vr} as a continuous-time point process, to describe phenomena whereby past events increase the probability of the occurrence of future events. A HP can be characterised in terms of a conditional intensity function that is comprised of a baseline component representing independent event arrivals and a self-exciting component accounting for the increased excitation that arises due to past events. A key feature of the self-exciting component is the triggering kernel, which describes the temporal behaviour of the excitation due to past events as a function of time elapsed. This triggering kernel is often ascribed a parametric form, such as exponential or power law. 

Several authors \citep{Browning:2021, White:2012hy, Linderman:2015tk, Mohler:2013hy} have adapted the HP to a discrete-time resolution, to allow for a broader range of problems to be solved using this framework. While they have many attractive qualities and occur frequently in practice, discrete-time Hawkes processes (DTHPs) are less well studied than their continuous-time counterparts.  The framework around these processes is a simplification of their continuous-time counterparts, but often event count data are only available at discrete, and often regular, time intervals. Thus, these processes are useful to avoid introducing artificial noise into the data by imputing counts onto a continuous-time resolution.  Regarding the approximation errors introduced due to the discretisation of the time domain, \citep{Kirchner:2017es} provides a comprehensive analysis. Similarly to continuous-time HPs, the DTHP enjoys a causal interpretation, with past events triggering future events, and in the multivariate case with nodes influencing other nodes.

In this study we propose a flexible model to estimate the temporal excitation pattern of self-exciting processes, specifically for DTHPs. However, the model is easily transferable to instances of continuous-time HPs where a discretised triggering kernel is appropriate. The triggering kernel is represented as a random histogram function that does not assume a parametric form, and is estimated in such a way that both the number of change points and their respective locations are unknown. In particular, we employ reversible-jump Markov chain Monte Carlo (MCMC) \citep{Green:1995} to traverse parameter spaces of varying sizes. Further details can be found in Section \ref{sec:inference}. 

Several flexible Bayesian nonparametric models for continuous-time Hawkes processes have been proposed in the literature. A popular approach is the use of Gaussian process priors (GPPs). \cite{Zhou:2020em} and \cite{Zhou:2021em} develop efficient algorithms using GPPs to model both the baseline intensity and excitation kernel of the HP. \cite{Zhang:2020vi} similarly uses a GPP to estimate the triggering kernel only. However, these are not directly transferrable to modelling DTHPs as they produce continuous functions, whereas DTHPs require a triggering kernel that is defined on a discrete domain.
Dirichlet process priors (DPPs) have also been studied for Hawkes processes. \cite{Marwick:2020} considers a DPP for the excitation kernel in the context of extreme value theory, and on the baseline rate using a hierarchical Dirichlet process prior. 

Bayesian nonparametric random histogram kernels for continuous-time Hawkes processes have been studied previously by \cite{Donnet:2020} and \cite{Sulem:2021}. However, often data are only available at discrete time intervals, so other models must be considered to avoid artificially imputing the data. In this work, we extend the model first presented in \cite{Donnet:2020} to DTHPs by proposing a trans-dimensional histogram function for the triggering kernel, for which the locations of change points are integer-valued. This condition on the change points is necessary to allow for the estimation of a magnitude parameter, which describes the expected number of offspring events from a single parent event. 

The outline of this article is as follows. A description of the proposed model is provided. We demonstrate its use in recovering the temporal excitation pattern of DTHPs through a comprehensive simulation study that considers the model under a range of scenarios with varying observational lengths and prior settings. We then consider several case studies using real-world data, modelling the interaction functions for COVID-19 cases between France and Italy, and terrorist activity in parts of Southeast Asia. These case studies were selected to demonstrate a range of contrasting dynamics, comprising event types from both high and low intensity processes. The case studies are also an exemplar of how these models are extensible to other, in particular multivariate, applications. Lastly, we discuss the outcomes of this analysis and their implications, limitations and future work. 

\section{Methods} \label{sec:methods}

\subsection{Discrete-time Hawkes processes}

Similar to its continuous-time counterpart, the DTHP is characterised in terms of its conditional intensity function $\lambda(t|\mathcal{H}_{t-1})$, which represents the expected number of events to occur in the process at time $t$ conditional on $\mathcal{H}_{t-1}$, the history up to but not including $t$. In a $K$-dimensional setting, the DTHP $\bm{N(t)} = (N^1(t), \dots, N^K(t))$ is a multivariate counting process comprised of $K$ distinct components, where $N^k(t)$, $k=1,\dots,K$, describes the number of events to occur in the $k^\text{th}$ dimension at time $t$. The corresponding conditional means, $\lambda^k(t|\mathcal{H}_{t-1}) \ \forall k$, are given by,  

\begin{align}
	\lambda^k(t|\mathcal{H}_{t-1}) &= E\{N^k(t) - N^k(t-1) | H_{t-1}\} \nonumber  \\ 
		&= \mu^k + \sum_{l=1}^K \alpha^{lk} \sum_{i:t_i^l<t} y_{t_i^l} g^{lk}(t-t_i^l) \label{eq:cond_int_dis} \ 
\end{align}
where $\mu^k$ describes the arrival rate of exogenous events in the $k^\text{th}$ process, $t_i^l<t$ and $y_{t_i^l}$ represents the observed number of events on day $t_i$ in the $l^\text{th}$ dimension. The second term overall describes the excitation behaviour and is comprised of two parameters. 
The triggering kernel $g^{lk}(t-t_i^l)$ describes the influence of past events from the $l^\text{th}$ process on the intensity of the $k^\text{th}$ process, as a function of time elapsed. The non-negative magnitude parameter $\alpha^{lk} \in \mathbb{R}_{ \ge 0} \ \forall l,k$ is then the expected number of offspring events generated in the $k^\text{th}$ dimension by a single parent event in the $l^\text{th}$ dimension \citep{White:2012hy}. 

The DTHP also allows for an additional layer of flexibility in which the underlying data generating process is specified. In this study the number of events in the $k^\text{th}$ dimension at time $t$, $y^l_{t}$, is assumed to be Poisson with mean according to (\ref{eq:cond_int_dis}), which is conditional on past events. The likelihood function, $L$, for this process is then,

\begin{align}
	L \propto \prod_{k=1}^K \prod_{t=1}^T \bigg[ (\mu^k + \sum_{l=1}^K   \alpha^{lk} \sum_{i:t_i^l<t} y_{t_i^l} g^{lk}(t-t_i^l))^{y_t^k} \exp(-(\mu^k + \sum_{l=1}^K   \alpha^{lk} \sum_{i:t_i^l<t} y_{t_i^l} g^{lk}(t-t_i^l)))   \bigg].
\end{align}

\subsection{Random histogram kernel}

We consider a flexible random histogram kernel defined on $(0,s_{\text{max}}), \  s_{\text{max}} \in  \mathbb{Z_+},$ for each excitation kernel of the DTHP. To ensure the excitation kernels are a proper density, in that they sum to 1, the histograms are normalised to enable inference to be performed on the magnitude parameters $\alpha^{lk} \ \forall l,k$. The triggering kernel describing the influence of the $l^\text{th}$ dimension onto the $k^\text{th}$ dimension, with a given number of components $J^{lk} \in \mathbb{Z_+}$, then has the form,

\begin{align}
	g^{lk}(t-t_i^l) &= \sum_{j=1}^{J^{lk}}  \frac{\theta_j^{lk}}{\sum_{h=1}^{J^{lk}} (s_h^{lk} - s_{h-1}^{lk}) \theta_h^{lk}} \mathbb{I}_{t-t_i^l \in I_j^{lk}}	 \label{eq:lambda} \\
	\text{where } \bm{s}^{lk} &= (s_0^{lk}=0, s_1^{lk}, \dots, s_J^{lk}=s_\text{max}), \ \ \ &s_h^{lk} \in \mathbb{Z}_+, \ \forall h \nonumber \\
	\bm{\theta}^{lk} &= (\theta_1^{lk} = 1,\theta_2^{lk} = \gamma_1^{lk}, \dots, \theta_J^{lk} =\gamma_{J-1}^{lk}), \ \ \  &\gamma_j^{lk} \in \mathbb{R}_+, \ \forall j \nonumber \\
	I_j^{lk} &= [s^{lk}_{j-1},s^{lk}_j] \ \forall j.
\end{align}

In other words, the heights of the histogram function are estimated relative to the first step, which is fixed at 1 and subsequently normalised. This step is necessary to ensure that the combination of height parameters required to produce a particular histogram is unique, since inference is performed on the unnormalised heights. The number of components in each histogram, $J^{lk}$, is permitted to vary through reversible-jump MCMC with a birth-death step \citep{Green:1995}. For a K-dimensional DTHP under the random histogram kernel, the conditional intensity function of the $k^\text{th}$ dimension is of the form,

\begin{align}
	\lambda^k(t|\mathcal{H}_{t-1}) 	&= \mu^k + \sum_{l=1}^K  \alpha^{lk} \sum_{i:t_i^l<t} y_{t_i^l}  \sum_{j=1}^{J^{lk}}  \frac{\theta_j^{lk}}{\sum_{h=1}^{J^{lk}} (s_h^{lk} - s_{h-1}^{lk}) \theta_h^{lk}} \mathbb{I}_{t-t_i^l \in I_j^{lk}}.
\end{align}

\section{Inference}\label{sec:inference}

Implementation and inference for the proposed model are described in the context of a comprehensive simulation study. To simplify notation, this section will assume a univariate DTHP. However, it is readily generalisable to higher dimensions, as illustrated in Section \ref{sec:methods}.

\subsection{Overview of simulation study} \label{sec:sim_study}

Numerous simulations were conducted under a range of scenarios to test the performance of the proposed model. Three prior settings were considered; these are referred to as informative, relatively informative and quite uninformative. Additionally for each prior setting, four datasets were considered with the length of the observational intervals ranging from 50 to 500 days. Repeated realisations of the process were generated for each pairwise prior and observational interval length setting. Although these data were generated independently for the various time series lengths, they use the same model parameters resulting in independent realisations of the process. By considering a combination of prior settings and durations, we obtain an indication of how closely the prior needs to reflect the true parameters to adequately infer their values, and how a longer time series affects this outcome.  

\subsection{Prior distributions}

All of the continuous model parameters are inherently non-negative. Thus, we perform inference on the natural logarithm of the model parameters to explore the unconstrained parameter space. Priors on each of the continuous parameters, namely the logarithm of the baseline parameter $\bar{\mu}$, the logarithm of the magnitude parameter $\bar{\alpha}$, and the logarithm of the unnormalised histogram heights $\bar{\gamma}$, are chosen to reflect the three levels of prior information specified previously. 

In particular, the informative prior scenario places narrow normal priors on each of these parameters. For the static parameters, i.e. the baseline and magnitude parameters, these priors are centred around the true parameter values. The prior for the histogram heights are centred around the average of the true histogram height values, denoted $\gamma_{\text{avg}}$, since the number of components in the histogram can vary at each iteration of the algorithm. In the relatively informative scenario standard normal priors are assigned to all continuous variables, and uniform priors with wide support for the quite uninformative scenario. The hyperparameter settings for each of these scenarios are specified in Table \ref{tab:scenarios}.  


\begin{table}[H]
\centering
\begin{tabular}{ c|ccc } 
 \hline
& I  & RI & QUI \\
 \hline
$\bar{\mu}$  & $N(\log(\mu_{\text{true}})-0.5^2, 0.5)$ & $N(0, 1)$ &$U(-5,5)$ \\
$\bar{\alpha}$ & $N(\log(\alpha_{\text{true}})-0.5^2, 0.5)$  & $N(0, 1)$ &$U(-5,5)$  \\
$\bar{\gamma}$ & $N(\log(\gamma_{\text{avg}})-0.5^2, 0.5)$ & $N(0, 1)$ & $U(-5,5)$ \\
 \hline
\end{tabular}
 \caption{Prior specifications for continuous variables under the informative (I), relatively informative (RI) and quite uninformative (QUI) scenarios.}
 \label{tab:scenarios}
\end{table}




The priors on the discrete variables remained fixed throughout our analysis. For the change points of the histogram function $\bm{s}$, all configurations for a given $J$ and $s_{\text{max}}$ are assumed equally likely apriori as in \cite{Bodin:2012}. The number of components, $J$, ranges from $1$ to $s_\text{max}$, and all possible values are assigned equal prior mass. Hence the priors on the number of components and the knot points are defined as,

\begin{align*}
    \Pi(J=x) &= \frac{1}{s_\text{max}} \ \forall x \in \{1, \dots ,s_\text{max}\} \\
    \Pi(\bm{s} | J) &= \frac{J! (s_\text{max} - J)!}{s_\text{max}!}. 
\end{align*}

\subsection{Proposal distributions}

At each iteration the current histogram is updated for a given number of components, $J$. The continuous parameters $\bar{\mu}$, $\bar{\alpha}$ and $\bar{\gamma_j}, j \in \{1,\dots,J-1 \} $ are updated via random walk Metropolis Hastings proposals. Each intermediate change point $s_j \in \{s_1,...,s_{J-1}\}$ with proposal $s_j^*$ has proposal density,

\begin{equation*}
	q(s_j^*|\bm{s}) = \frac{1}{n_\text{vacant}}, j=1,\dots,J-1 
\end{equation*}
 where $n_\text{vacant}$ is the number of integers in the intervals adjacent to $s_j$ with no current knot points. That is, the number of integers contained in the open intervals $(s_{j-1},s_j)$ and $(s_j,s_{j+1})$. This procedure shifts the locations of the knot points in the histogram while keeping $J$ fixed.
 

 
After updating the histogram for a fixed $J$, a birth-death proposal is used to increase or decrease $J$. A split-merge proposal was also considered, but not selected for the final model due to a lack of improvement when accounting for the additional computational expense. Further discussion regarding the exploration of the split-merge step is in Section \ref{sec:mv_alt} of this article and in the supplementary materials.

To propose a birth move with probability $p_\text{b}$ moving from $J=x$ to $J=x+1$, a new knot point, $s^*$, is first proposed from the set of vacant knot points, with probability $\frac{1}{s_\text{max} - x}$. A new height $\gamma^*$ is then proposed from a normal density $N(m_{\gamma^*}, 0.1)$, where $m_{\gamma^*}$ is the mean of the existing histogram heights $\gamma_j$. Conversely, a death move is chosen with probability $p_\text{d} = 1-p_\text{b}$ to move from $J=x+1$ to $J=x$. A knot point to remove from $\bm{s}$, namely $s^*$, is proposed with probability $\frac{1}{x}$ and the corresponding $\gamma_{s^*}$ is removed. The acceptance probability for a birth move, $A$, is then,
\begin{align*}
	A = \frac{p_\text{d} \frac{1}{x}}{p_\text{b} \frac{1}{s_\text{max}-x} N(\gamma_{s^*}; m_{\gamma^*}, 0.1)} \times \text{posterior ratio}
\end{align*}
and the acceptance probability for the death move is $A^{-1}$.


\section{Results}

We describe a univariate simulation study, first fitting the proposed model with three parallel chains, and then considering the averaged result over 90 parallel chains from various initialisations. A comparison to a parametric kernel is also performed at this stage. Following this, we extend the problem to the multivariate setting and discuss alternative models considered in the development of this article. Lastly, we fit the proposed model to two substantive case studies: namely characterising the spread of COVID-19 between France and Italy, and terrorist activity across regions in Southeast Asia.
%
%
%
%

\subsection{Univariate simulation study} \label{sec:uv_sim}

\begin{figure}[H]
	\centering
	\includegraphics[width=0.4\textwidth]{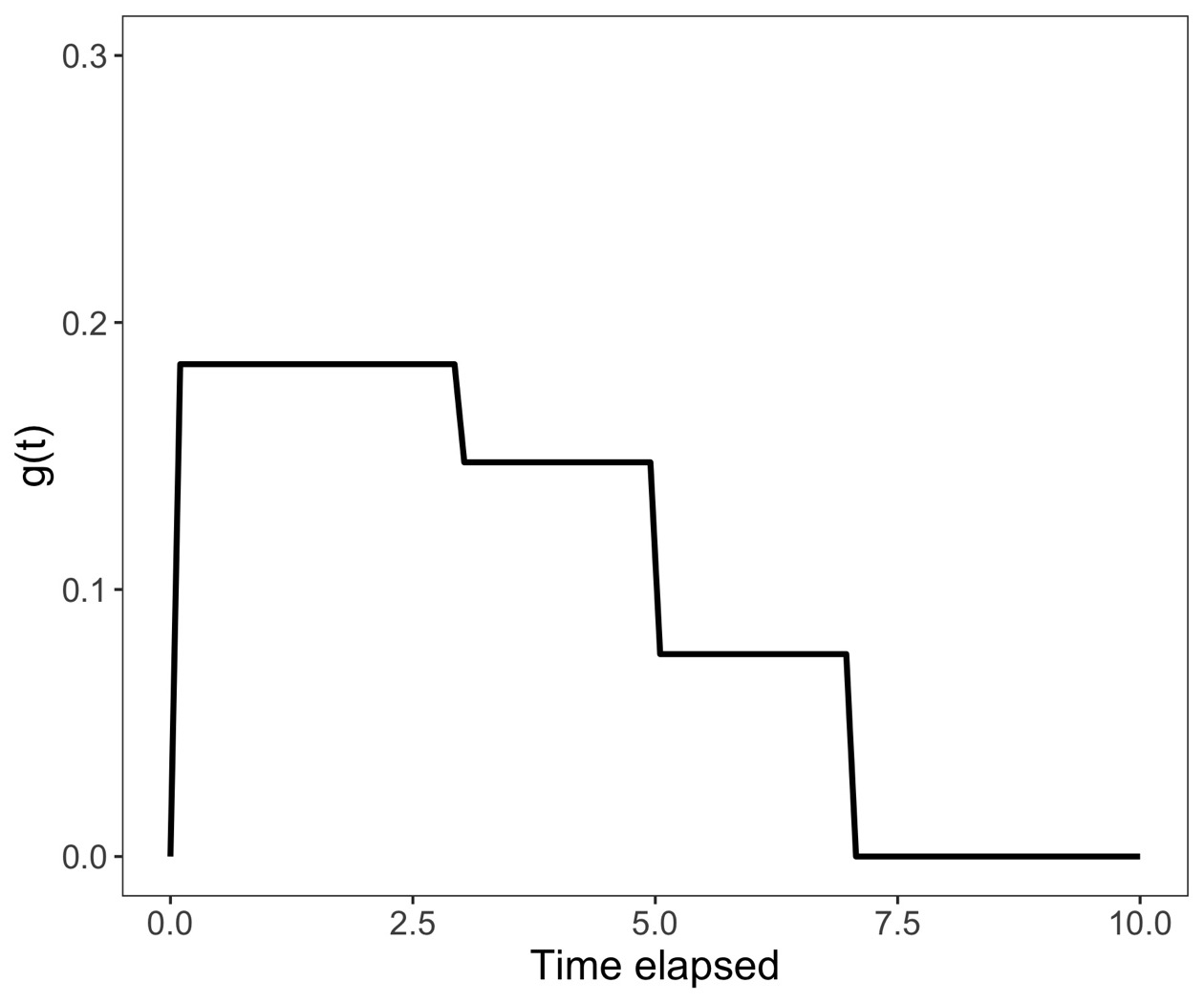}
	\caption{True excitation function. This represents the temporal excitation behaviours as a function of time elapsed since an event occurrence.}
	\label{fig:true_exc}
\end{figure}

Consider a univariate DTHP with $\lambda(t)$ given by,
\begin{align}
	\lambda(t) &= \mu +  \alpha \sum_{i:t_i<t} \bigg( y_{t_i} \sum_{j=1}^J  \frac{\theta_j}{\sum_{h=1}^J (s_h - s_{h-1}) \theta_h} \mathbb{I}_{t-t_i \in I_j}	\bigg) \label{eq:lambda} 
\end{align}

where $\mu = 1, \alpha = 0.9$ and the true triggering kernel with $J=3$ is shown in Figure \ref{fig:true_exc}.

A realisation of the process was simulated for each combination of prior setting and time series length, described in Section \ref{sec:inference}, and this was repeated three times. Only one such realisation for each scenario is presented here, as similar conclusions were reached for all simulations that were performed. Furthermore, only the excitation kernels for time series lengths of 50 and 500 days are shown in the main article. The remaining results for intermediate time series lengths can be found in the supplementary material. 

For a given realisation of the process, three MCMC chains were run in parallel for 60,000 iterations. Posterior samples from these chains were then combined after discarding the first 30,000 samples from each. The chains were initialised from a flat histogram function. The static parameters $\mu$ and $\alpha$ (those that retain their interpretation regardless of current number of components in the histogram) were drawn from the prior. Convergence of the posterior distribution for the histogram kernel was assessed by comparing the posterior distributions from independent chains. The static parameters were assessed by inspecting trace plots, posterior densities and autocorrelation plots from the posterior samples.

\begin{figure}[H]
    \centering
    \begin{subfigure}{0.5\textwidth}
	\centering
    \begin{subfigure}[b]{0.45\textwidth}
        \centering
        \includegraphics[width=\textwidth]{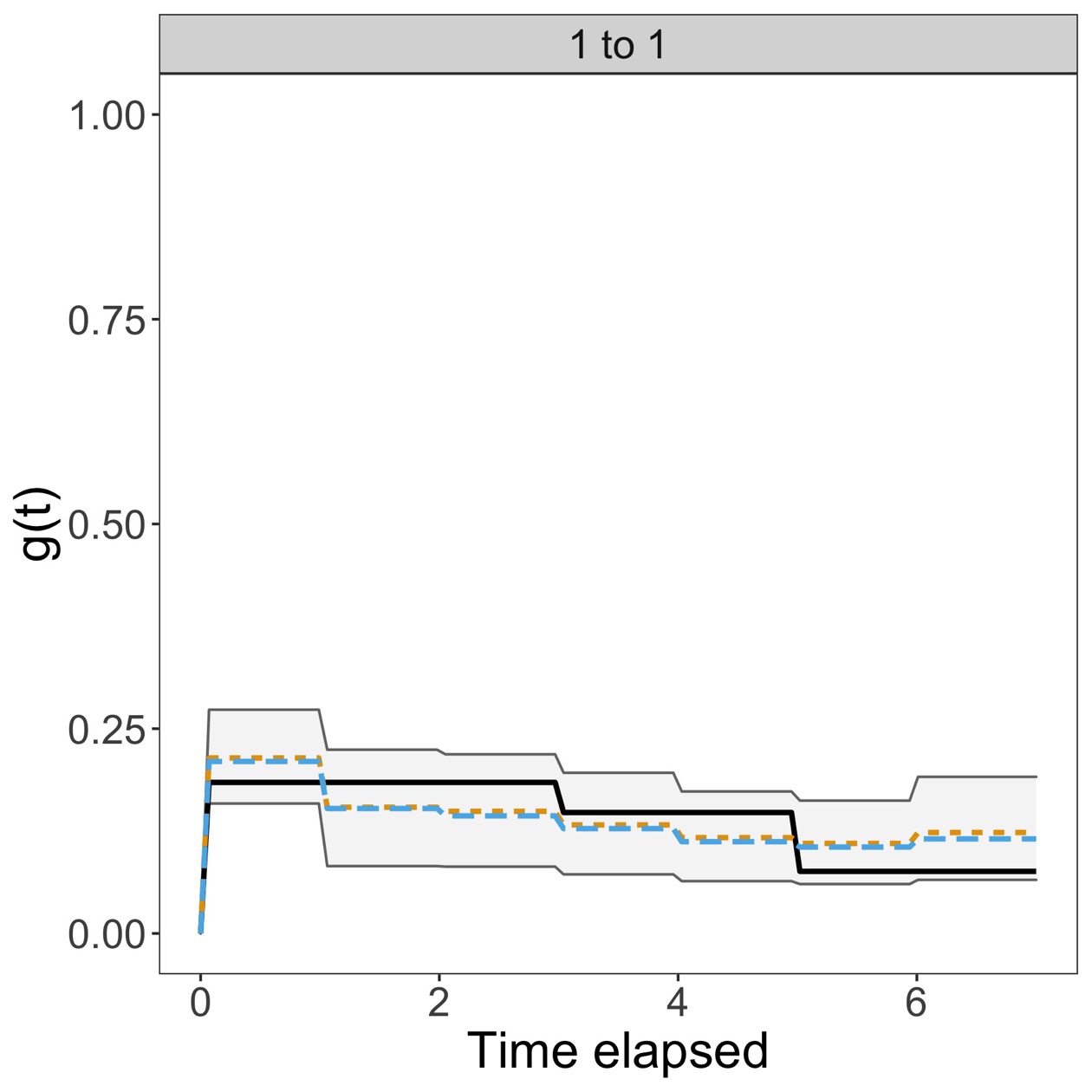}
        \caption{50 days. Total event count: 290}
    \end{subfigure}
    \hfill
    \begin{subfigure}[b]{0.45\textwidth}
        \centering
        \includegraphics[width=\textwidth]{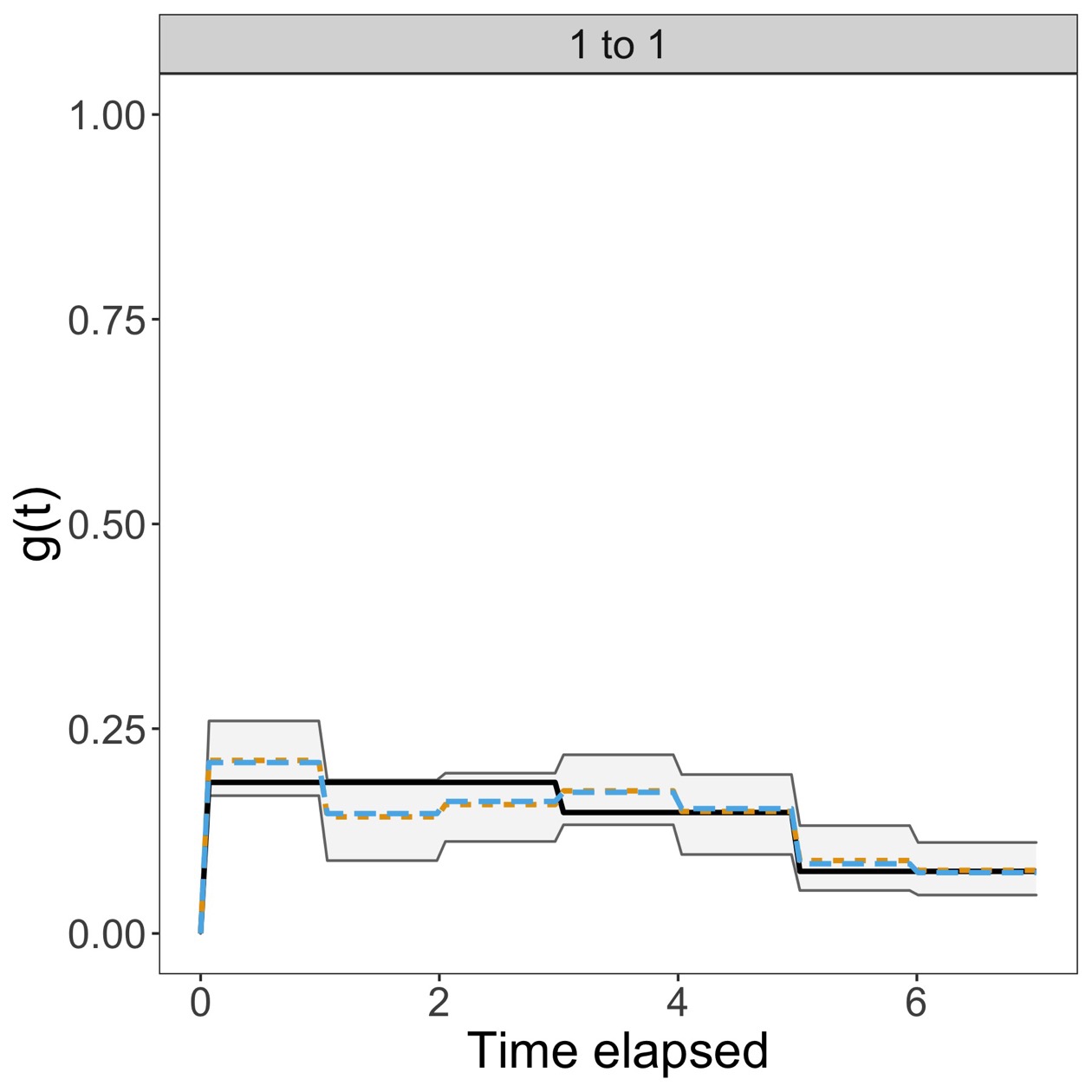}
        \caption{500 days. Total event count: 5583}
    \end{subfigure}
	\end{subfigure}    
	\caption[] 
	{\small Estimated excitation function under the informative prior setting.\\ \textbf{Solid black line:} true histogram function. \textbf{Dashed orange line:} posterior mean.\\ \textbf{Dashed blue line:} posterior median. \textbf{Grey ribbon:} 80\% posterior interval.}
	\label{fig:1d_inf}
\end{figure}

\begin{figure}[H]
    \centering
    \begin{subfigure}{0.5\textwidth}
    \begin{subfigure}[b]{0.45\textwidth}
        \centering
        \includegraphics[width=\textwidth]{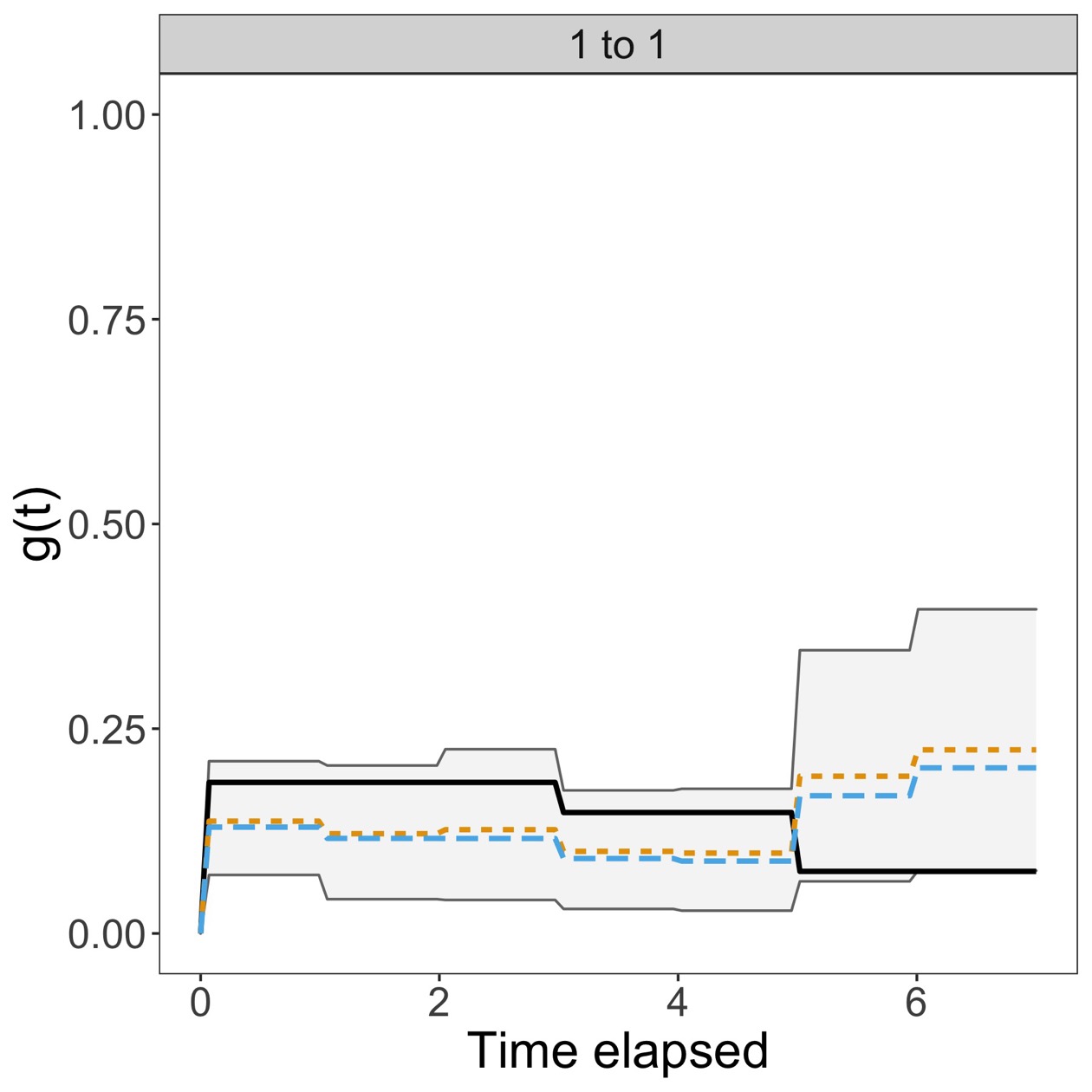}
        \caption{50 days. Total event count: 206}
    \end{subfigure}
    \hfill
    \begin{subfigure}[b]{0.45\textwidth}
        \centering
        \includegraphics[width=\textwidth]{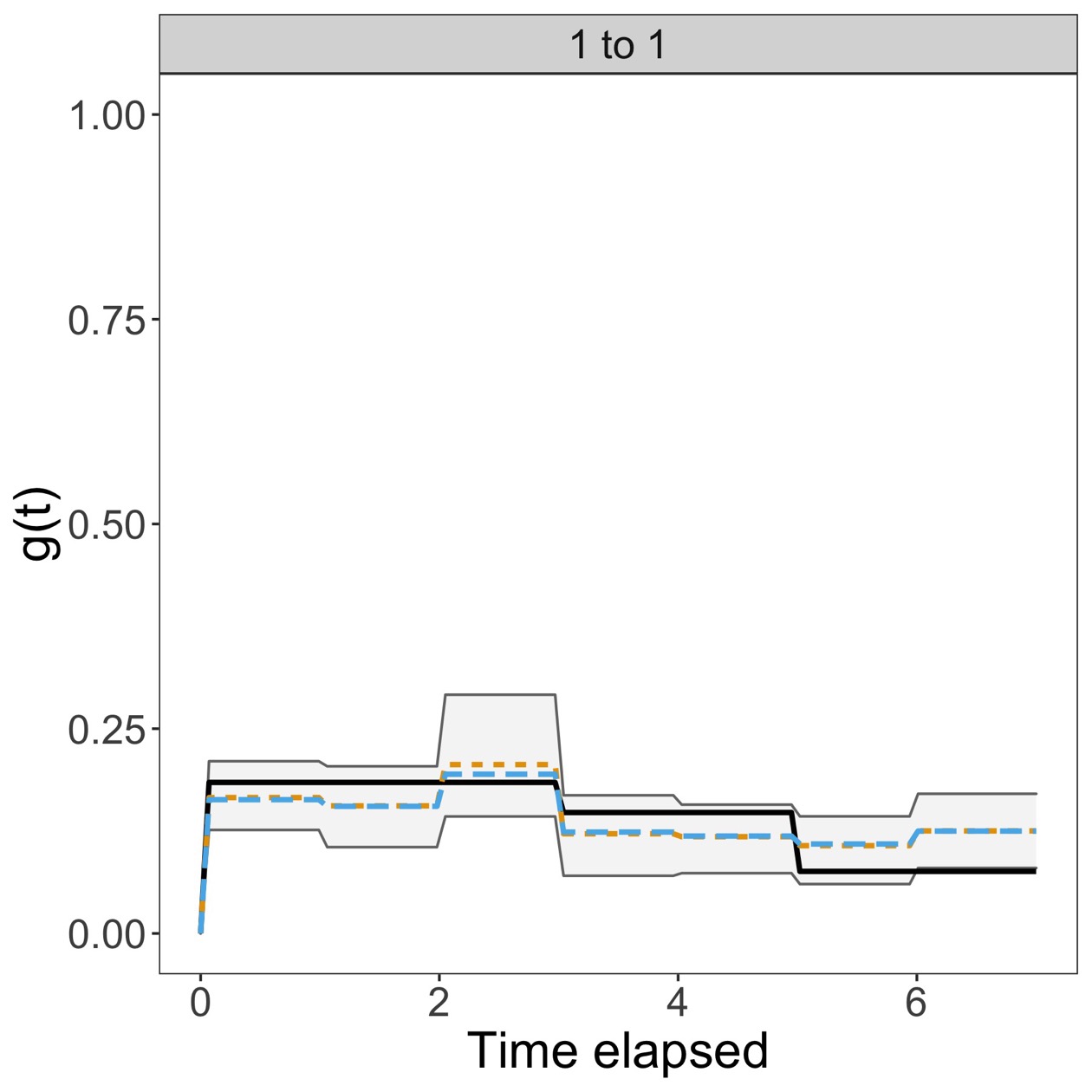}
        \caption{500 days. Total event count: 4347}
    \end{subfigure}
	\end{subfigure}    
	\caption[] 
	{\small Estimated excitation function under the relatively informative prior setting.\\ \textbf{Solid black line:} true histogram function. \textbf{Dashed orange line:} posterior mean.\\ \textbf{Dashed blue line:} posterior median. \textbf{Grey ribbon:} 80\% posterior interval.}
	\label{fig:1d_relinf}
\end{figure}

\begin{figure}[H]
    \centering
    \begin{subfigure}{0.5\textwidth}
    \begin{subfigure}[b]{0.45\textwidth}
        \centering
        \includegraphics[width=\textwidth]{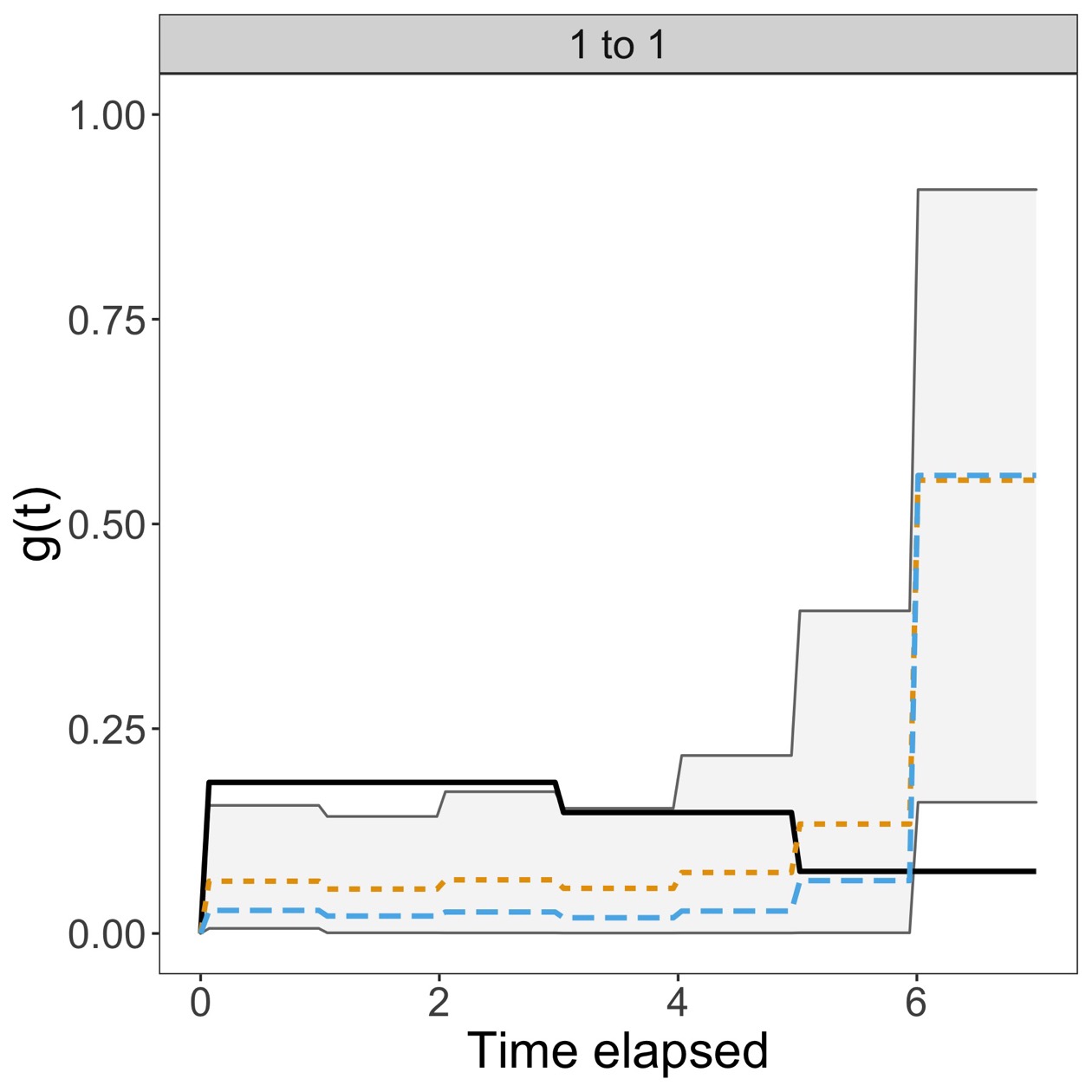}
        \caption{50 days. Total event count: 289}
    \end{subfigure}
    \hfill
    \begin{subfigure}[b]{0.45\textwidth}
        \centering
        \includegraphics[width=\textwidth]{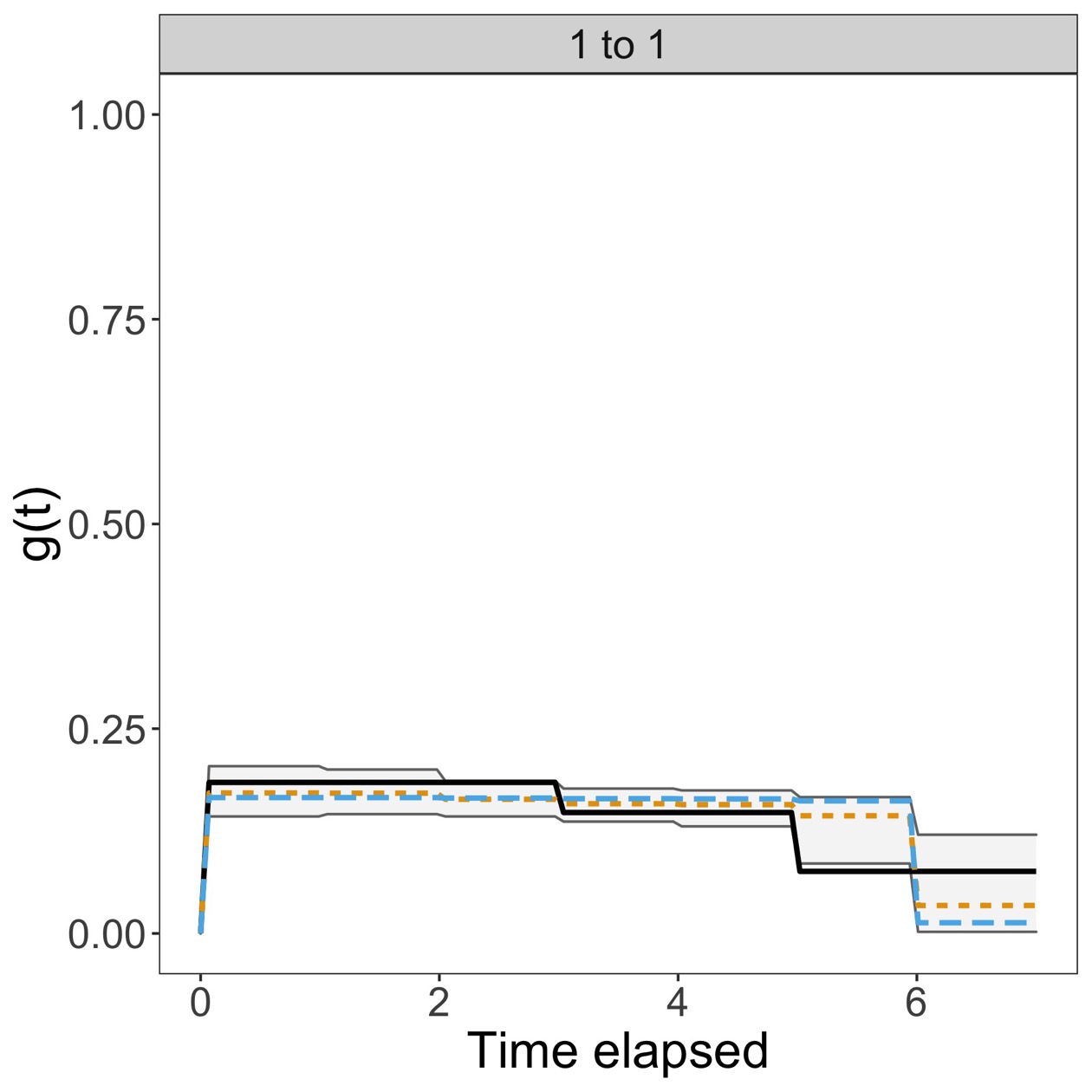}
        \caption{500 days. Total event count: 4692}
    \end{subfigure}
	\end{subfigure}    
	\caption[] 
	{\small Estimated excitation function under the quite uninformative prior setting. \\ \textbf{Solid black line:} true histogram function. \textbf{Dashed orange line:} posterior mean.\\ \textbf{Dashed blue line:} posterior median. \textbf{Grey ribbon:} 80\% posterior interval.}
	\label{fig:1d_uninf}
\end{figure}

\begin{figure}[H]
    \centering
    \begin{subfigure}{\textwidth}
	\centering
    \begin{subfigure}[b]{0.23\textwidth}
        \centering
        \includegraphics[width=\textwidth]{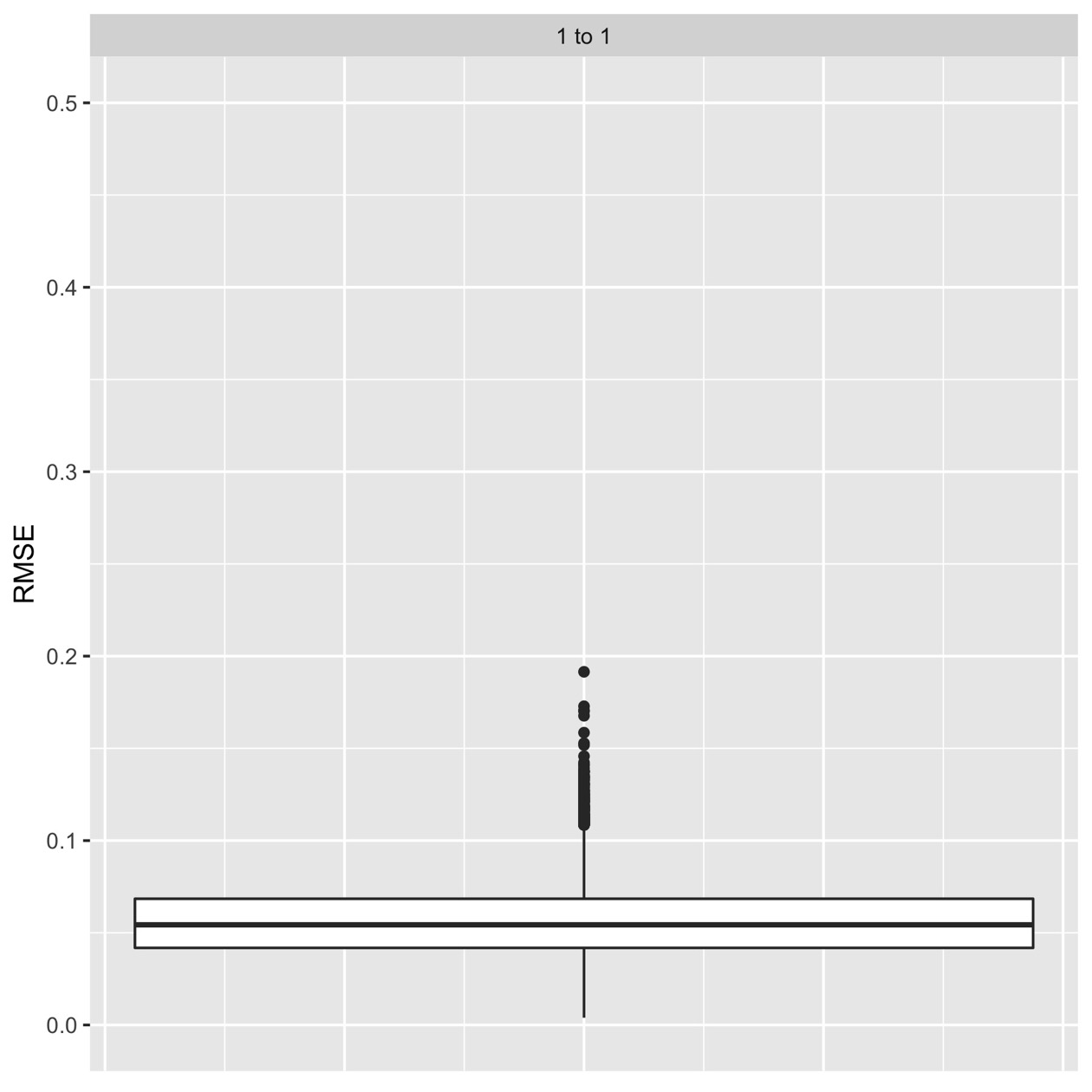}
        \caption{50 days}
    \end{subfigure}
    \hfill
    \begin{subfigure}[b]{0.23\textwidth}
        \centering
        \includegraphics[width=\textwidth]{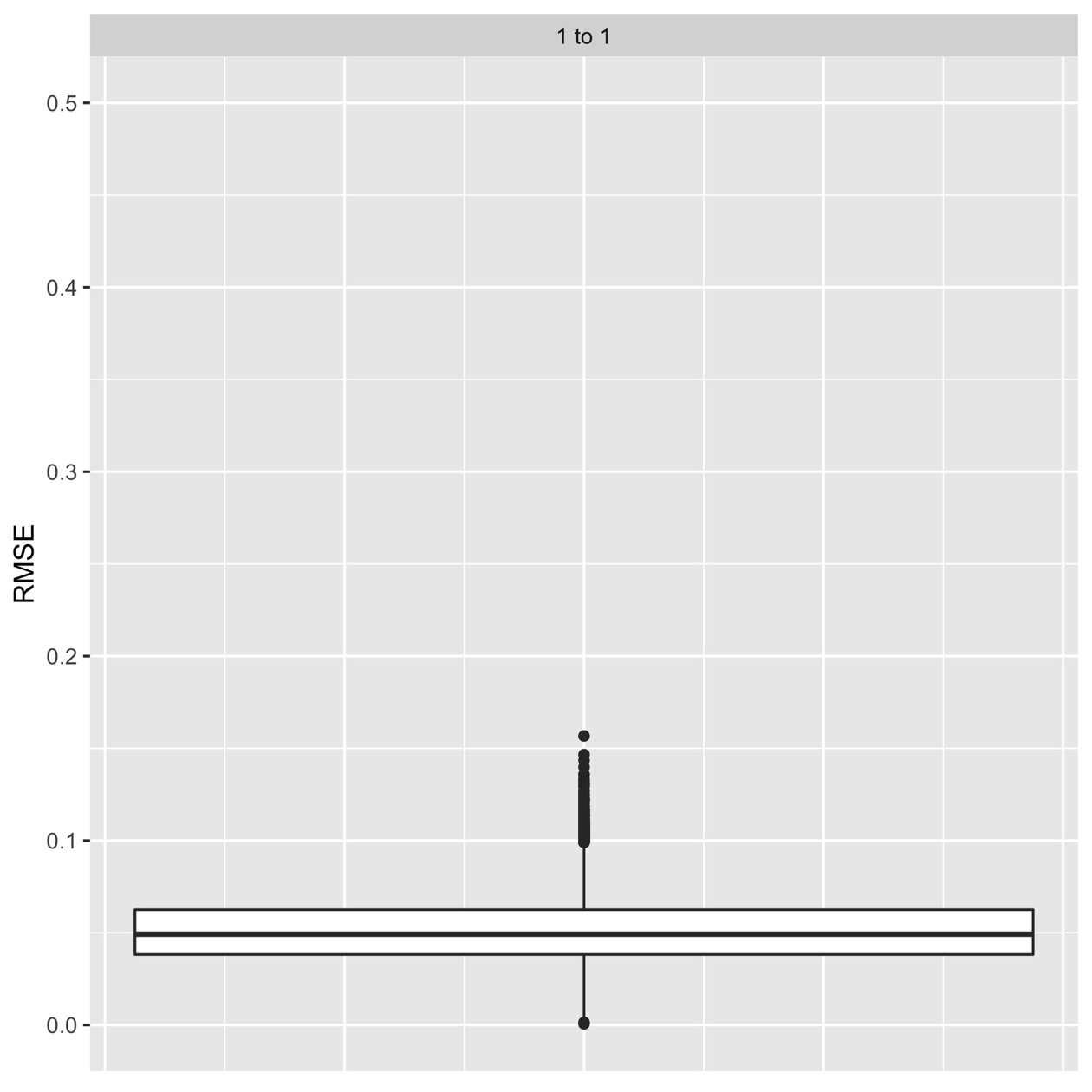}
        \caption{100 days}
    \end{subfigure}
    \hfill
    \begin{subfigure}[b]{0.23\textwidth}
        \centering
        \includegraphics[width=\textwidth]{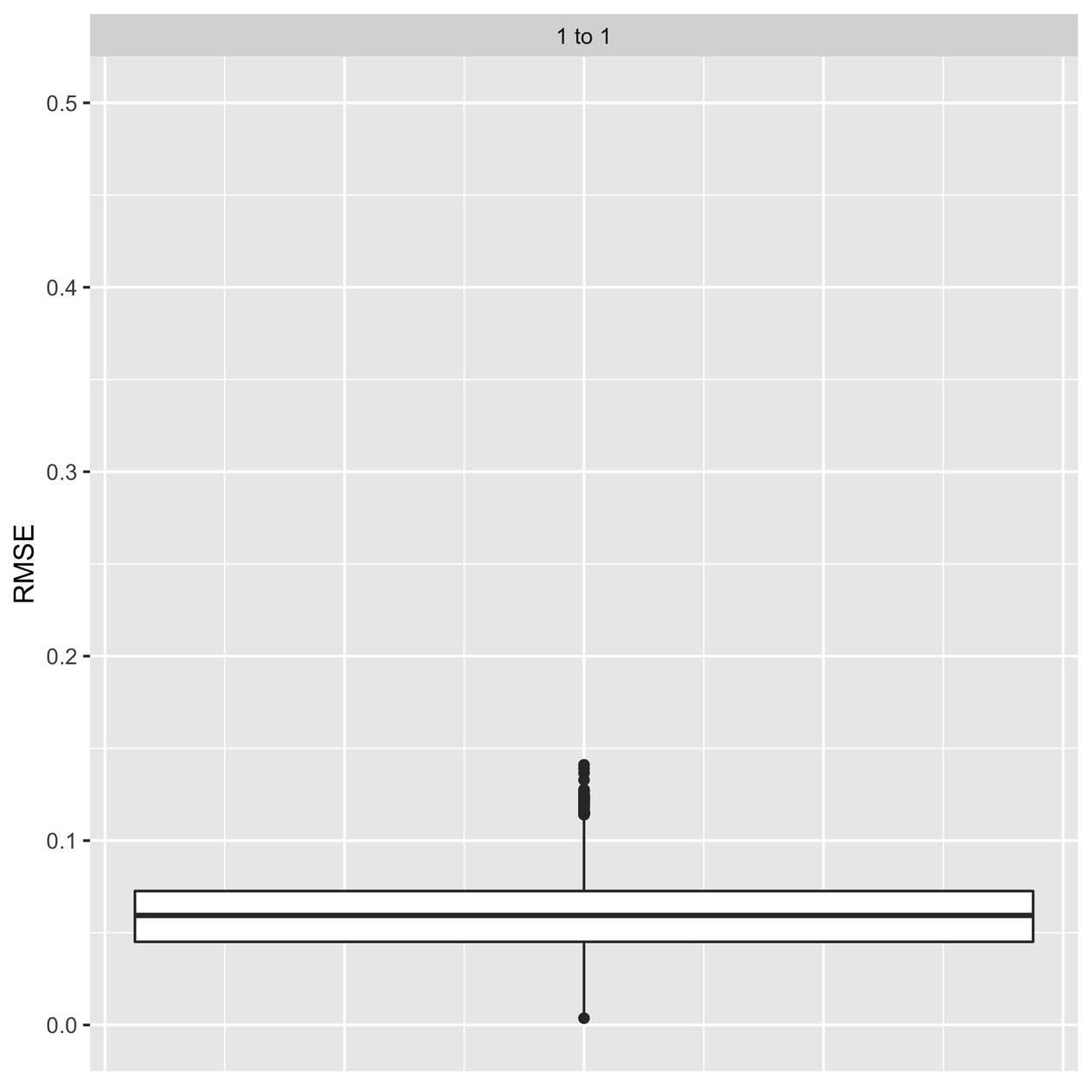}
        \caption{200 days}
    \end{subfigure}
    \hfill
    \begin{subfigure}[b]{0.23\textwidth}
        \centering
        \includegraphics[width=\textwidth]{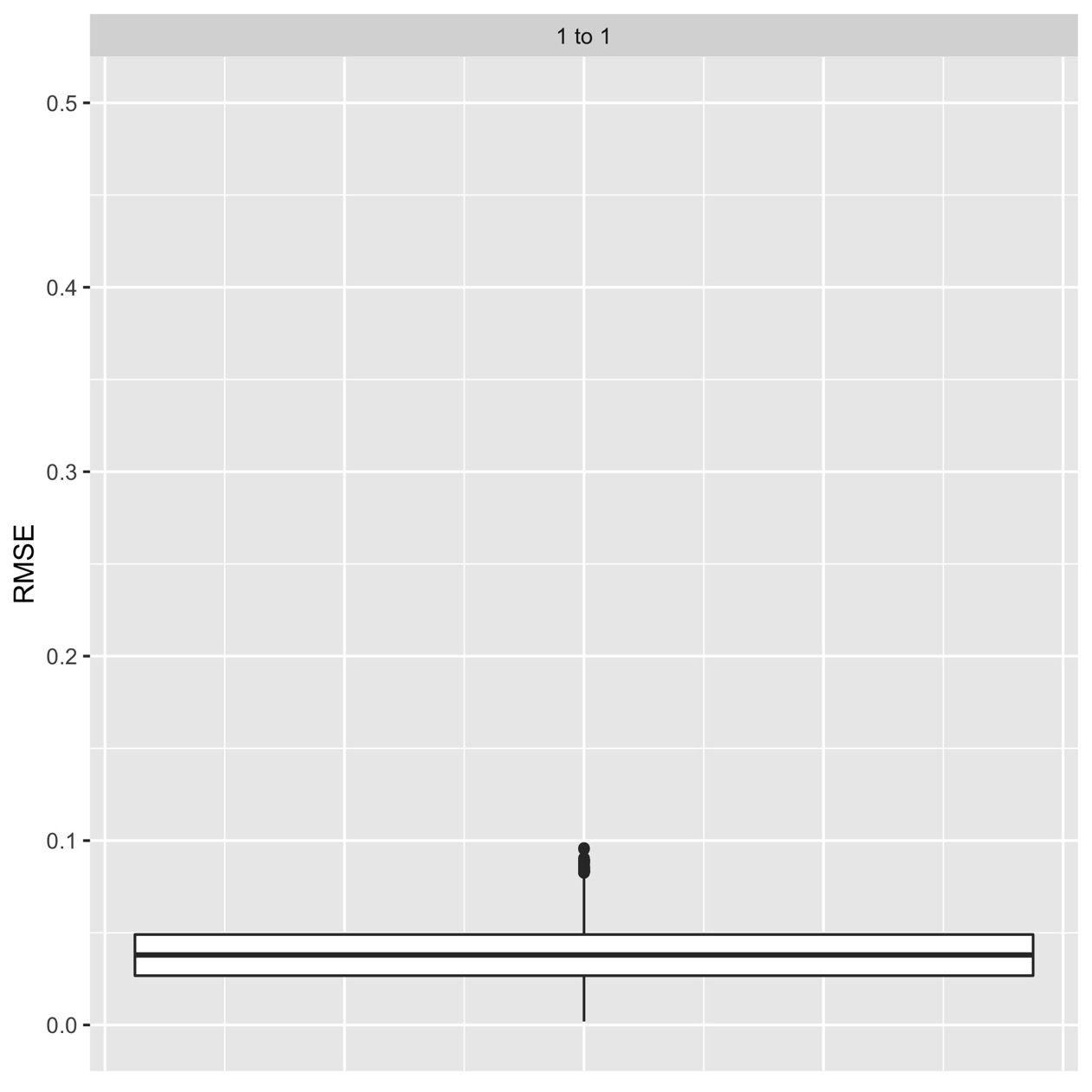}
        \caption{500 days}
    \end{subfigure}
	\end{subfigure}    
	\caption[] 
	{\small Boxplots of the root mean squared error (comparing the estimated triggering kernel for each posterior sample to the true histogram) for the informative prior setting.}
	\label{fig:rmse_1d_inf}
\end{figure}

\begin{figure}[H]
    \centering
    \begin{subfigure}{\textwidth}
    \begin{subfigure}[b]{0.23\textwidth}
        \centering
        \includegraphics[width=\textwidth]{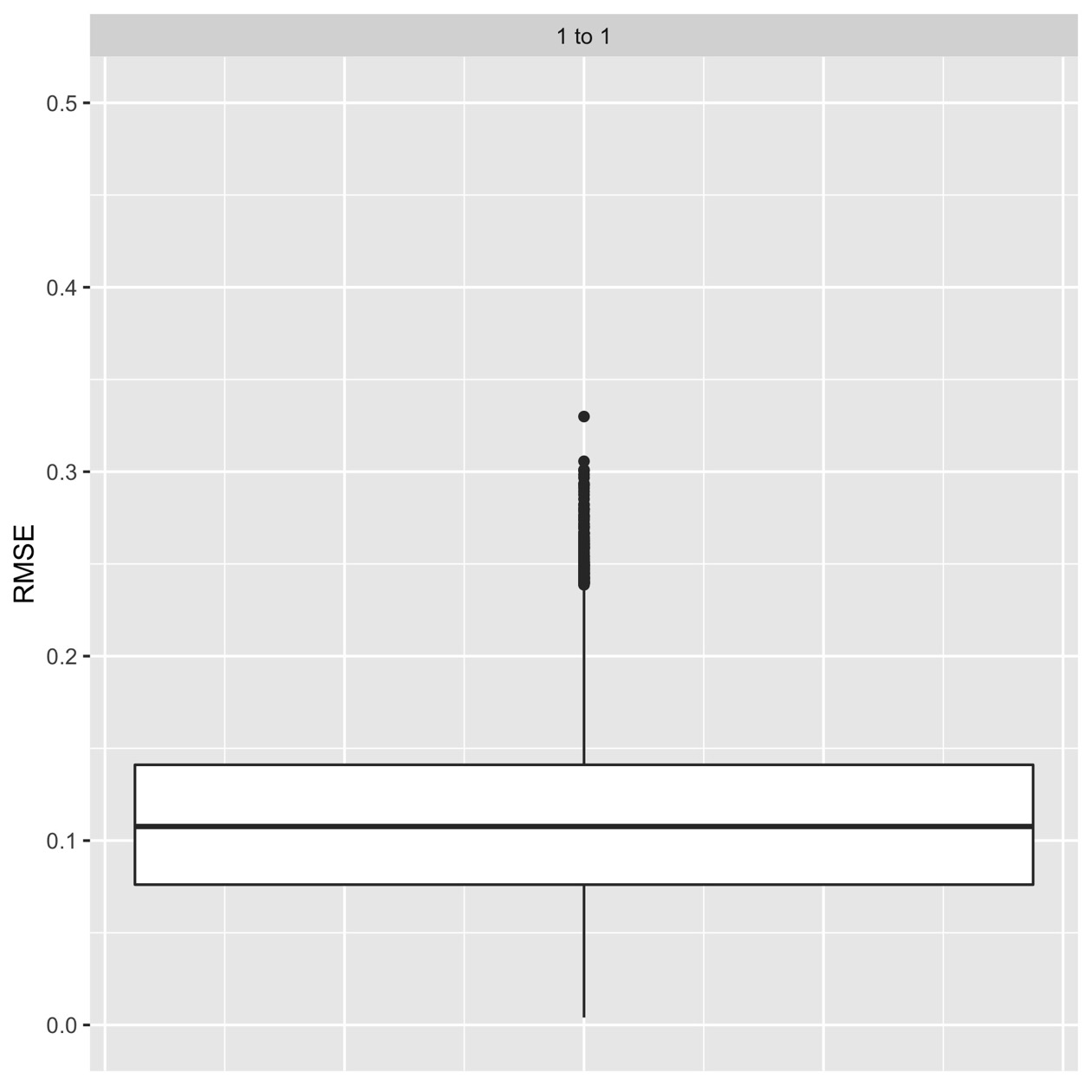}
        \caption{50 days}
    \end{subfigure}
    \hfill
    \begin{subfigure}[b]{0.23\textwidth}
        \centering
        \includegraphics[width=\textwidth]{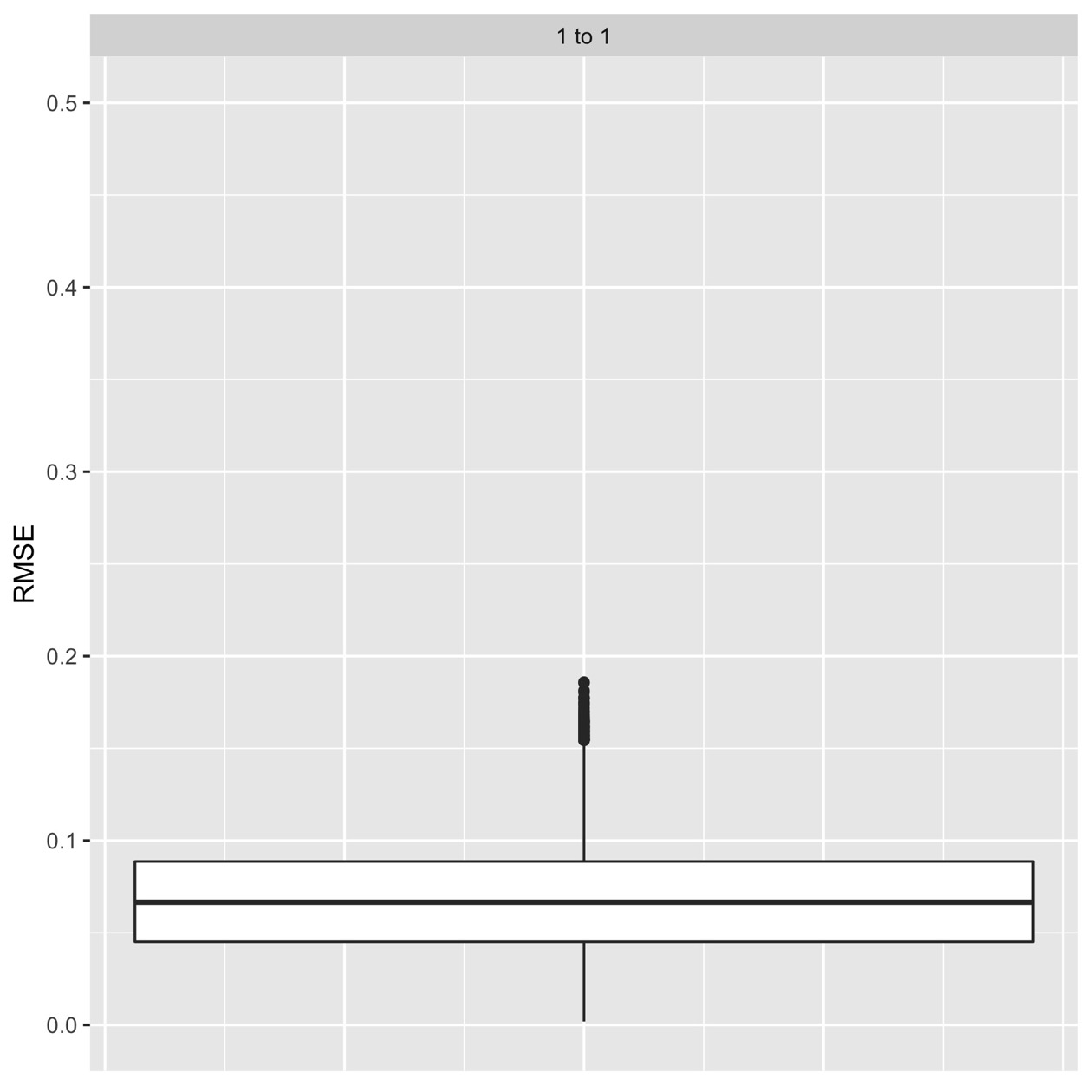}
        \caption{100 days}
    \end{subfigure}
    \hfill
    \begin{subfigure}[b]{0.23\textwidth}
        \centering
        \includegraphics[width=\textwidth]{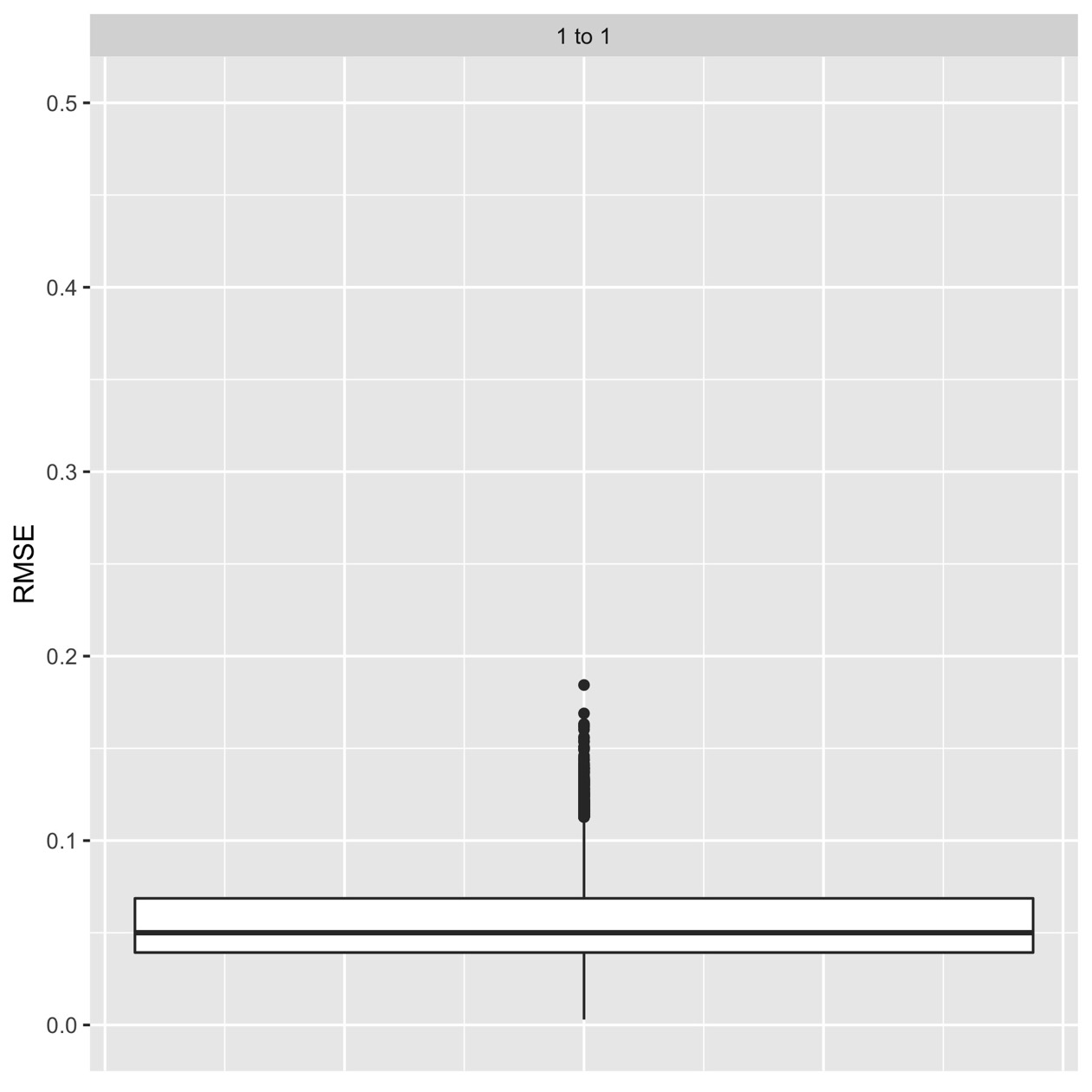}
        \caption{200 days}
    \end{subfigure}
    \hfill
    \begin{subfigure}[b]{0.23\textwidth}
        \centering
        \includegraphics[width=\textwidth]{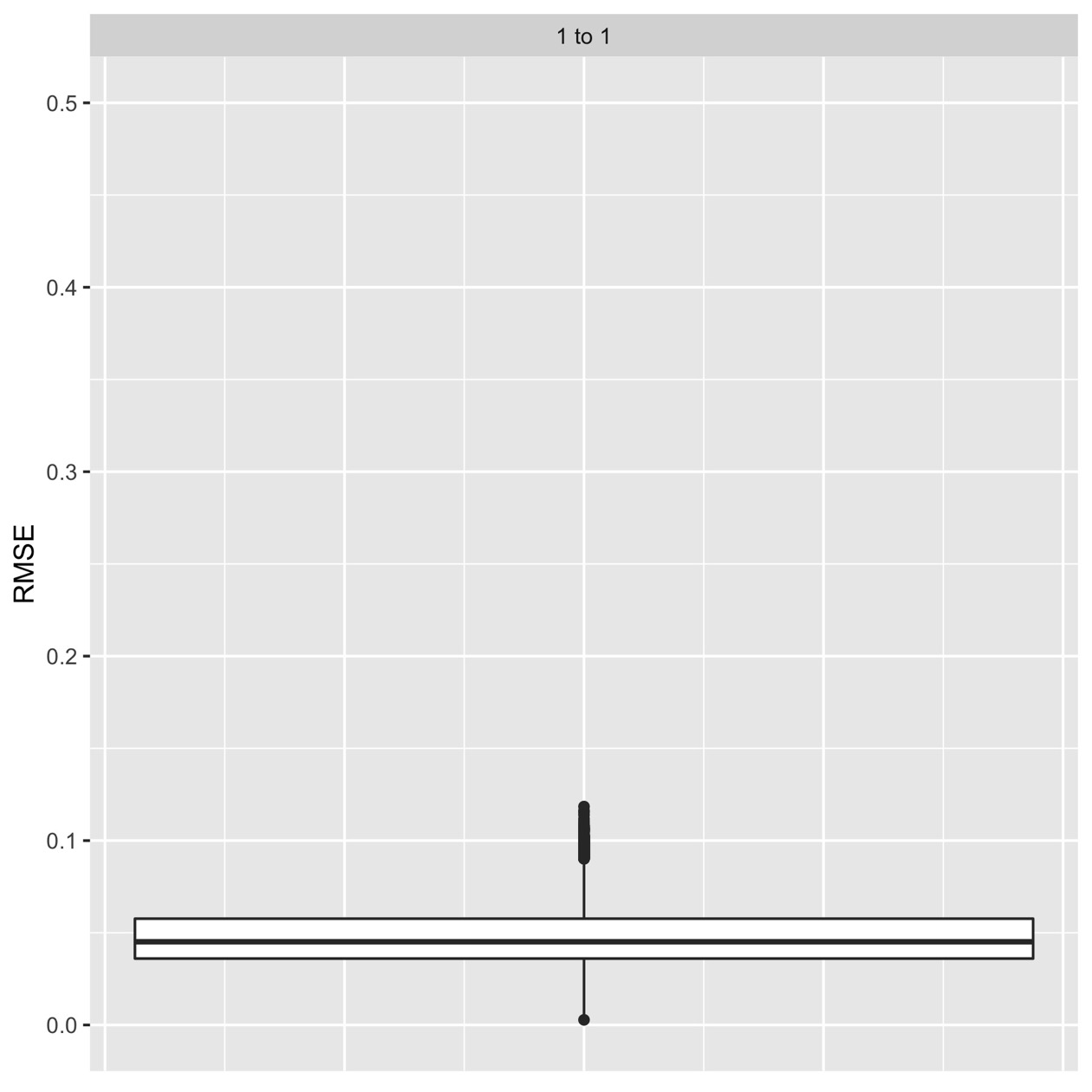}
        \caption{500 days}
    \end{subfigure}
	\end{subfigure}    
	\caption[] 
	{\small Boxplots of the root mean squared error (comparing the estimated triggering kernel for each posterior sample to the true histogram) for the relatively informative prior setting.}
	\label{fig:rmse_1d_relinf}
\end{figure}

\begin{figure}[H]
    \centering
    \begin{subfigure}{\textwidth}
    \begin{subfigure}[b]{0.23\textwidth}
        \centering
        \includegraphics[width=\textwidth]{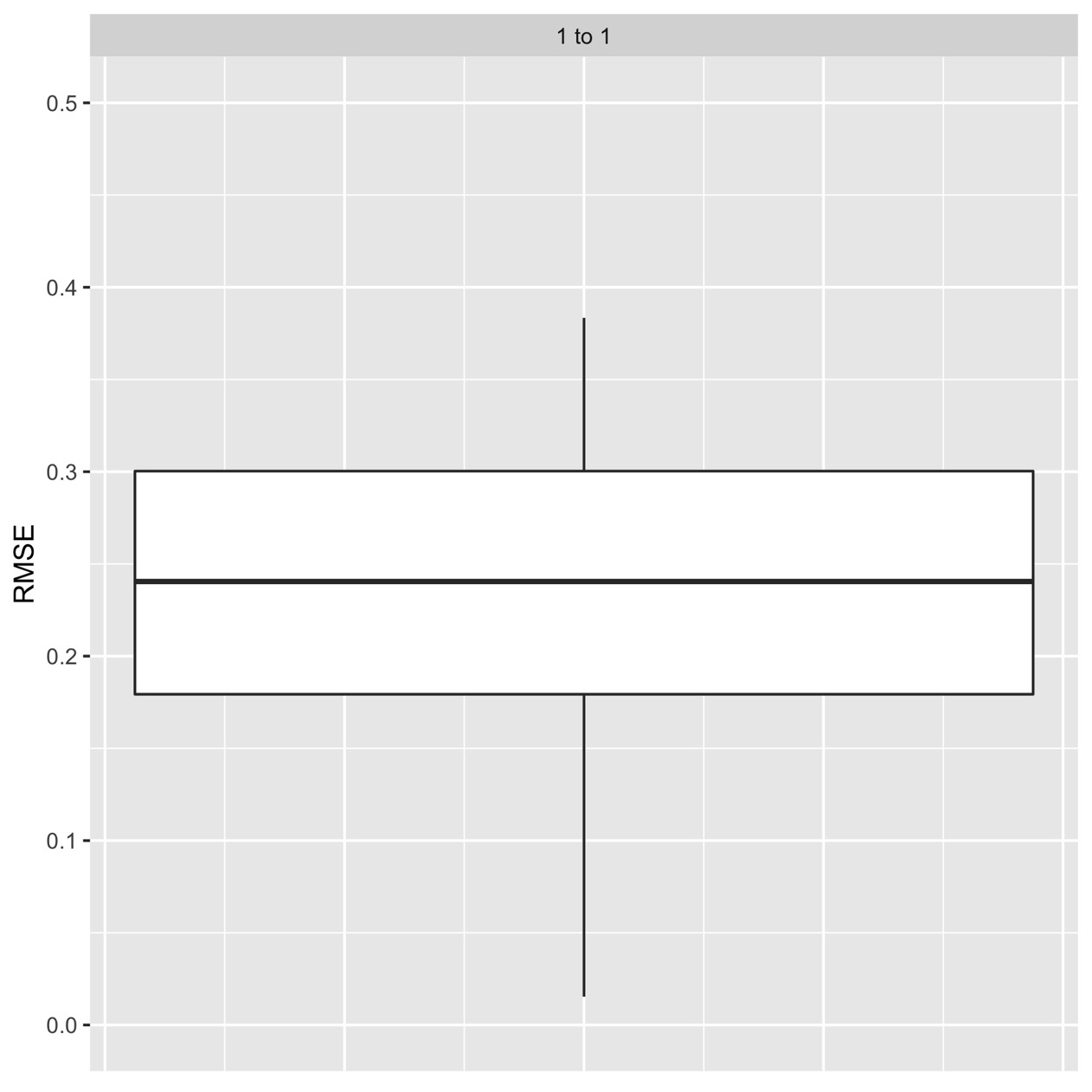}
        \caption{50 days}
    \end{subfigure}
    \hfill
    \begin{subfigure}[b]{0.23\textwidth}
        \centering
        \includegraphics[width=\textwidth]{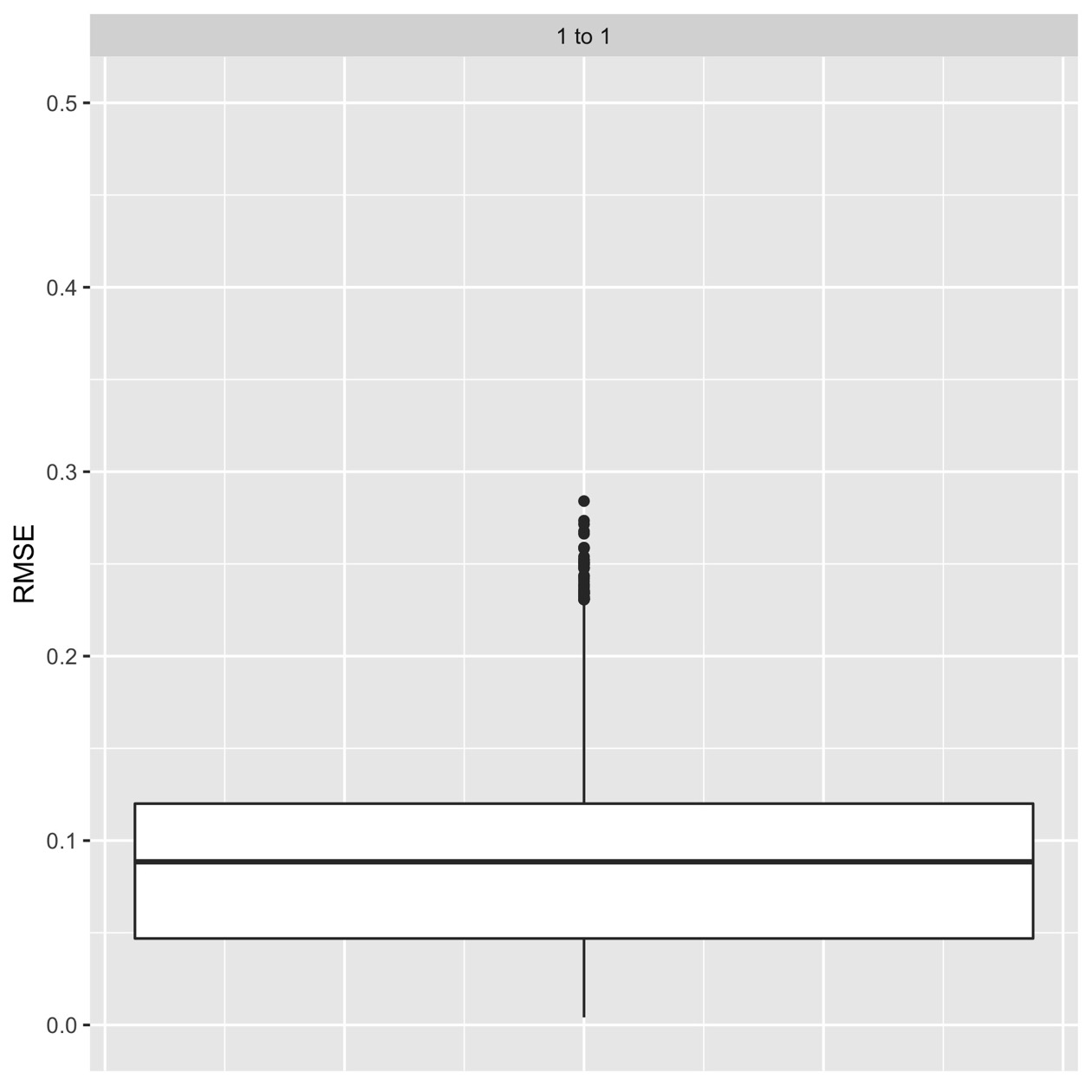}
        \caption{100 days}
    \end{subfigure}
    \hfill
    \begin{subfigure}[b]{0.23\textwidth}
        \centering
        \includegraphics[width=\textwidth]{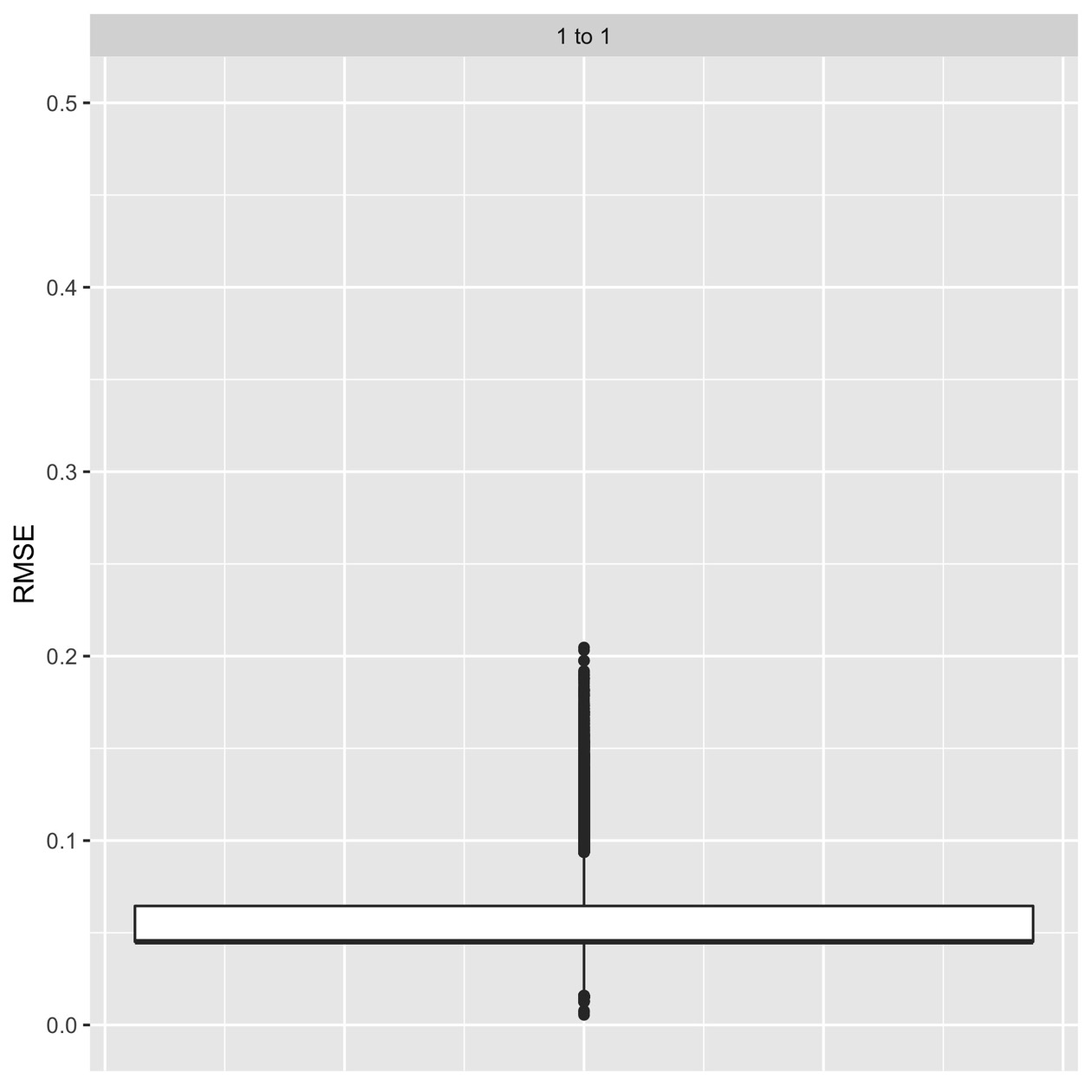}
        \caption{200 days}
    \end{subfigure}
    \hfill
    \begin{subfigure}[b]{0.23\textwidth}
        \centering
        \includegraphics[width=\textwidth]{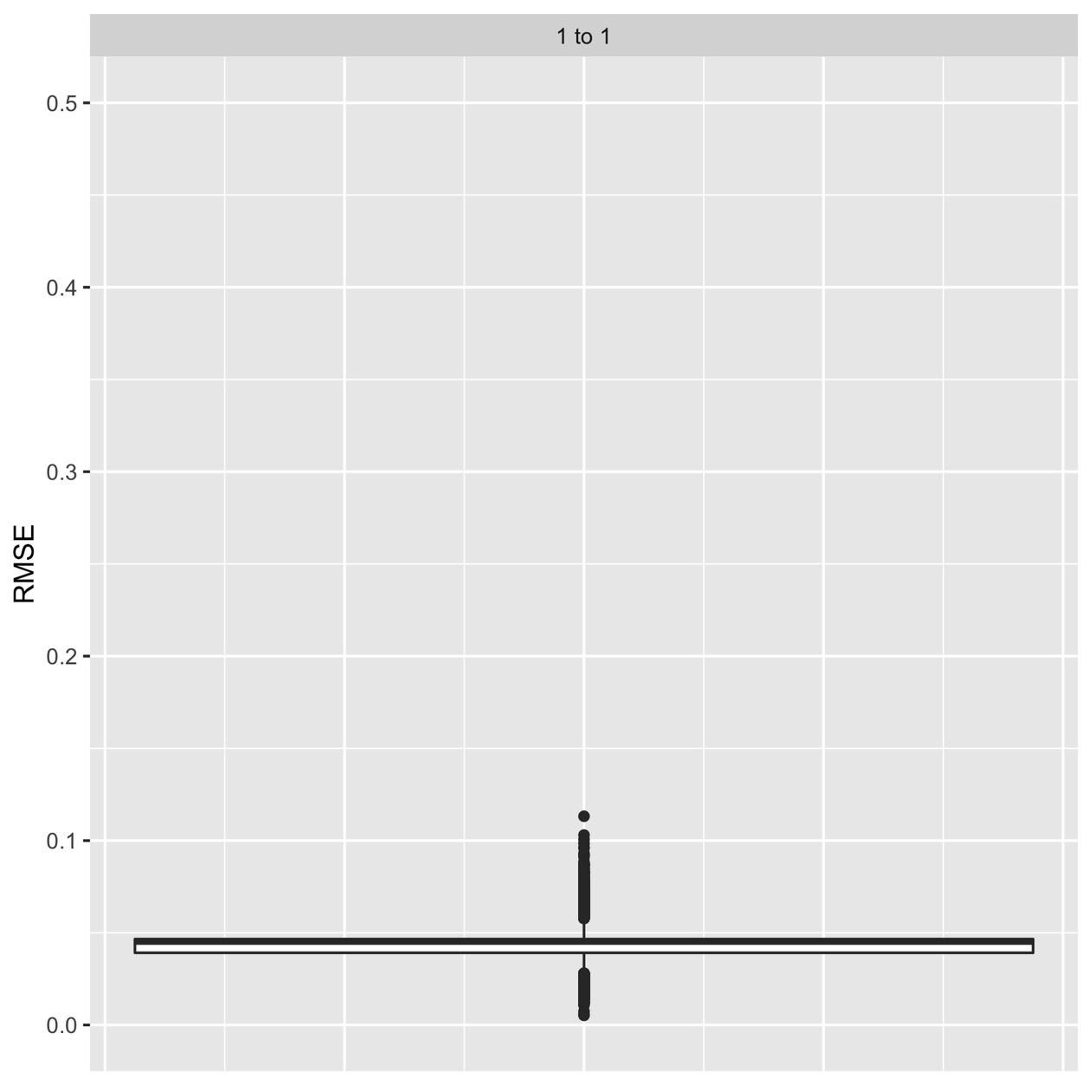}
        \caption{500 days}
    \end{subfigure}
	\end{subfigure}    
	\caption[] 
	{\small Boxplots of the root mean squared error (comparing the estimated triggering kernel for each posterior sample to the true histogram) for the quite uninformative prior setting.}
	\label{fig:rmse_1d_uninf}
\end{figure}

Figures \ref{fig:1d_inf} - \ref{fig:1d_uninf} present the estimated histogram functions for each combination of prior setting and time series length. Furthermore, Figures \ref{fig:rmse_1d_inf} - \ref{fig:rmse_1d_uninf} present boxplots of the root mean squared error between the estimated triggering kernel for each posterior sample and the true kernel. The proposed model demonstrates good performance for the informative and relatively informative prior settings, with the true histogram function predominantly contained within the 80\% posterior intervals. The RMSE does not change significantly in the informative prior setting. In the latter, more relaxed prior setting, posterior accuracy improves and RMSE reduces as the time series length increases. Completely relaxing the priors for the quite uninformative prior setting results in a significant deterioration in model performance and convergence of the MCMC algorithm when the time series length is short. However, this also improves when a longer time series is considered. Moreover, it is perhaps surprising that the credible region in this scenario is smaller than for the informative prior setting. However, this is likely attributed to randomness in the algorithm, since this phenomenon did not occur when other independent simulations were performed using this setting. In summary, a more informative prior is required for shorter time series.

The static parameters tell the same story. Good mixing and convergence of the MCMC algorithm is noted in the scenarios involving informative and relatively informative priors. 
We also see a deterioration in performance for the quite uninformative scenario, with slow mixing and indications that the chains have not converged. Longer chains were not explored since reversible-jump MCMC is computationally expensive; thus when moving to higher dimensions longer chains are not a realistic option. As a result, inference under the quite uninformative setting was deemed unreliable. Further details on MCMC convergence diagnostics for the static parameters are provided in the supplementary material. 

\subsubsection{Repeated samples}
 
The previous analysis revealed that while the true histogram functions are generally within the 80\% posterior intervals, the variability in these intervals often displays unexpected behaviour, particularly for shorter time series. One would expect the intervals generally to be monotonically decreasing due to the shape of the true excitation function; however, this is often not the case. This can be attributed to random error. Thus we repeat the analysis, increasing the number of parallel chains to 90 for a given realisation of the process.These chains are then combined by taking the last 30,000 posterior samples (out of 60,000 samples).

\begin{figure}[H]
    \centering
    \begin{subfigure}{0.5\textwidth}
    \begin{subfigure}[b]{0.45\textwidth}
        \centering
        \includegraphics[width=\textwidth]{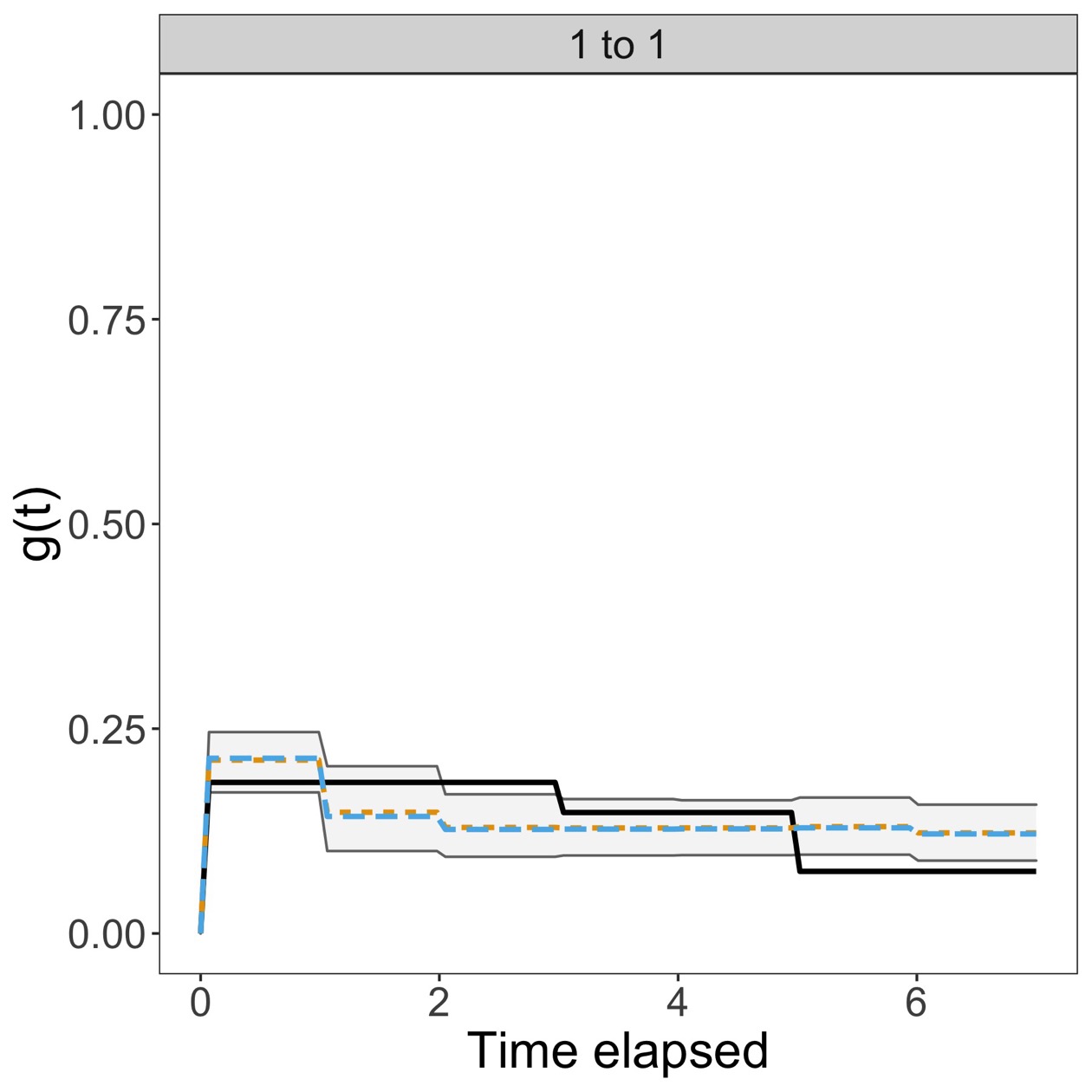}
        \caption{50 days}
    \end{subfigure}
    \hfill
    \begin{subfigure}[b]{0.45\textwidth}
        \centering
        \includegraphics[width=\textwidth]{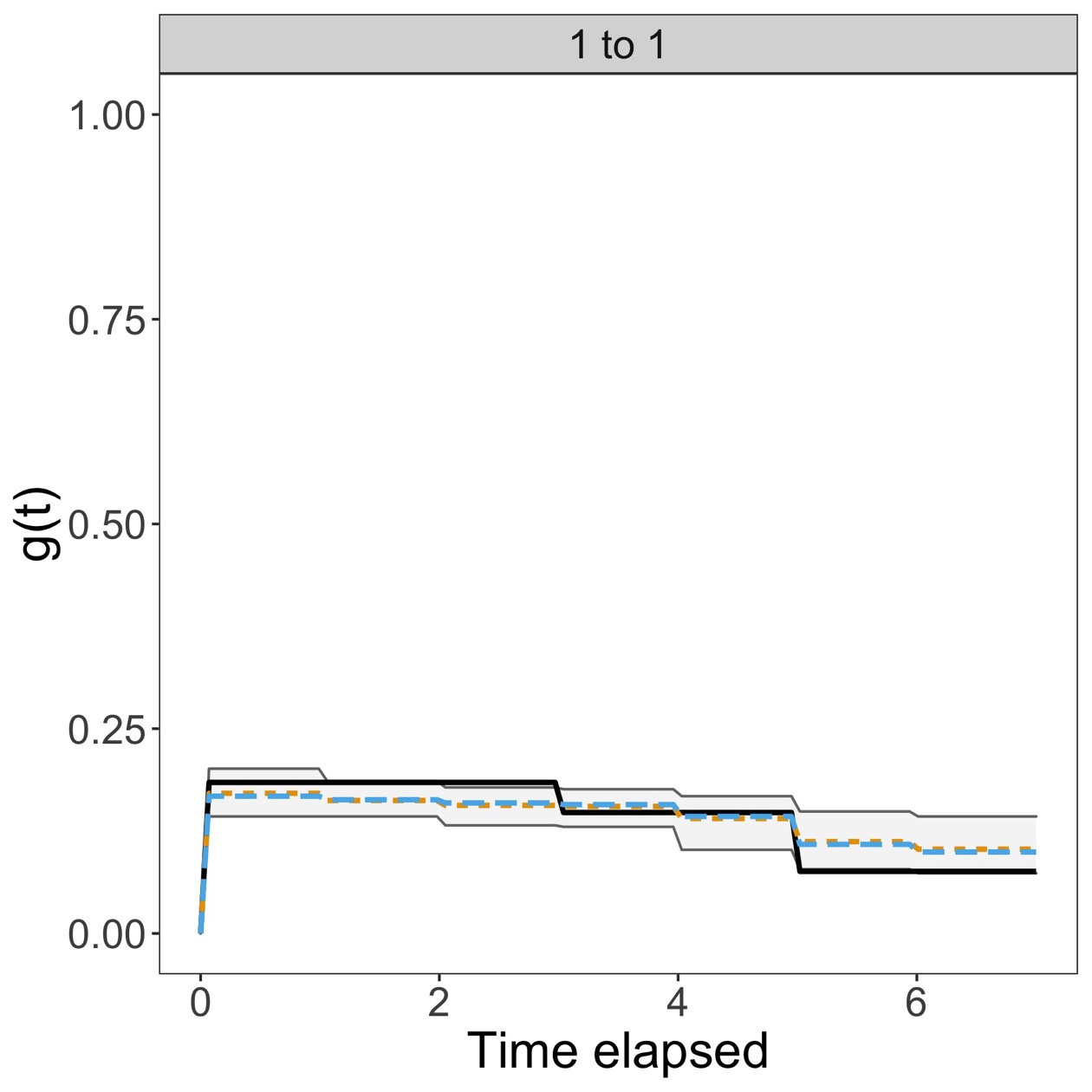}
        \caption{500 days}
    \end{subfigure}
	\end{subfigure}    
	\caption[] 
	{\small Estimated excitation function under the informative prior setting with repeated samples.\\ \textbf{Solid black line:} true histogram function. \textbf{Dashed orange line:} posterior mean.\\ \textbf{Dashed blue line:} posterior median. \textbf{Grey ribbon:} 80\% posterior interval.}
	\label{fig:1d_inf_multiple}
\end{figure}

\begin{figure}[H]
    \centering
    \begin{subfigure}{0.5\textwidth}
    \begin{subfigure}[b]{0.45\textwidth}
        \centering
        \includegraphics[width=\textwidth]{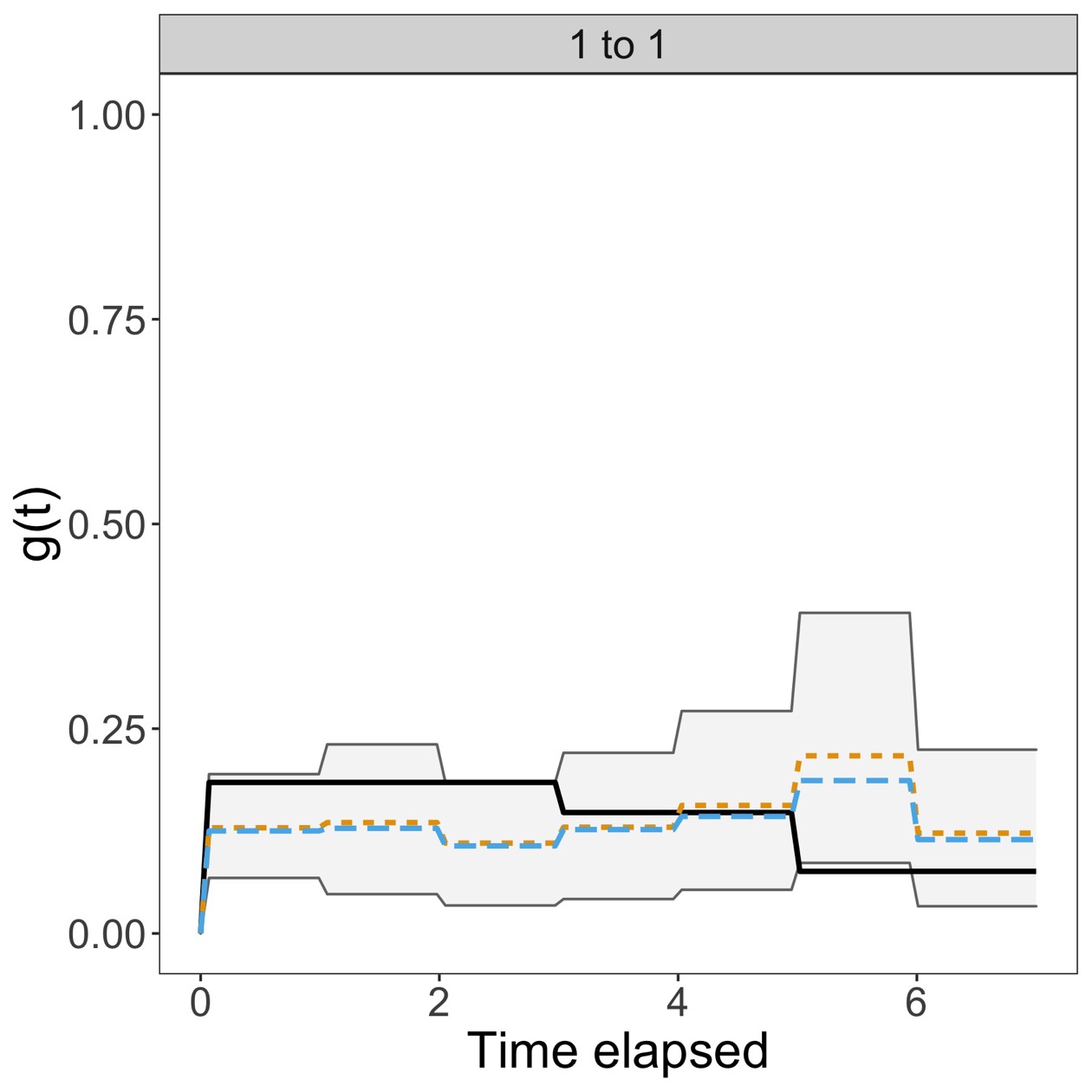}
        \caption{50 days}
    \end{subfigure}
    \hfill
    \begin{subfigure}[b]{0.45\textwidth}
        \centering
        \includegraphics[width=\textwidth]{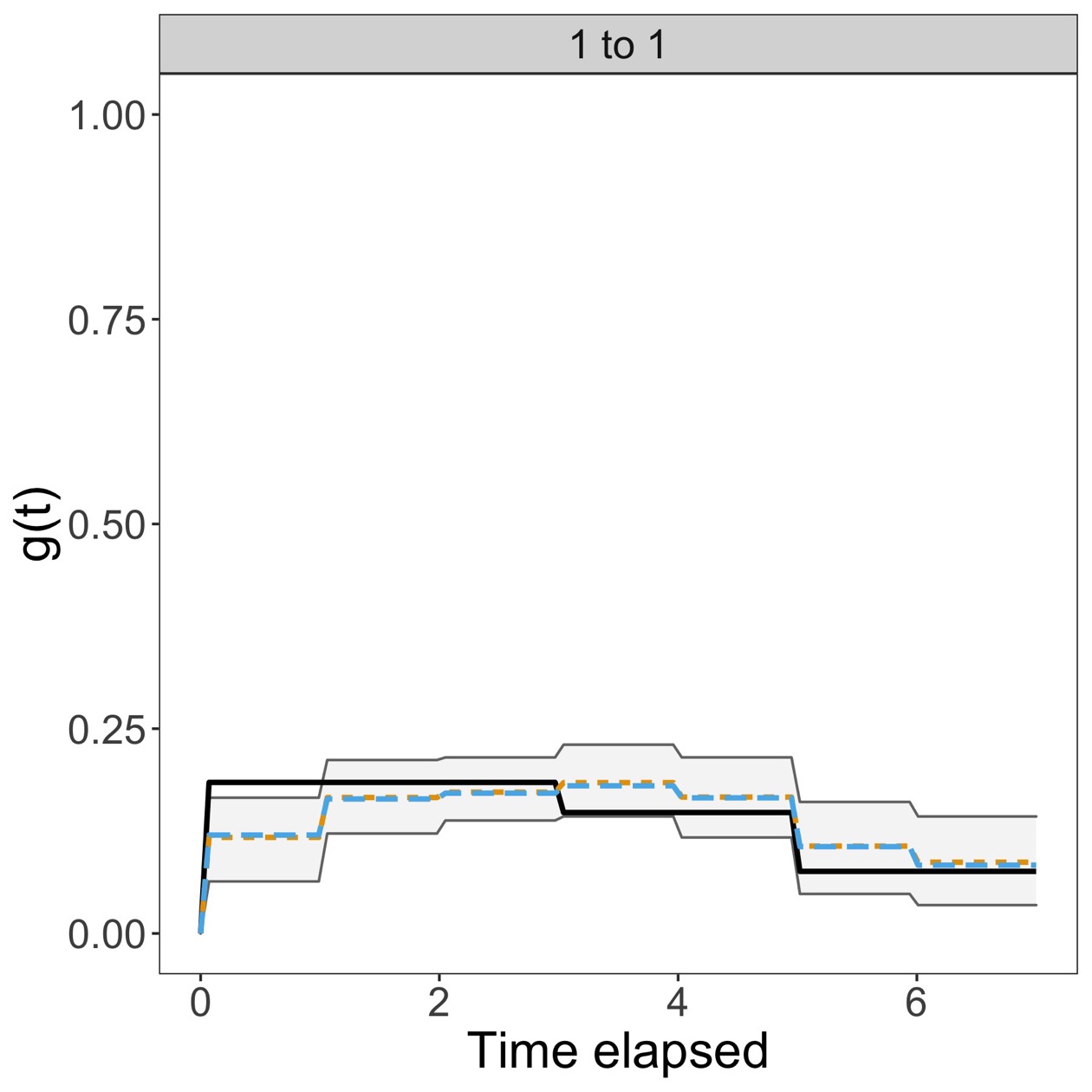}
        \caption{500 days}
    \end{subfigure}
	\end{subfigure}    
	\caption[] 
	{\small Estimated excitation function under the relatively informative prior setting with repeated samples.\\ \textbf{Solid black line:} true histogram function. \textbf{Dashed orange line:} posterior mean.\\ \textbf{Dashed blue line:} posterior median. \textbf{Grey ribbon:} 80\% posterior interval.}
	\label{fig:1d_relinf_multiple}
\end{figure}

\begin{figure}[H]
    \centering
    \begin{subfigure}{0.5\textwidth}
    \begin{subfigure}[b]{0.45\textwidth}
        \centering
        \includegraphics[width=\textwidth]{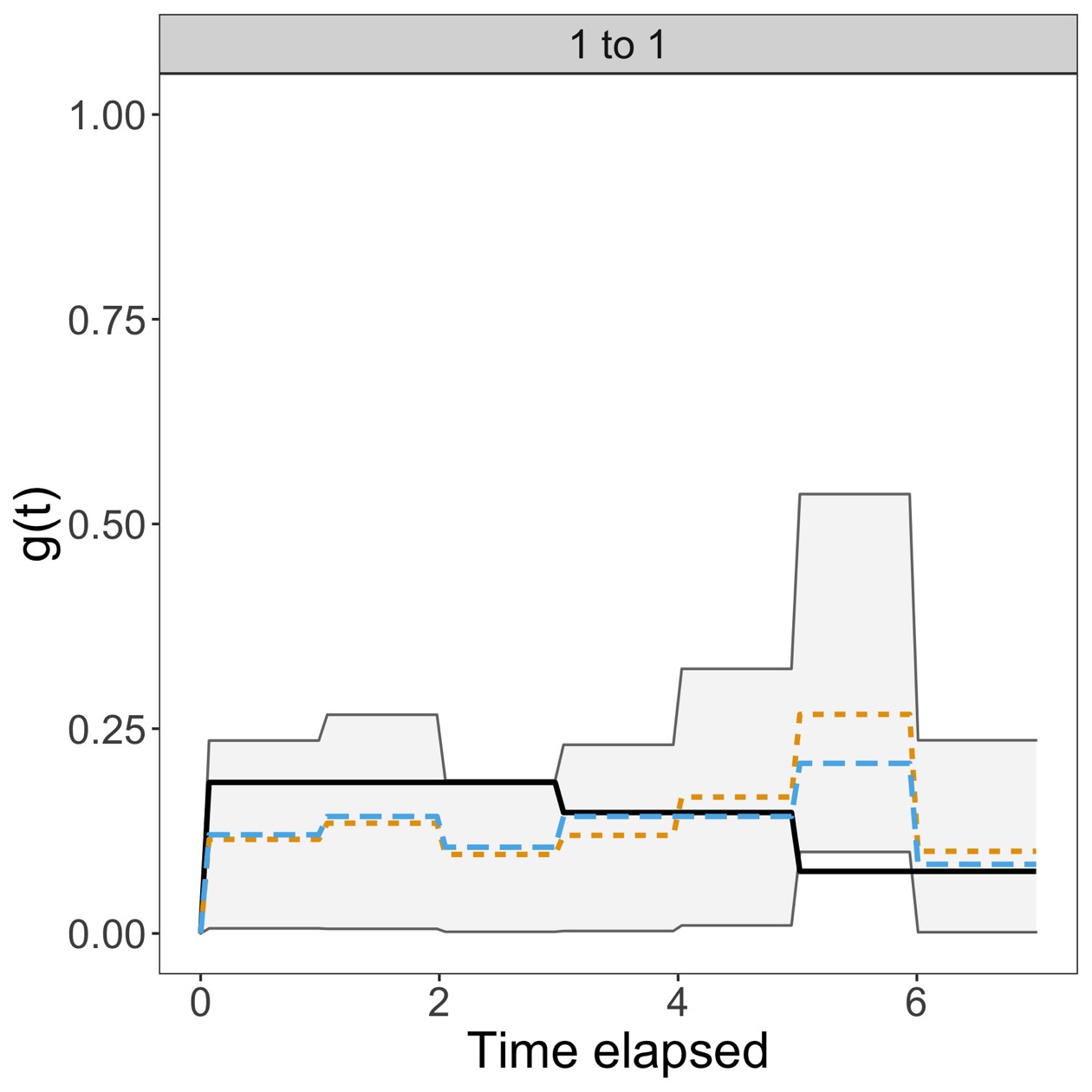}
        \caption{50 days}
    \end{subfigure}
    \hfill
    \begin{subfigure}[b]{0.45\textwidth}
        \centering
        \includegraphics[width=\textwidth]{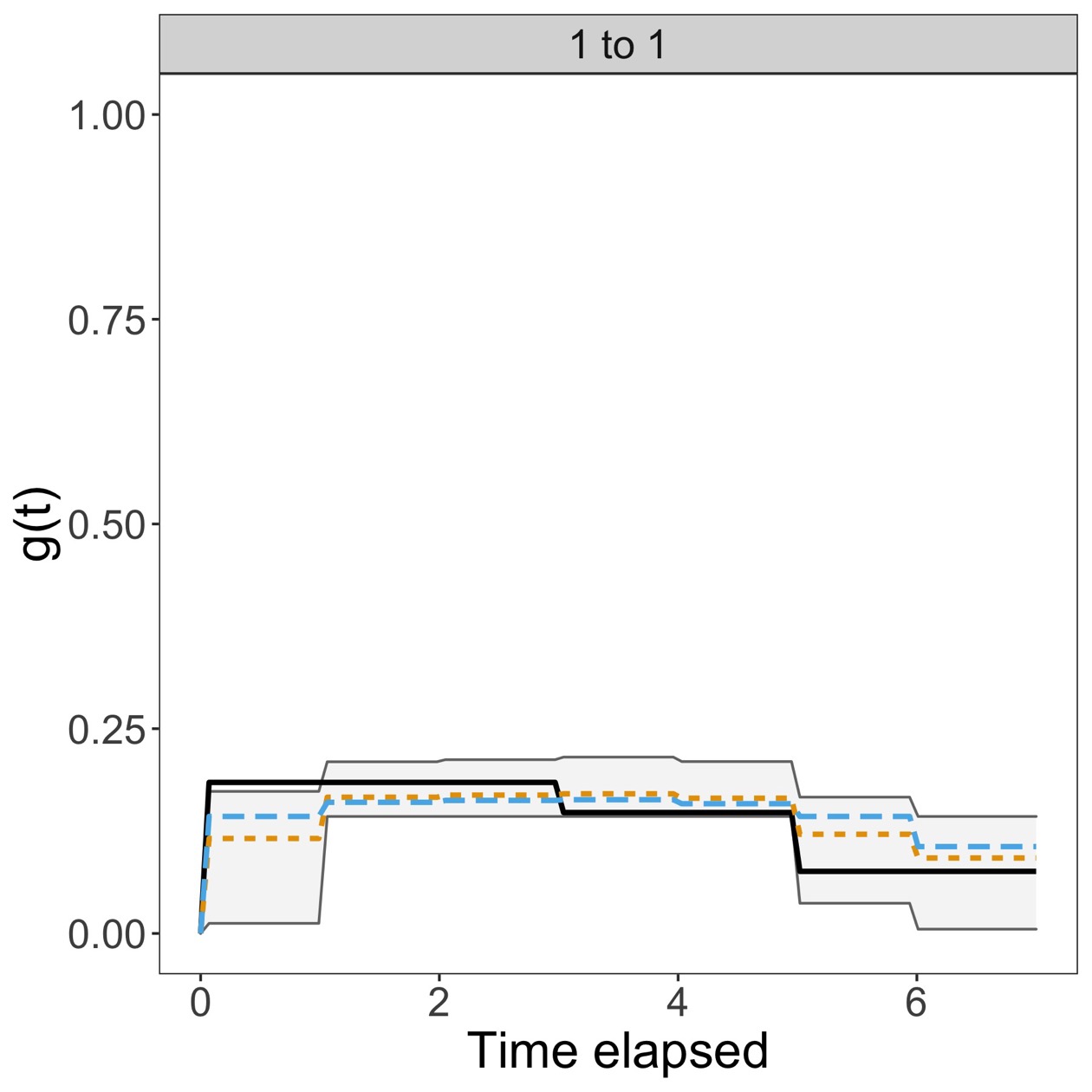}
        \caption{500 days}
    \end{subfigure}
	\end{subfigure}    
	\caption[] 
	{\small Estimated excitation function under the quite uninformative prior setting with repeated samples. \\ \textbf{Solid black line:} true histogram function. \textbf{Dashed orange line:} posterior mean.\\ \textbf{Dashed blue line:} posterior median. \textbf{Grey ribbon:} 80\% posterior interval.}
	\label{fig:1d_uninf_multiple}
\end{figure}

Figures \ref{fig:1d_inf_multiple} - \ref{fig:1d_uninf_multiple} display the estimated histograms from this analysis. Increasing the number of chains results in posterior distributions that better encompass the true parameters. Furthermore, as was expected, the posterior intervals are more consistent and predictable as the sample size or prior setting varies. 


 \subsubsection{Comparison to parametric triggering kernel}
 
 The flexibility afforded in the proposed model comes at the expense of computational efficiency. However in many real-world scenarios, it is likely that a simple parametric form for the triggering kernel will be insufficient to capture the true complexity within the process. Thus to motivate this work, we consider a comparison to a common parametric triggering kernel for a discrete parameter space, namely the geometric kernel given by,
 
 \begin{equation}
 	g(t-t_i) = \beta (1-\beta)^{(t-t_i-1)}
 \end{equation}
 where $\beta$ is the inverse mean of the geometric distribution. The kernel is truncated at $t-t_i = 7$ to align with the proposed histogram kernel. If the geometric kernel has no such truncation and does not have compact support, despite a simpler form, computational expense is significantly more than our proposed model.
 
Simulated data are generated based on three true triggering kernels: (1) a geometric kernel, (2) a histogram kernel that is monotonically decreasing in a fashion similar to that of a geometric kernel, and (3) a histogram kernel that peaks at the centre. For each of these three simulation settings, we fit the data first to the model with the proposed flexible histogram kernel, and then the model assuming the geometric kernel. The analysis is performed as described in Section \ref{sec:uv_sim} using the relatively informative prior setting.
  
\begin{figure}[H]
    \centering
    \begin{subfigure}{0.6\textwidth}
    \begin{subfigure}[b]{0.32\textwidth}
        \centering
        \includegraphics[width=\textwidth]{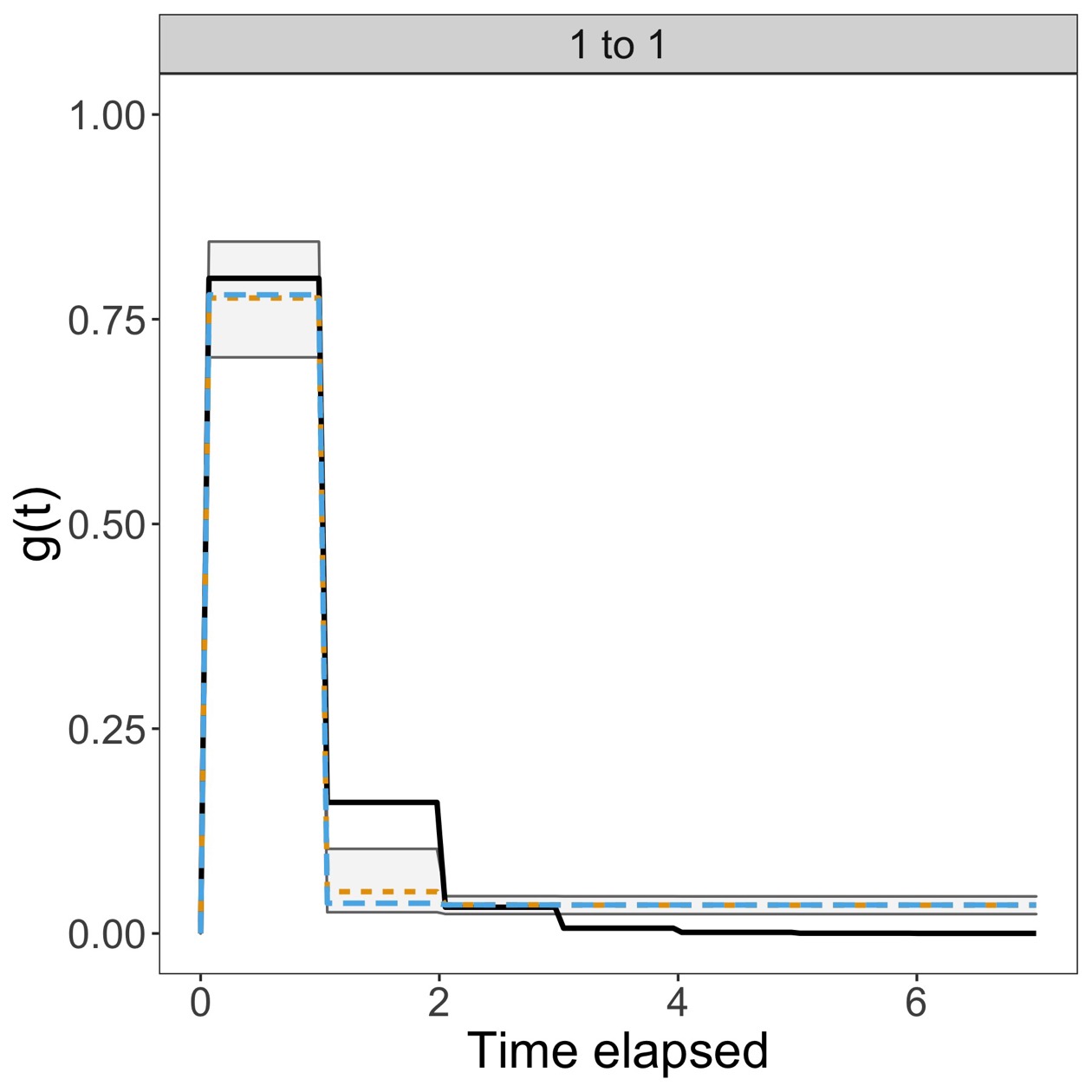}
        \caption{True kernel (1)}
    \end{subfigure}
    \hfill
    \begin{subfigure}[b]{0.32\textwidth}
        \centering
        \includegraphics[width=\textwidth]{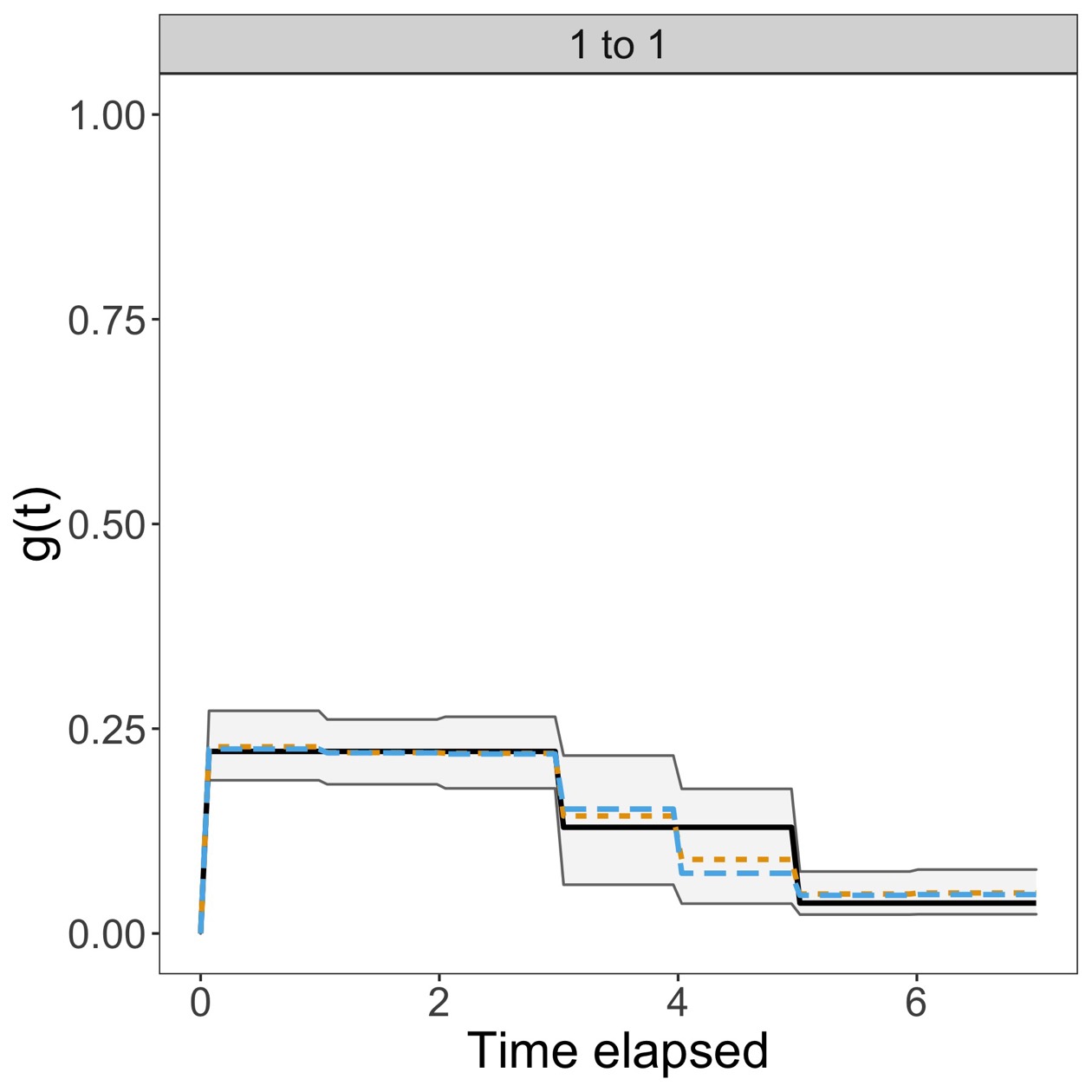}
        \caption{True kernel (2)}
    \end{subfigure}
    \hfill
    \begin{subfigure}[b]{0.32\textwidth}
        \centering
        \includegraphics[width=\textwidth]{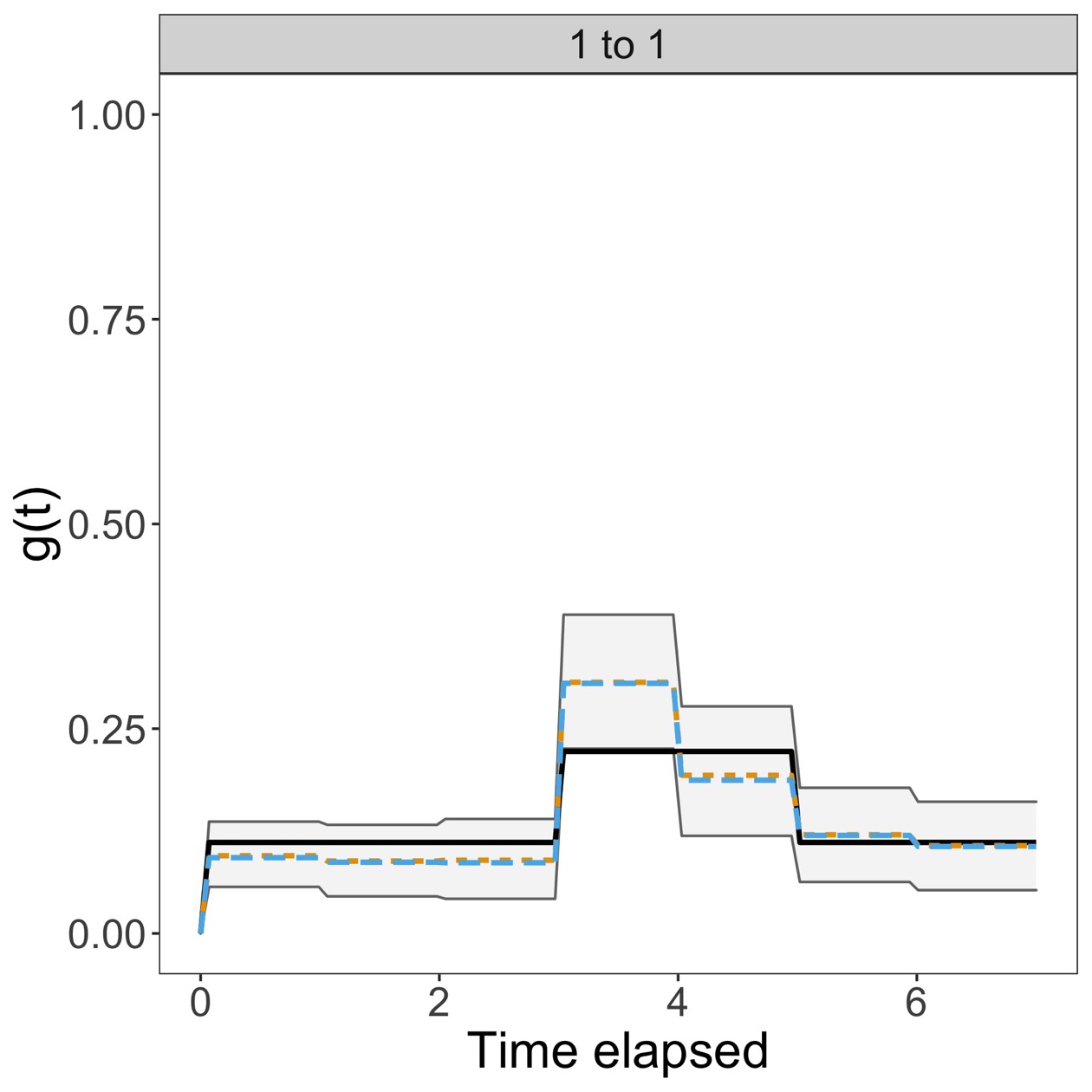}
        \caption{True kernel (3)}
    \end{subfigure}
	\end{subfigure}    
	\caption[] 
	{\small Estimated histogram excitation function. \\ \textbf{Solid black line:} true triggering kernel. \textbf{Dashed orange line:} posterior mean.\\ \textbf{Dashed blue line:} posterior median. \textbf{Grey ribbon:} 80\% posterior interval.}
	\label{fig:hist_model}
\end{figure}

\begin{figure}[H]
    \centering
    \begin{subfigure}{0.6\textwidth}
    \begin{subfigure}[b]{0.32\textwidth}
        \centering
        \includegraphics[width=\textwidth]{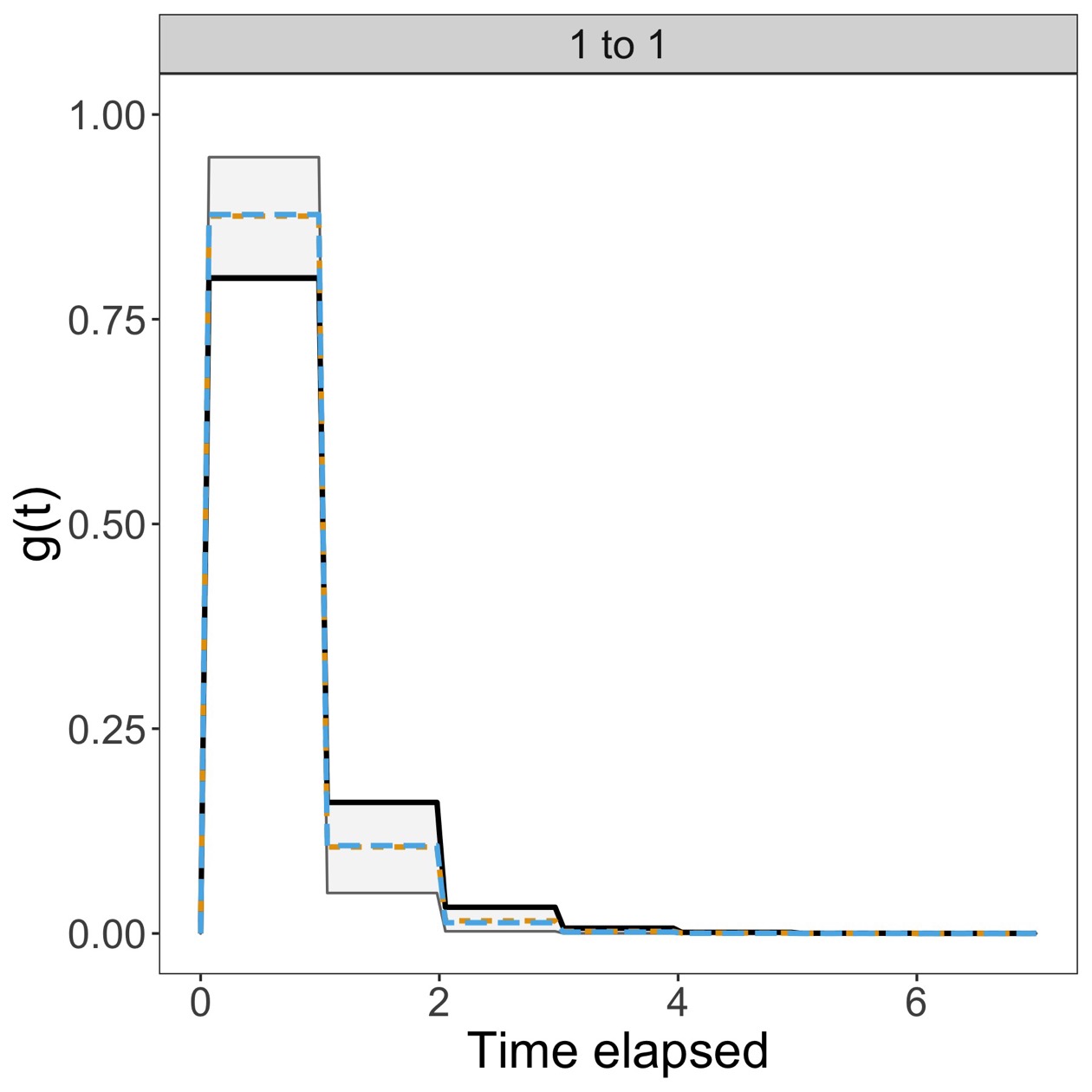}
        \caption{True kernel (1)}
    \end{subfigure}
    \hfill
    \begin{subfigure}[b]{0.32\textwidth}
        \centering
        \includegraphics[width=\textwidth]{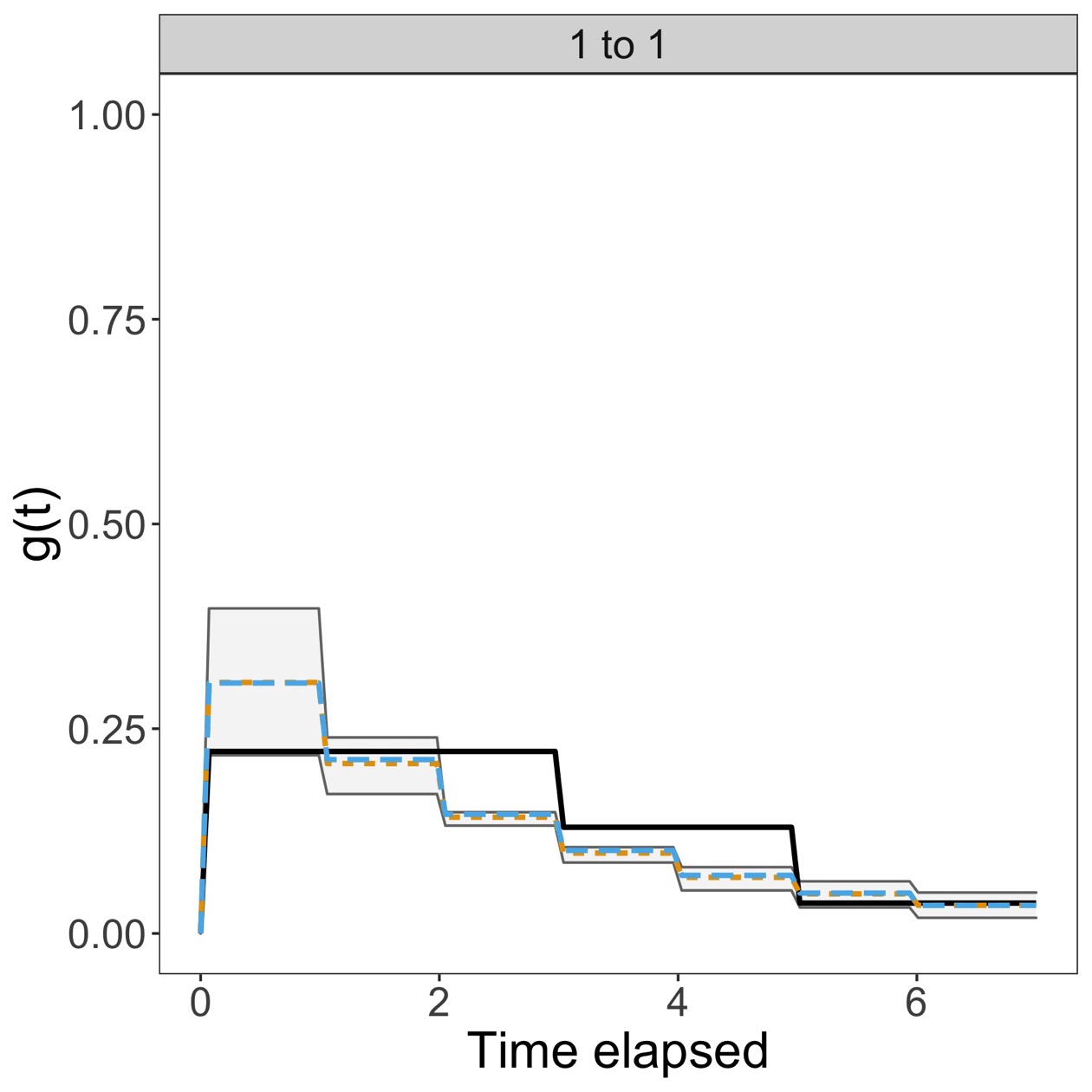}
        \caption{True kernel (2)}
    \end{subfigure}
    \hfill
    \begin{subfigure}[b]{0.32\textwidth}
        \centering
        \includegraphics[width=\textwidth]{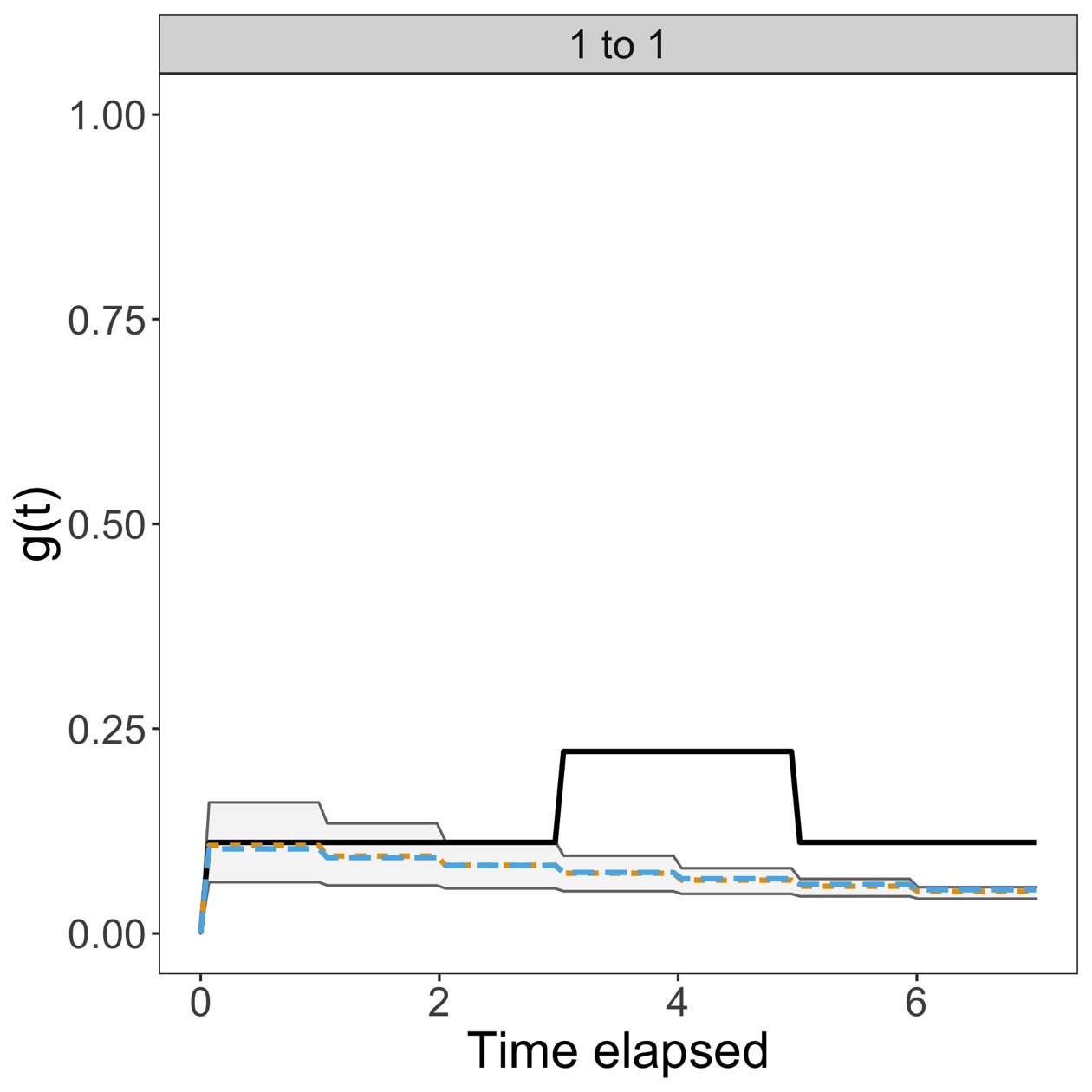}
        \caption{True kernel (3)}
    \end{subfigure}
	\end{subfigure}    
	\caption[] 
	{\small Estimated geometric excitation function. \\ \textbf{Solid black line:} true triggering kernel. \textbf{Dashed orange line:} posterior mean.\\ \textbf{Dashed blue line:} posterior median. \textbf{Grey ribbon:} 80\% posterior interval.}
	\label{fig:geom_model}
\end{figure}

\begin{figure}[H]
    \centering
    \begin{subfigure}{0.6\textwidth}
    \begin{subfigure}[b]{0.32\textwidth}
        \centering
        \includegraphics[width=\textwidth]{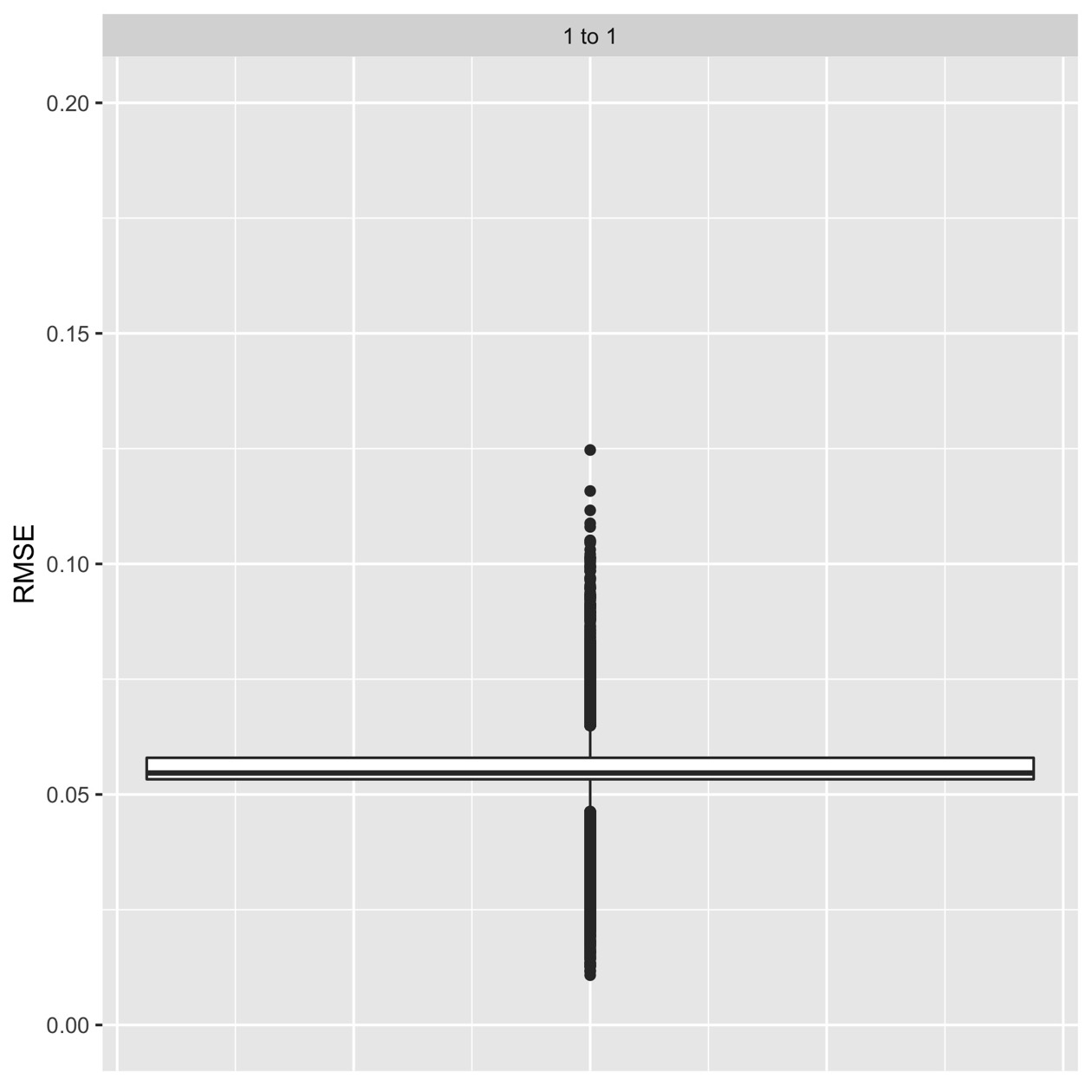}
        \caption{True kernel (1)}
    \end{subfigure}
    \hfill
    \begin{subfigure}[b]{0.32\textwidth}
        \centering
        \includegraphics[width=\textwidth]{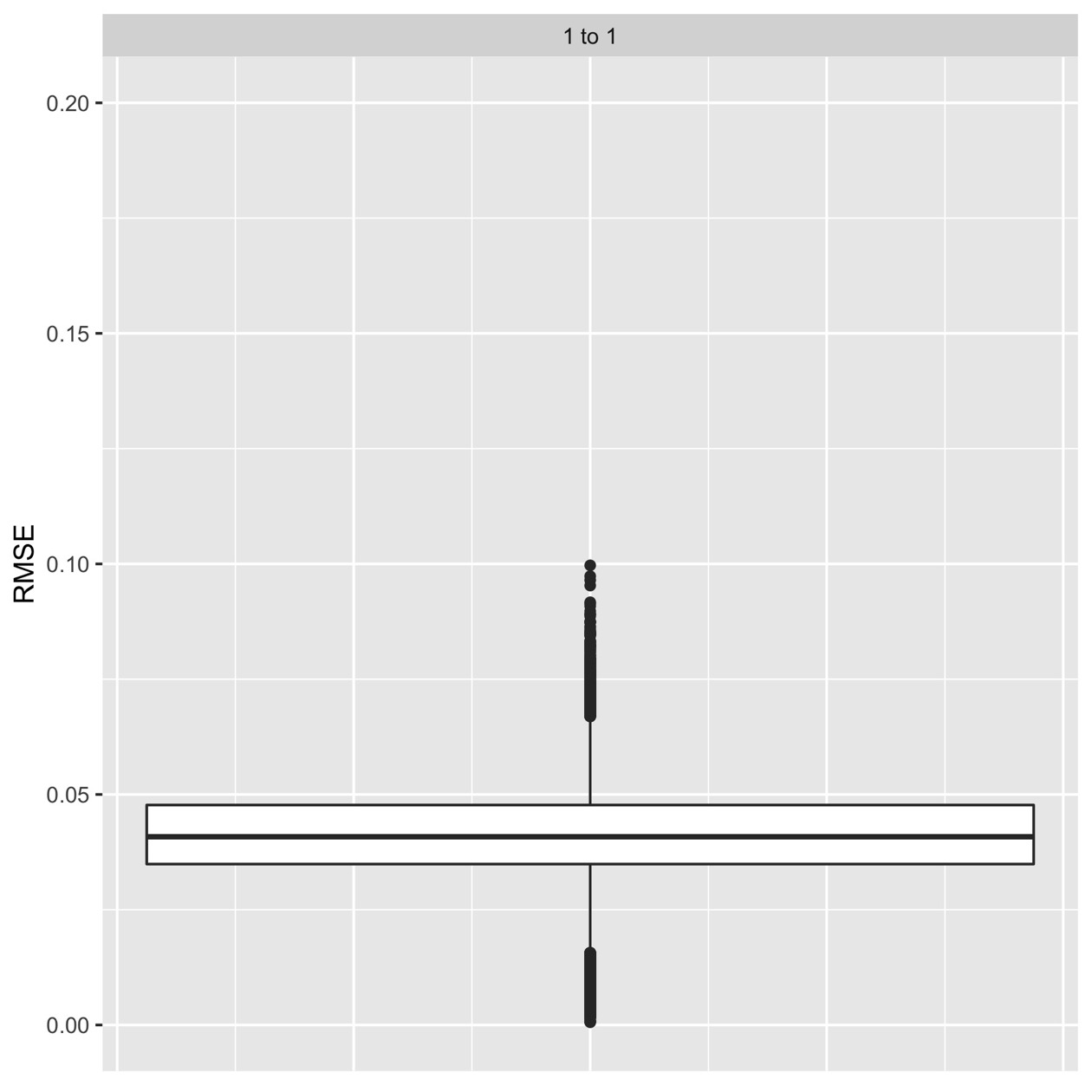}
        \caption{True kernel (2)}
    \end{subfigure}
    \hfill
    \begin{subfigure}[b]{0.32\textwidth}
        \centering
        \includegraphics[width=\textwidth]{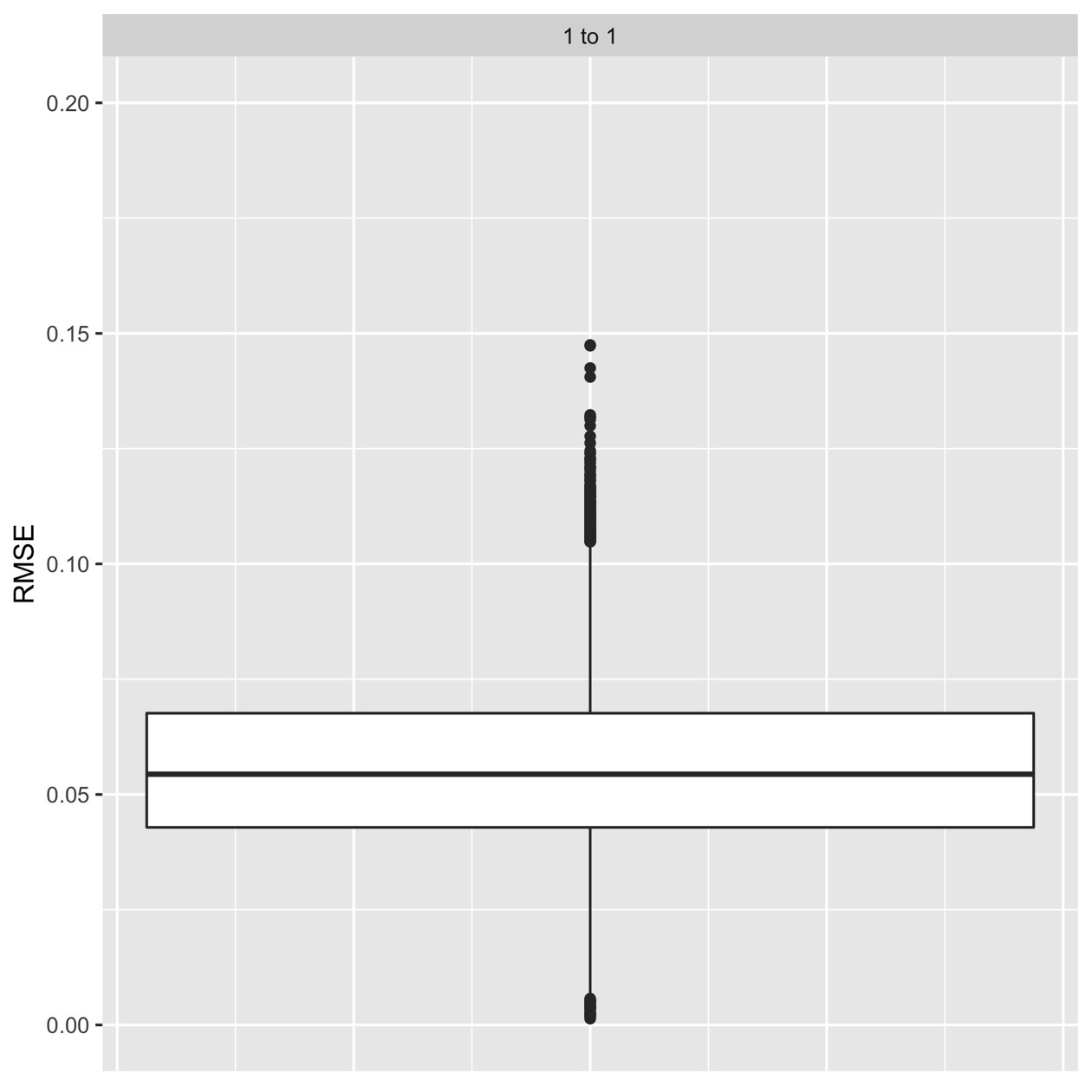}
        \caption{True kernel (3)}
    \end{subfigure}
	\end{subfigure}    
	\caption[] 
	{\small Under the histogram excitation function. \\ Boxplots of the root mean squared error (comparing the estimated triggering kernel for each posterior sample to the true function).}
	\label{fig:rmse_hist_model}
\end{figure}

\begin{figure}[H]
    \centering
    \begin{subfigure}{0.6\textwidth}
    \begin{subfigure}[b]{0.32\textwidth}
        \centering
        \includegraphics[width=\textwidth]{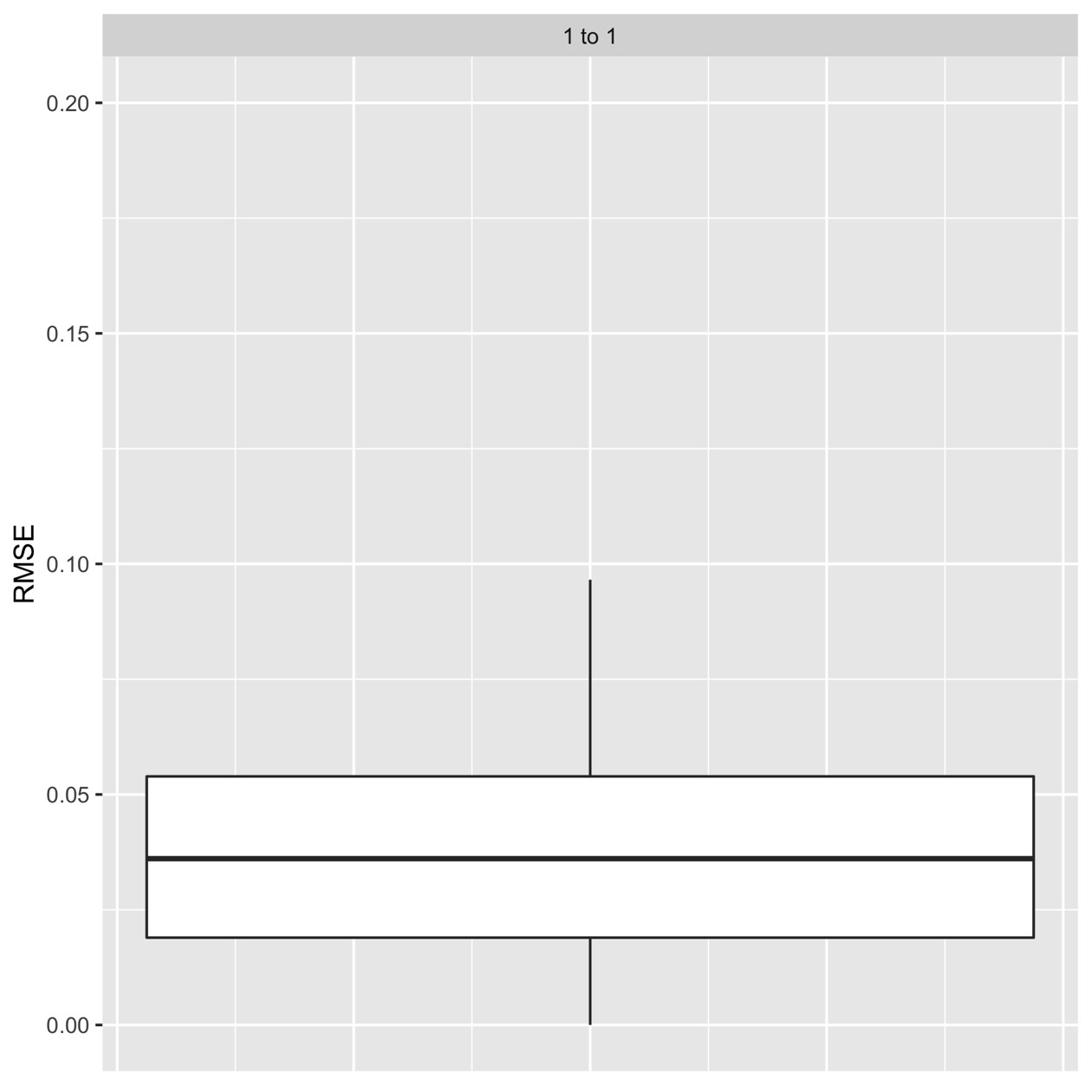}
        \caption{True kernel (1)}
    \end{subfigure}
    \hfill
    \begin{subfigure}[b]{0.32\textwidth}
        \centering
        \includegraphics[width=\textwidth]{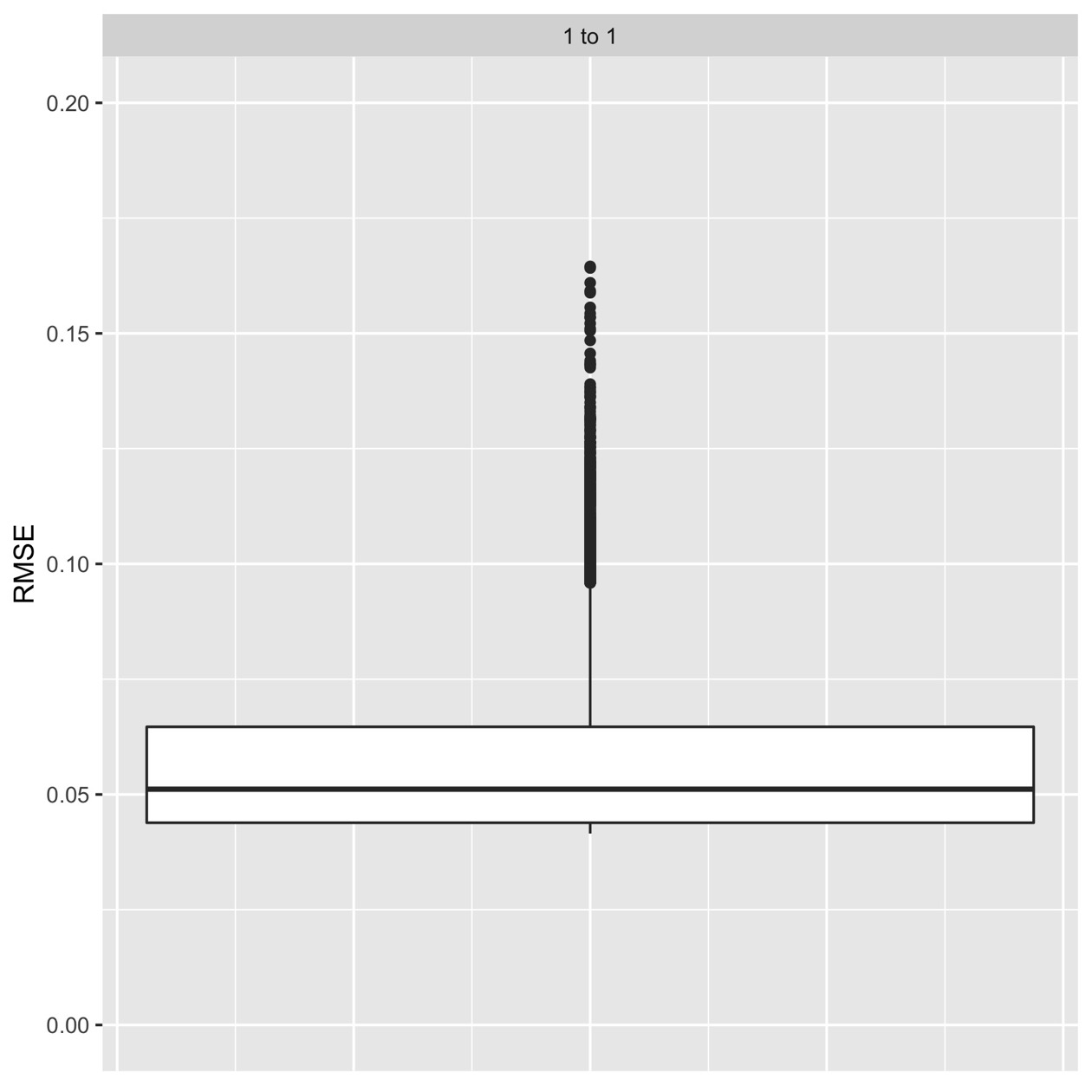}
        \caption{True kernel (2)}
    \end{subfigure}
    \hfill
    \begin{subfigure}[b]{0.32\textwidth}
        \centering
        \includegraphics[width=\textwidth]{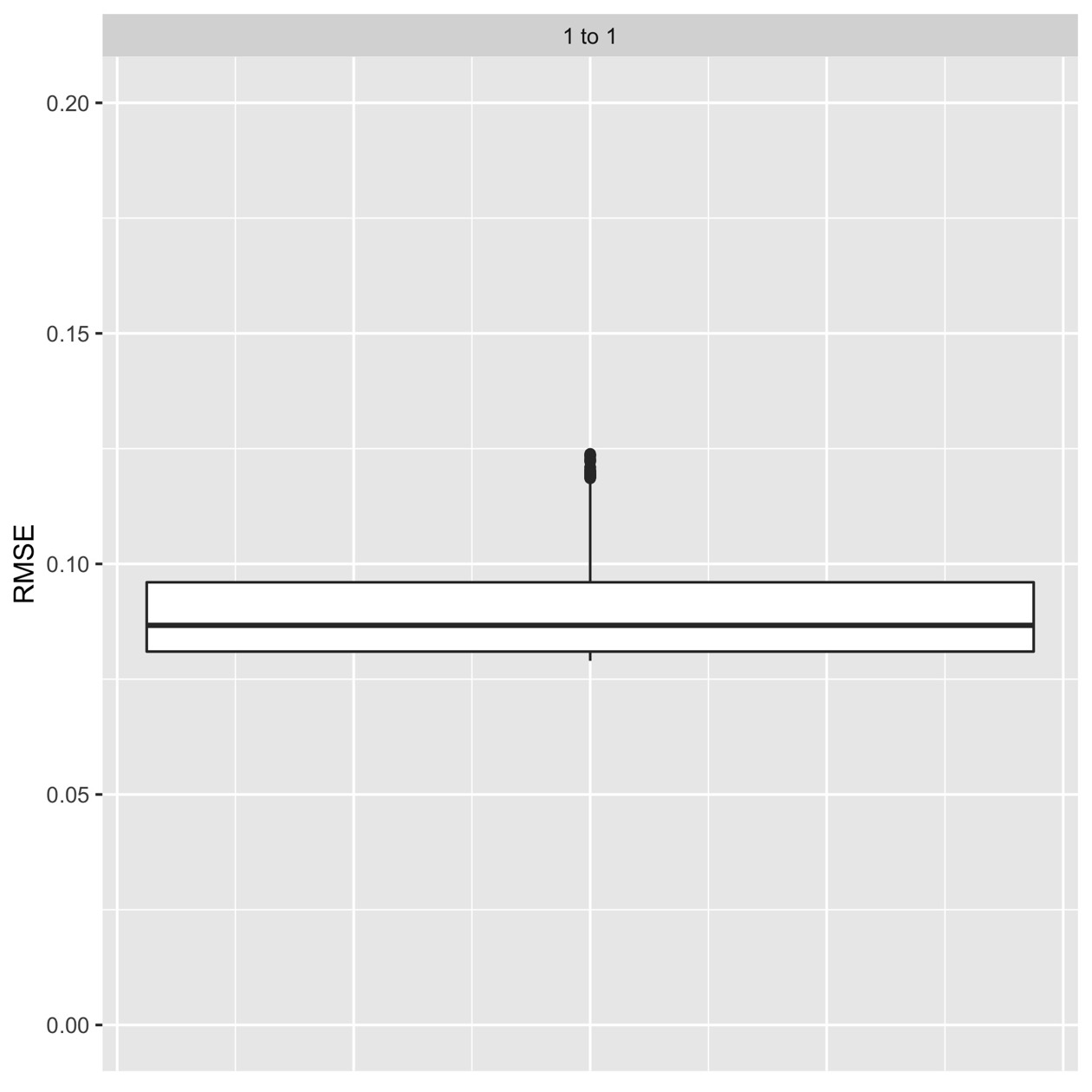}
        \caption{True kernel (3)}
    \end{subfigure}
	\end{subfigure}    
	\caption[] 
	{\small Under the geometric excitation function. \\ Boxplots of the root mean squared error (comparing the estimated triggering kernel for each posterior sample to the true function).}
	\label{fig:rmse_geom_model}
\end{figure}

 Figures \ref{fig:hist_model} and \ref{fig:geom_model} display this comparison. Figures \ref{fig:rmse_hist_model} and \ref{fig:rmse_geom_model} quantify the difference between the estimated posterior distribution of the kernel and the true kernel via the root mean squared error. For the first scenario with a geometric simulator kernel, both the geometric and histogram models recover the true kernel well. While the median RMSE for our histogram model is larger compared to the geometric model, the latter has a wide interquartile range for which our model's interquartile range largely lies within. The second and third simulation scenarios, which both assume histogram kernels, are recovered well by the histogram model. However, the geometric kernel is unable to capture the pattern of the simulations under a histogram kernel, even when the true kernel is monotonically decreasing. This is further reinforced when considering box plots of the RMSE, which show that our model achieves significantly lower RMSE in both of these scenarios compared to the geometric model.

\subsection{Multivariate simulation study}

We now consider a multivariate discrete-time Hawkes process. The histogram functions used to simulate the data are identical to those in the univariate scenario. However to ensure the process is stable the magnitude parameters are adjusted down such that $\alpha^{lk}=0.2 \ \forall \ l,k$.

In this section results are presented for a bivariate process that demonstrates the multivariate case and is an exemplar for extending to higher dimensions when sufficient data and computational resources are available. Moreover, higher-dimensional processes are discussed in Section \ref{sec:mv_alt}. For this study, again three independent realisations were generated using the same model parameters for each combination of prior setting and time series length described in Section \ref{sec:inference}. Additionally, we consider time series lengths of up to 2000 days to account for the increased dimension of the parameter space. Again, only one such realisation for each scenario is presented here and only time series lengths of 50 and 500 days are shown. The results for the remaining time series lengths can be found in the supplementary material. For a given simulation, the inference is performed and initialised as outlined in Section \ref{sec:uv_sim}.

\begin{figure}[H]
    \centering
    \begin{subfigure}{0.8\textwidth}
    \begin{subfigure}[b]{0.45\textwidth}
        \centering
        \includegraphics[width=\textwidth]{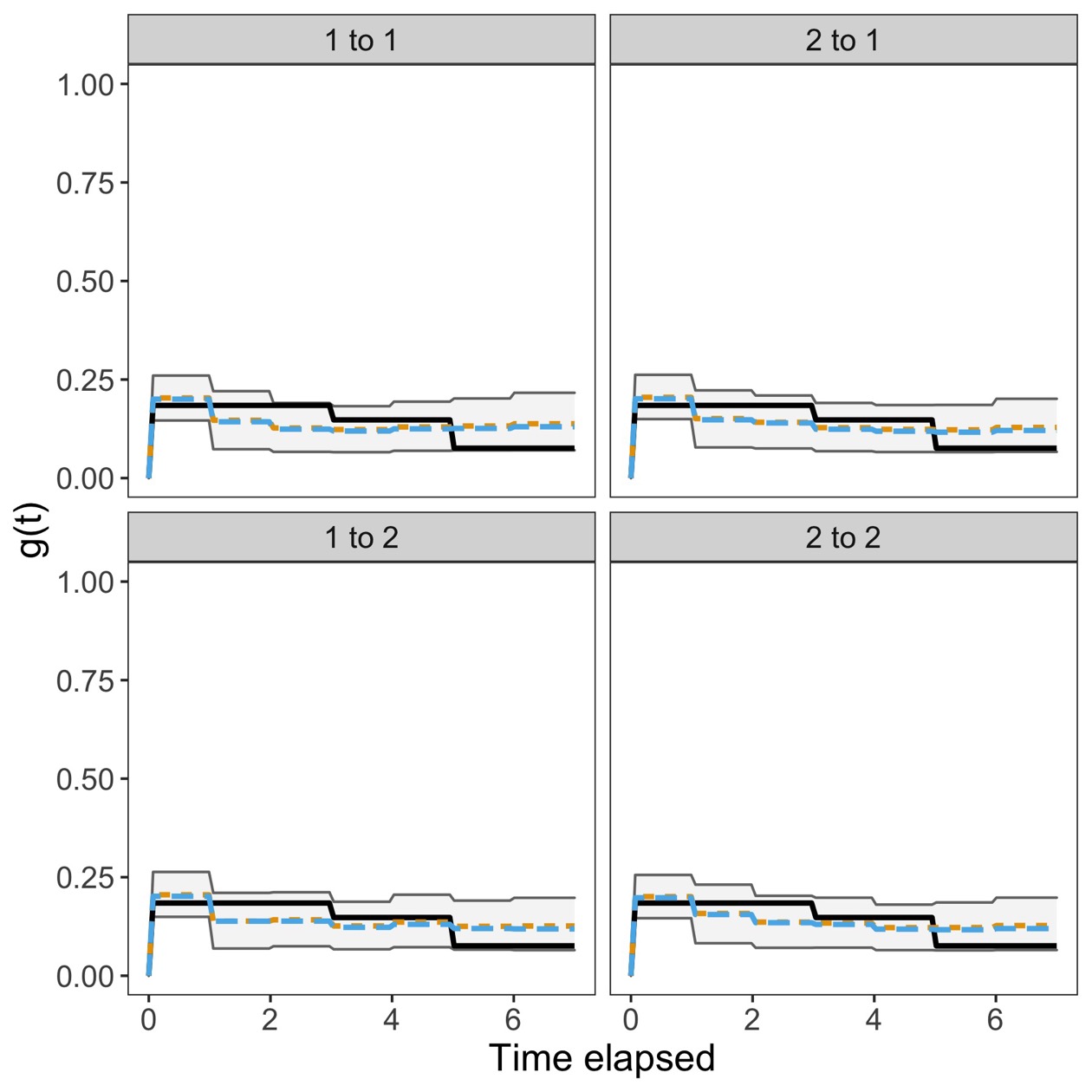}
        \caption{50 days. Total event count: 174}
    \end{subfigure}
    \hfill
    \begin{subfigure}[b]{0.45\textwidth}
        \centering
        \includegraphics[width=\textwidth]{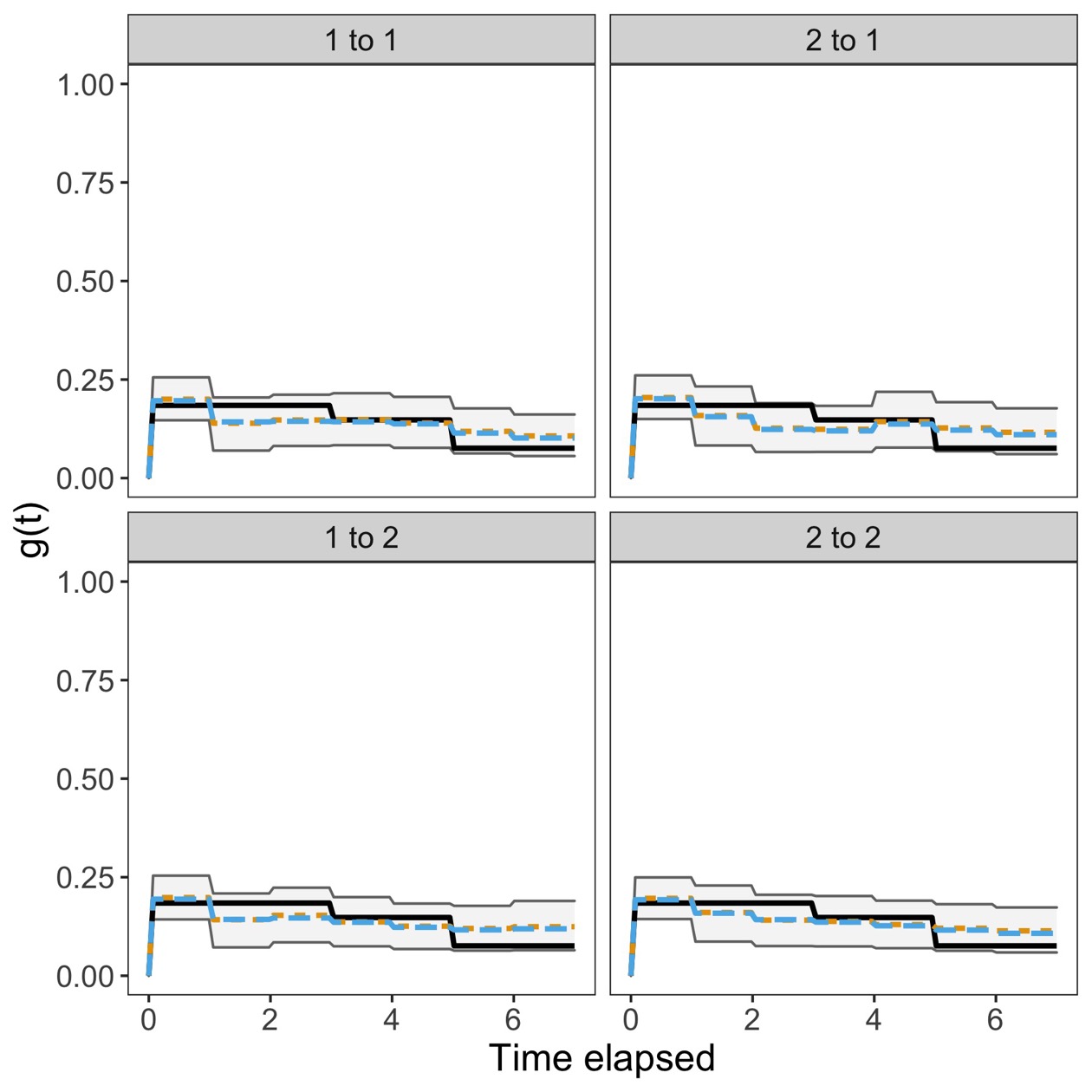}
        \caption{500 days. Total event count: 1587}
    \end{subfigure}
	\end{subfigure}    
	\caption[] 
	{\small Estimated excitation functions under the informative prior setting.\\ \textbf{Solid black line:} true histogram function. \textbf{Dashed orange line:} posterior mean.\\ \textbf{Dashed blue line:} posterior median. \textbf{Grey ribbon:} 80\% posterior interval.}
	\label{fig:2d_inf}
\end{figure}
\begin{figure}[H]
    \centering
    \begin{subfigure}{0.8\textwidth}
    \begin{subfigure}[b]{0.45\textwidth}
        \centering
        \includegraphics[width=\textwidth]{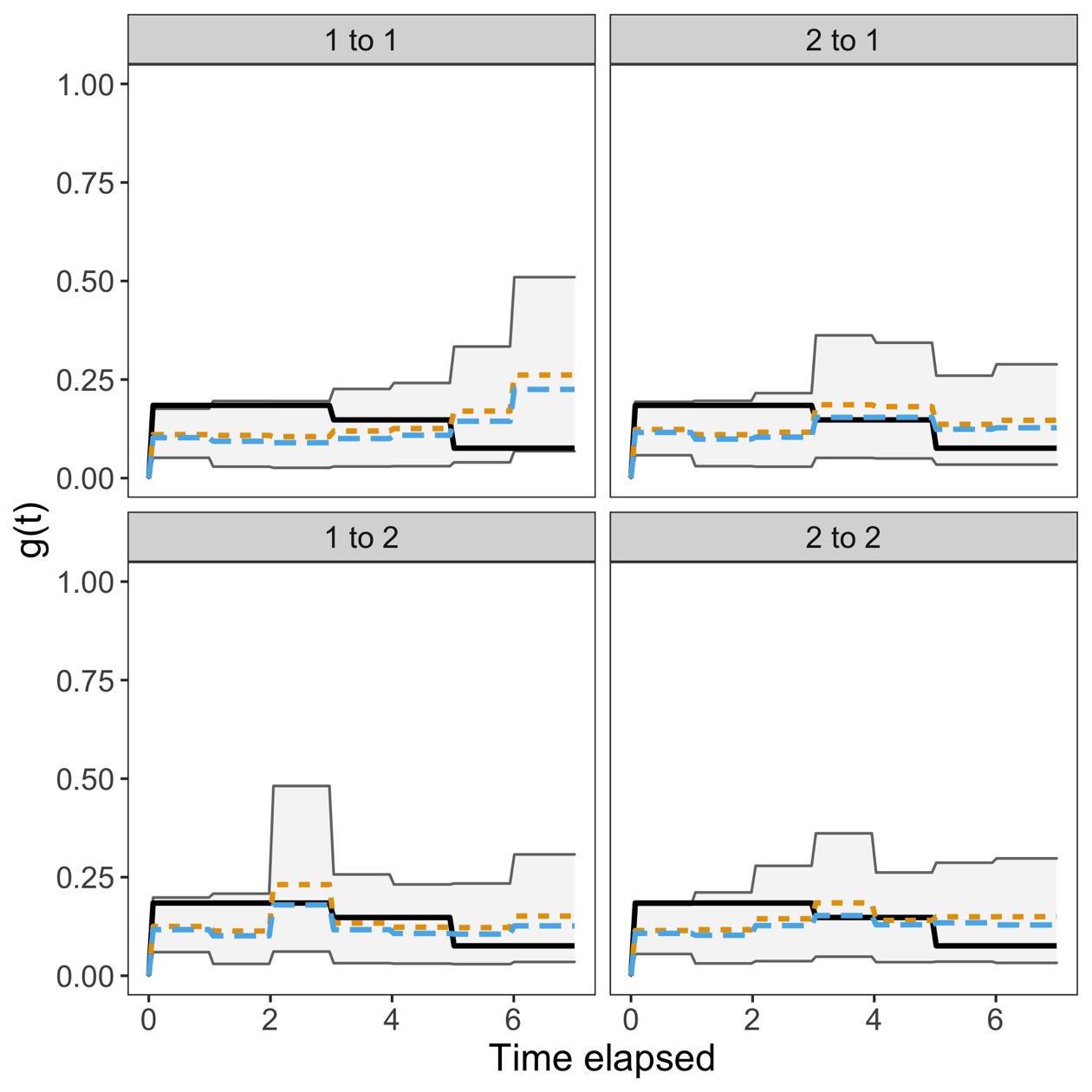}
        \caption{50 days. Total event count: 169}
    \end{subfigure}
    \hfill
    \begin{subfigure}[b]{0.45\textwidth}
        \centering
        \includegraphics[width=\textwidth]{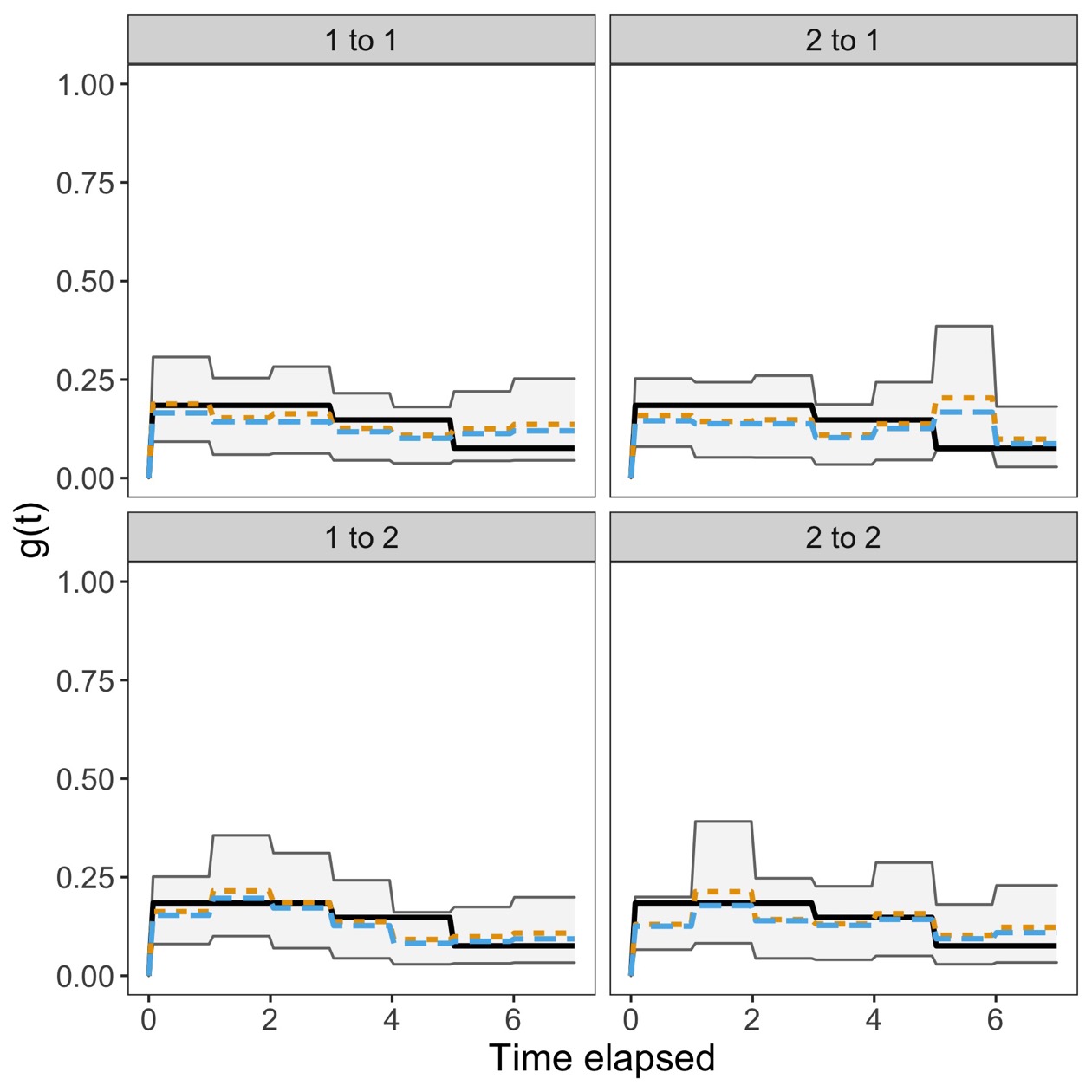}
        \caption{500 days. Total event count: 1721}
    \end{subfigure}
	\end{subfigure}    
	\caption[] 
	{\small Estimated excitation functions under the relatively informative prior setting.\\ \textbf{Solid black line:} true histogram function. \textbf{Dashed orange line:} posterior mean.\\ \textbf{Dashed blue line:} posterior median. \textbf{Grey ribbon:} 80\% posterior interval.}
	\label{fig:2d_relinf}
\end{figure}
\begin{figure}[H]
    \centering
    \begin{subfigure}{0.8\textwidth}
    \begin{subfigure}[b]{0.45\textwidth}
        \centering
        \includegraphics[width=\textwidth]{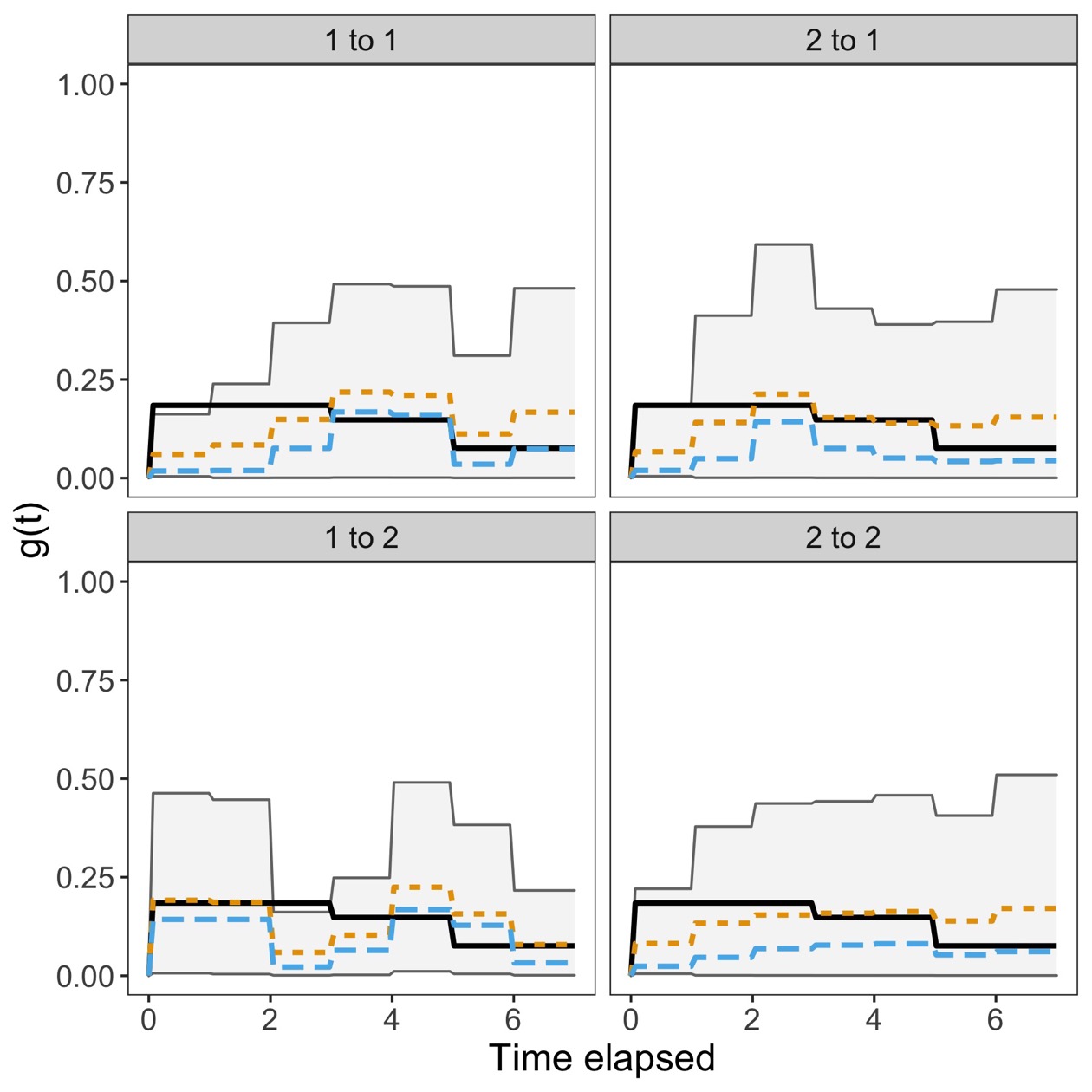}
        \caption{50 days. Total event count: 171}
    \end{subfigure}
    \hfill
    \begin{subfigure}[b]{0.45\textwidth}
        \centering
        \includegraphics[width=\textwidth]{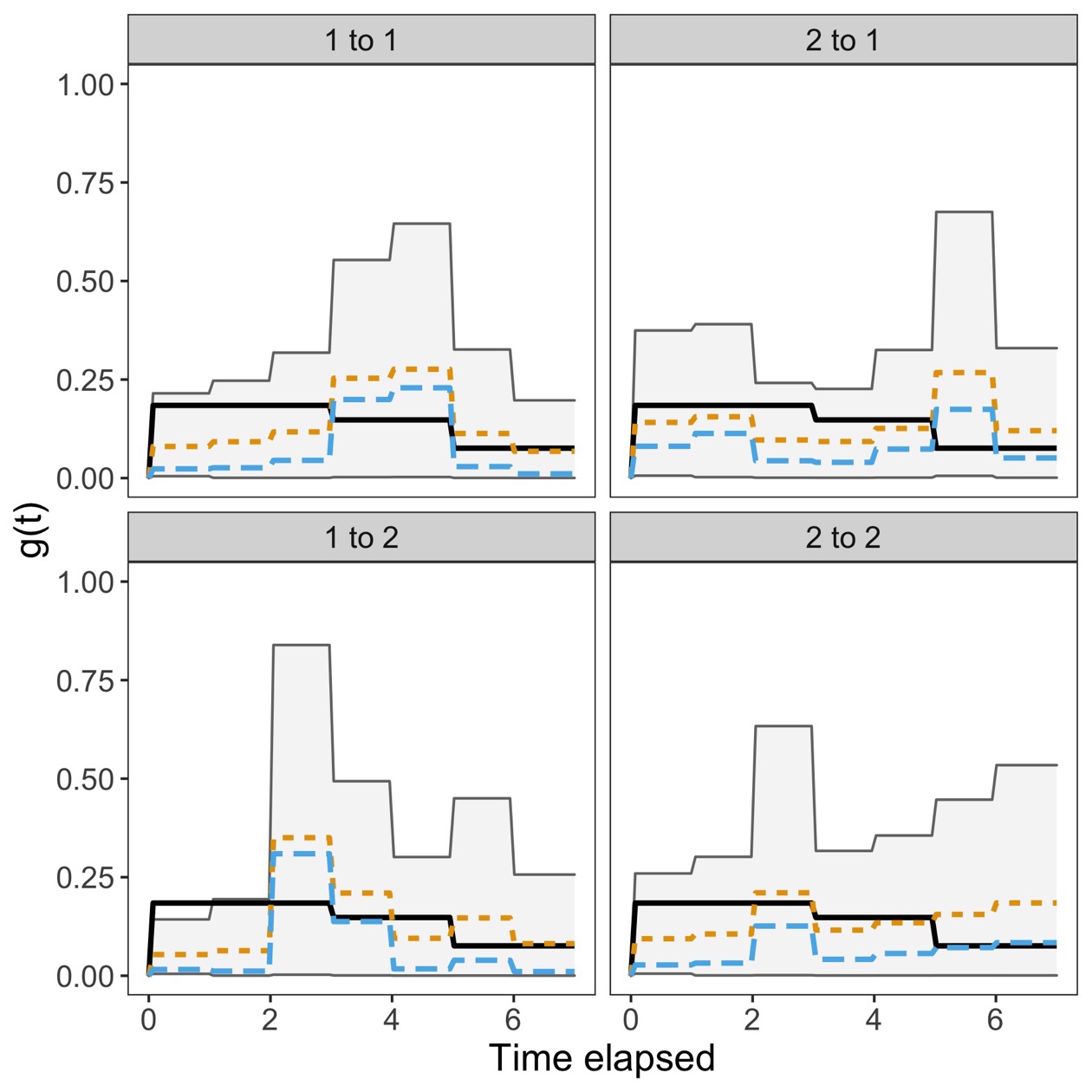}
        \caption{500 days. Total event count: 1580}
    \end{subfigure}
	\end{subfigure}    
	\caption[] 
	{\small Estimated excitation functions under the quite uninformative prior setting. \\ \textbf{Solid black line:} true histogram function. \textbf{Dashed orange line:} posterior mean.\\ \textbf{Dashed blue line:} posterior median. \textbf{Grey ribbon:} 80\% posterior interval.}
	\label{fig:2d_uninf}
\end{figure}

Figures \ref{fig:2d_inf} - \ref{fig:2d_uninf} present the estimated excitation functions for each of the prior settings and time series lengths. Overall, we reach similar conclusions to the univariate study. The simulated histogram is recovered very well using the informative prior setting. When relaxing the prior to the relatively informative prior setting, the posterior accuracy decreases initially. However increasing the number of days, and thus the sample size, improves this result. Additionally, the true parameters are predominantly within the 80\% posterior interval. The quite uninformative prior exhibits poor mixing and low acceptance rates. This led to poor MCMC convergence and thus we were unable to perform useful inference for these scenarios. As we have observed in our simulations, a longer time series has the potential to improve inferential accuracy. However, considering longer time series for this prior setting to obtain better results was computationally infeasible due to long runtimes of the reversible-jump MCMC and limited ability for parallelisation. 

\begin{figure}[H]
    \centering
    \begin{subfigure}{0.8\textwidth}
	\centering
    \begin{subfigure}[b]{0.3\textwidth}
        \centering
        \includegraphics[width=\textwidth]{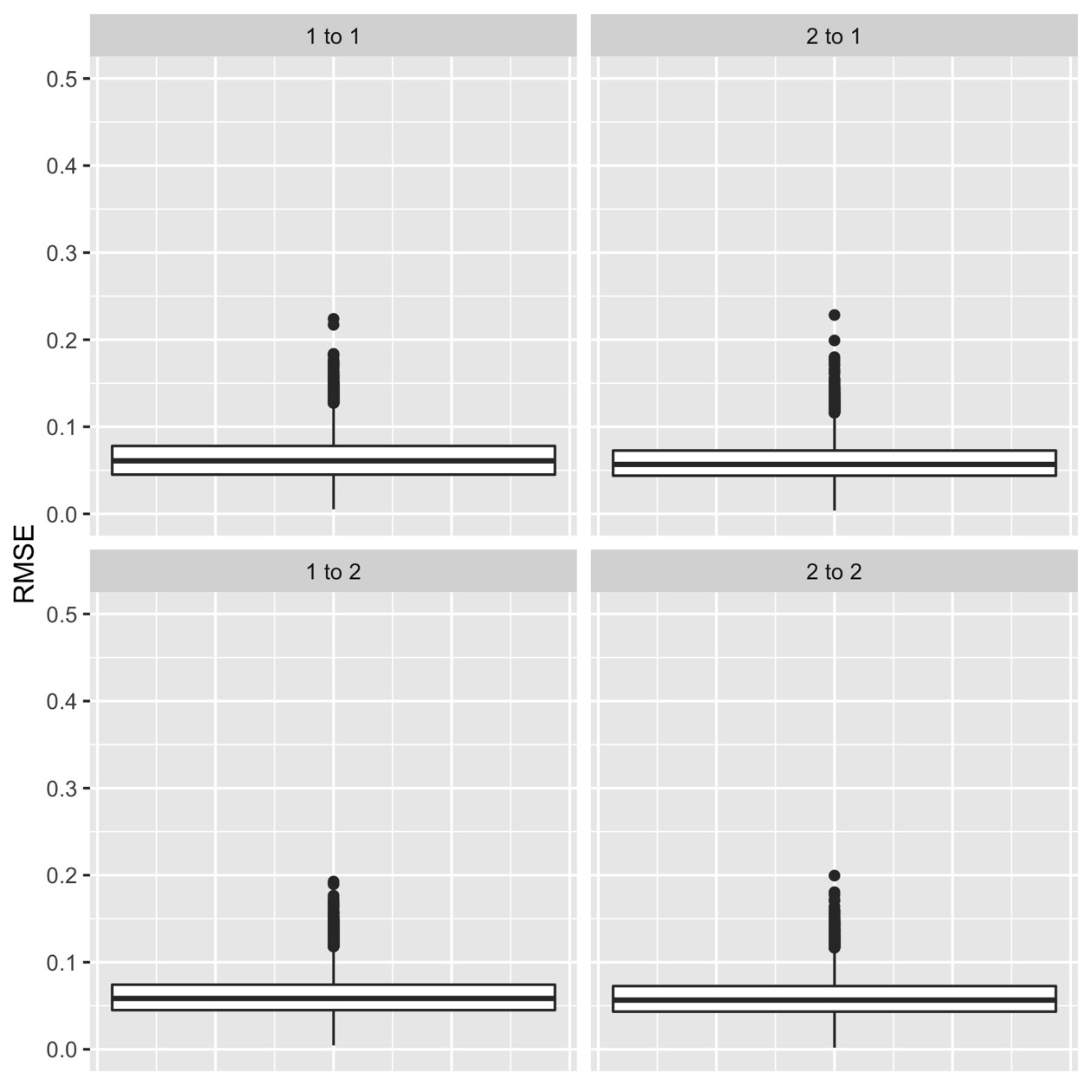}
        \caption{50 days}
    \end{subfigure}
    \hfill
    \begin{subfigure}[b]{0.3\textwidth}
        \centering
        \includegraphics[width=\textwidth]{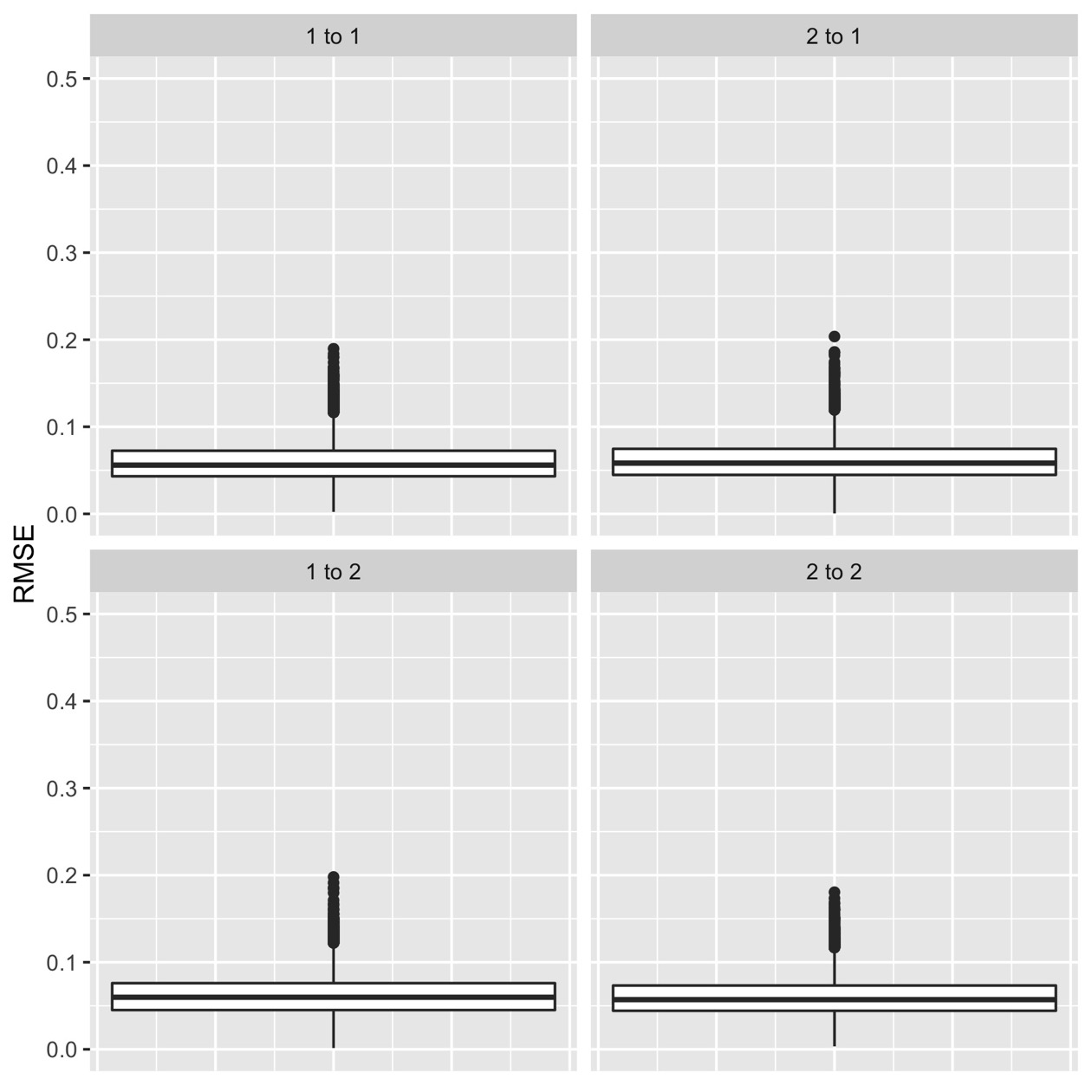}
        \caption{100 days}
    \end{subfigure}
    \hfill
    \begin{subfigure}[b]{0.3\textwidth}
        \centering
        \includegraphics[width=\textwidth]{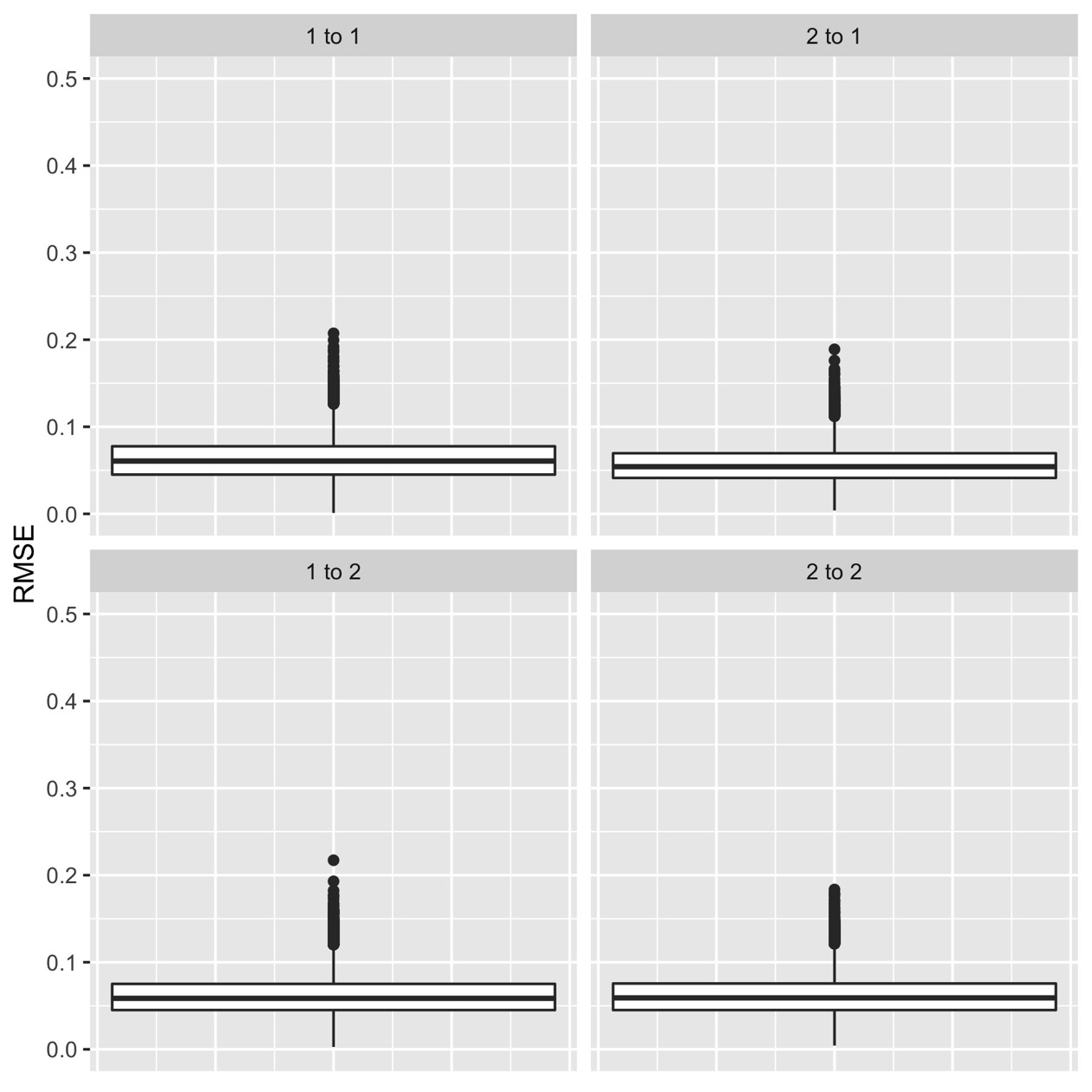}
        \caption{200 days}
    \end{subfigure}
    \hfill
    \begin{subfigure}[b]{0.3\textwidth}
        \centering
        \includegraphics[width=\textwidth]{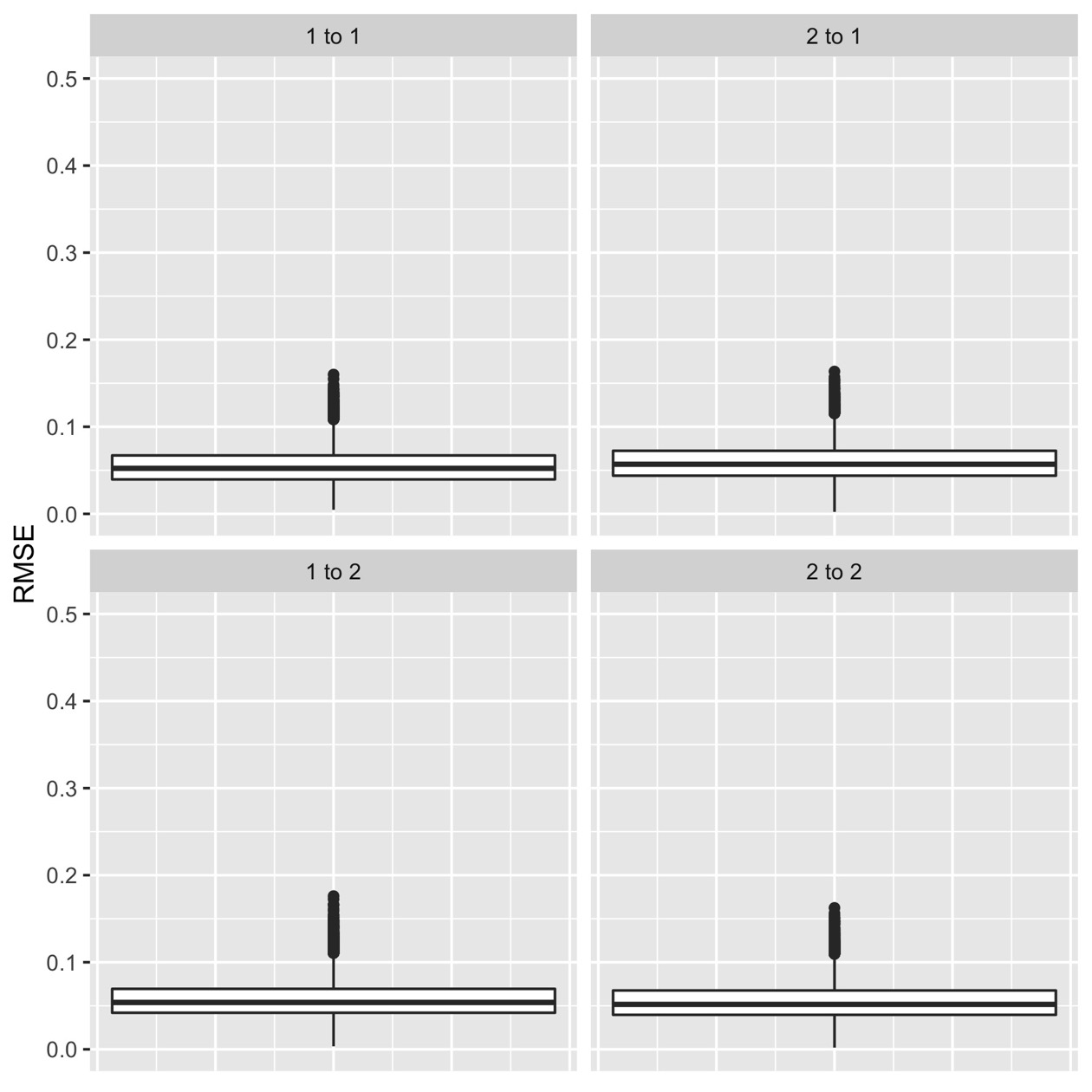}
        \caption{500 days}
    \end{subfigure}
            \hfill
    \begin{subfigure}[b]{0.3\textwidth}
        \centering
        \includegraphics[width=\textwidth]{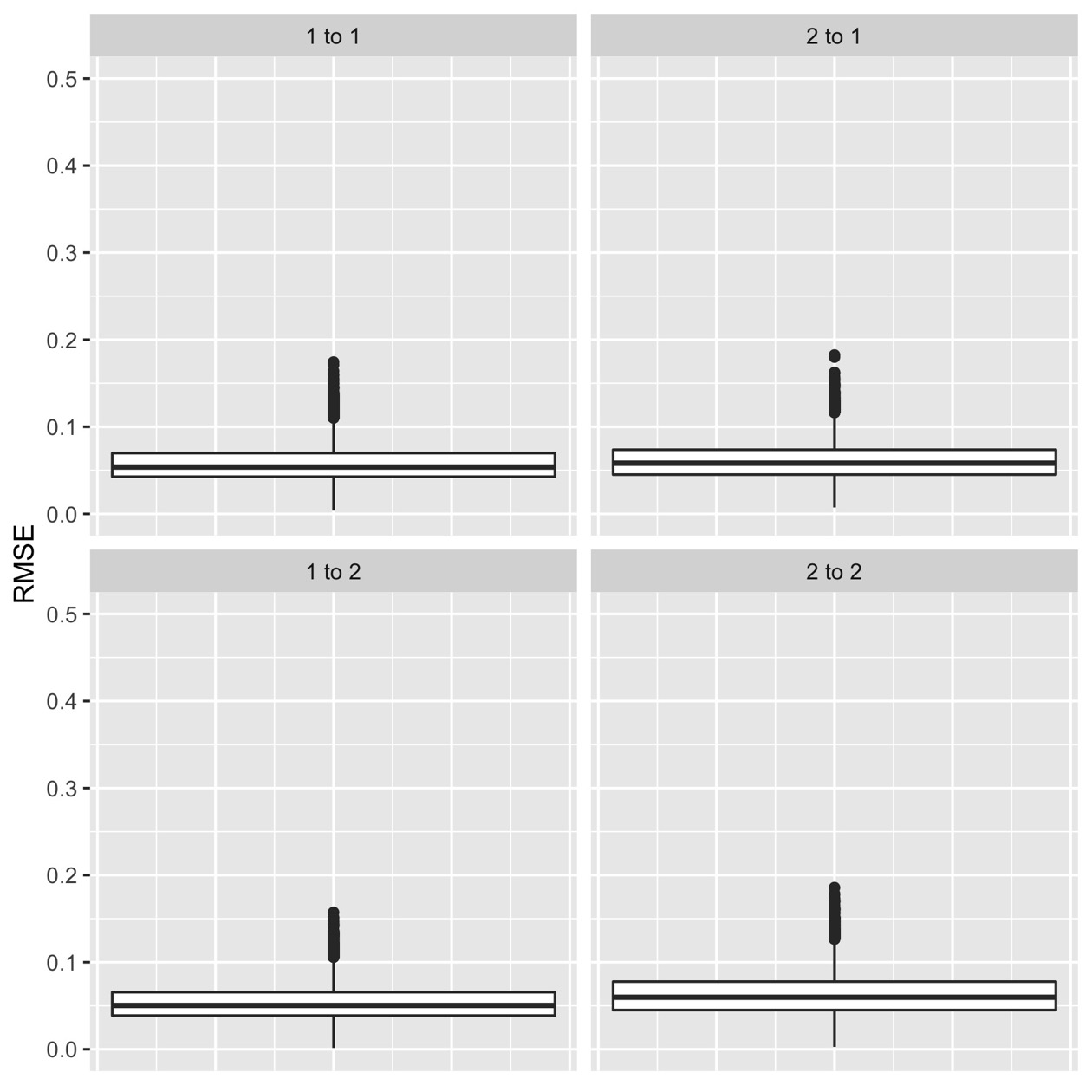}
        \caption{1000 days}
    \end{subfigure}
    \hfill
    \begin{subfigure}[b]{0.3\textwidth}
        \centering
        \includegraphics[width=\textwidth]{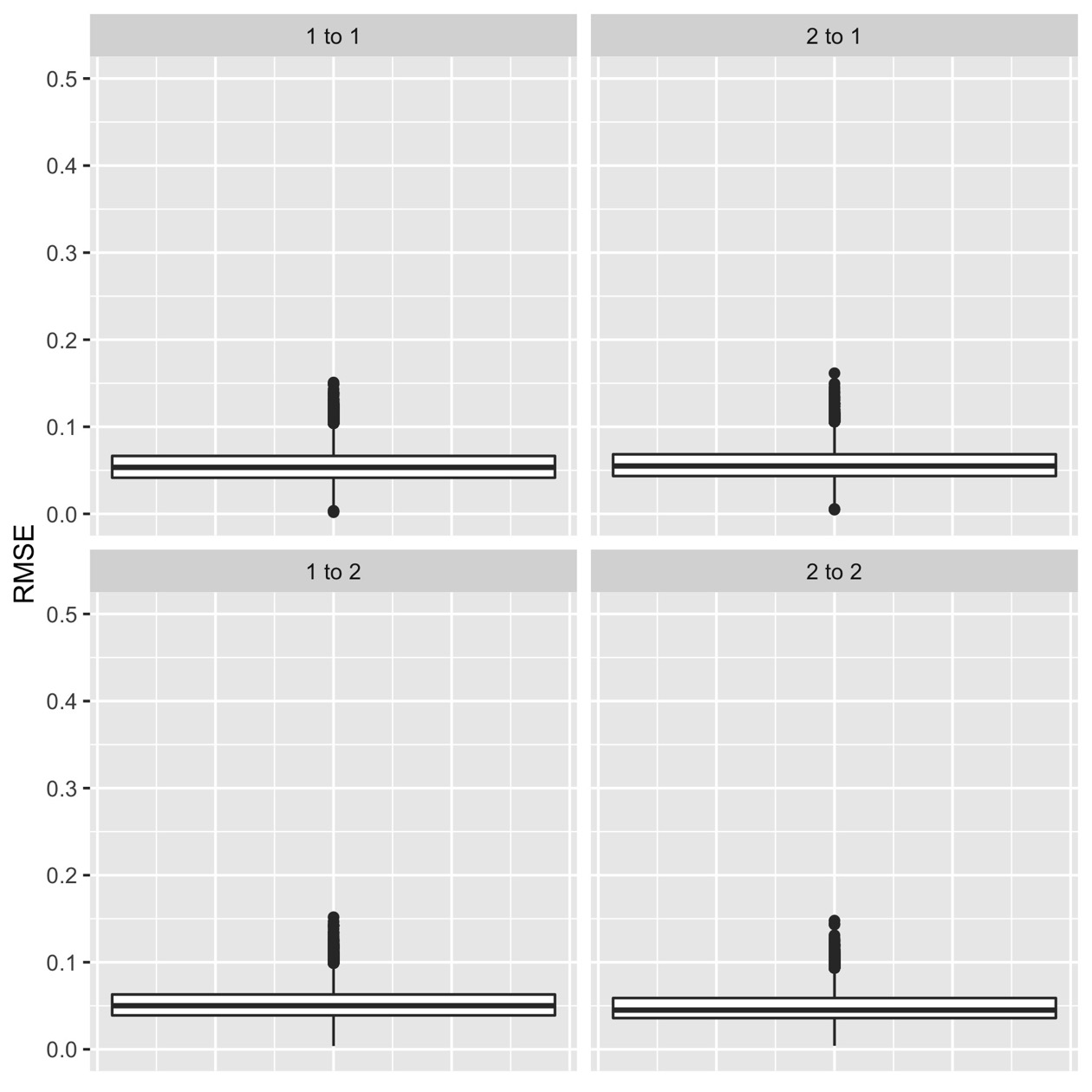}
        \caption{2000 days}
    \end{subfigure}
	\end{subfigure}    
	\caption[] 
	{\small Boxplots of the root mean squared error (comparing the estimated triggering kernel for each posterior sample to the true histogram) for the informative prior setting.}
	\label{fig:rmse_2d_inf}
\end{figure}

\begin{figure}[H]
    \centering
    \begin{subfigure}{0.8\textwidth}
    \begin{subfigure}[b]{0.3\textwidth}
        \centering
        \includegraphics[width=\textwidth]{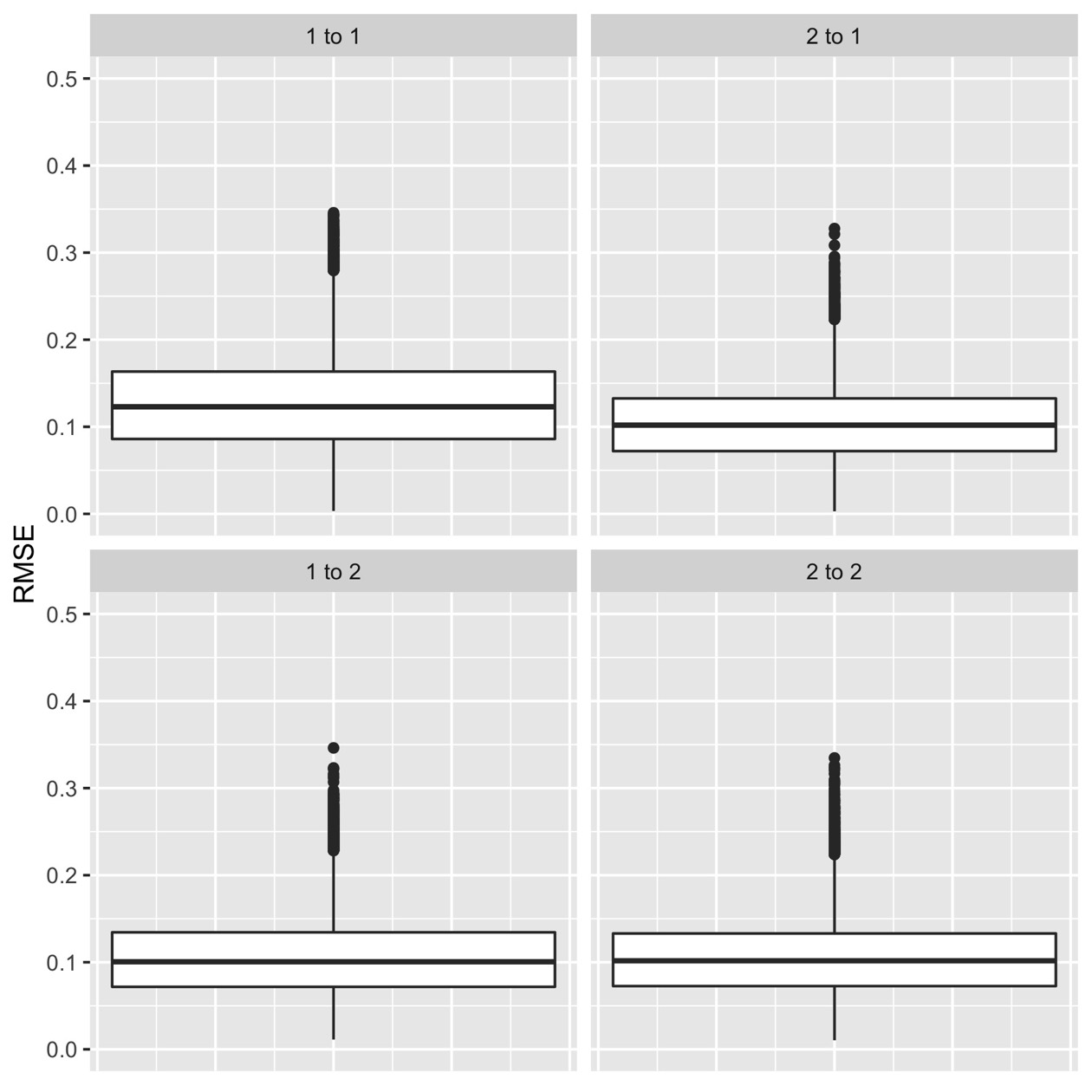}
        \caption{50 days}
    \end{subfigure}
    \hfill
    \begin{subfigure}[b]{0.3\textwidth}
        \centering
        \includegraphics[width=\textwidth]{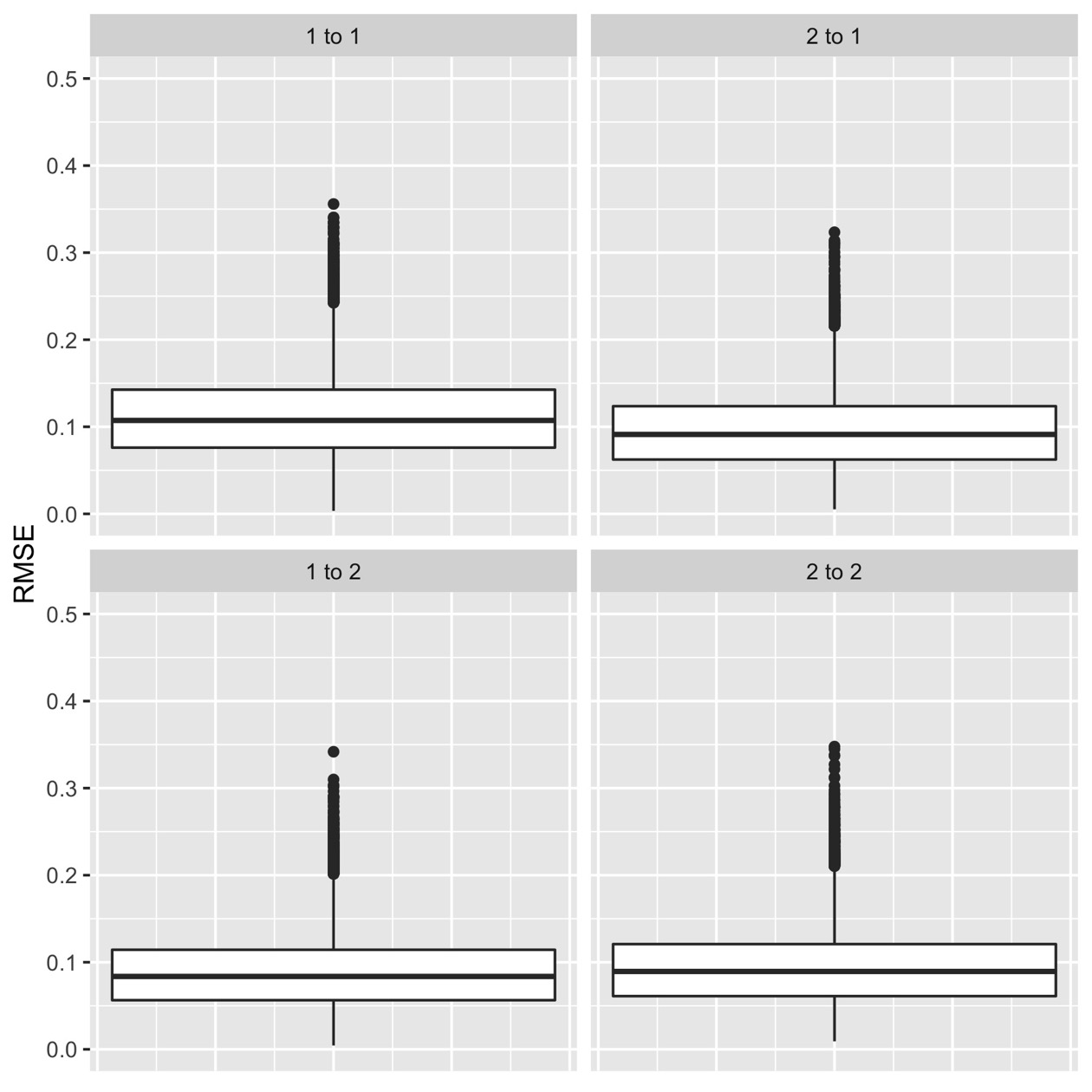}
        \caption{100 days}
    \end{subfigure}
    \hfill
    \begin{subfigure}[b]{0.3\textwidth}
        \centering
        \includegraphics[width=\textwidth]{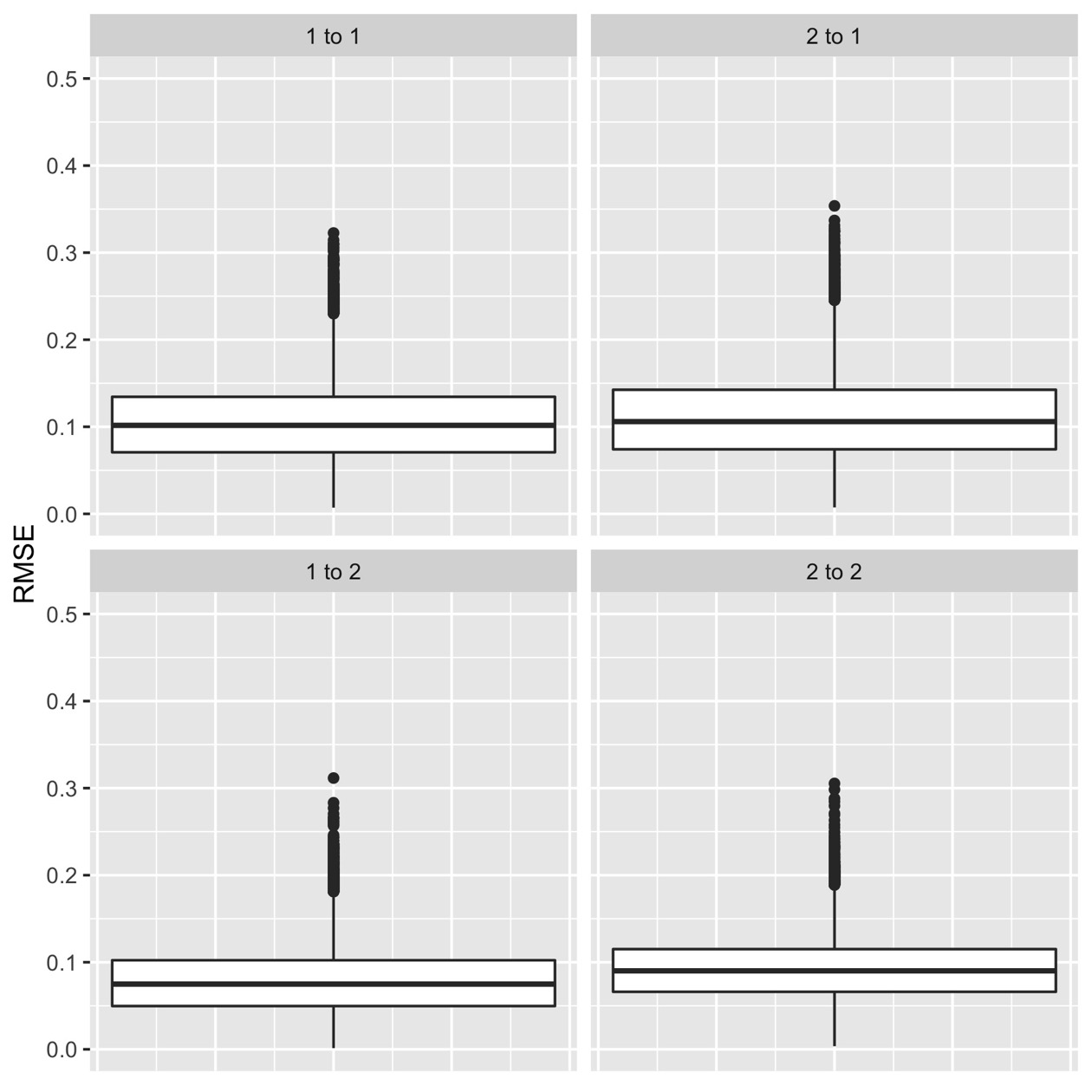}
        \caption{200 days}
    \end{subfigure}
    \hfill
    \begin{subfigure}[b]{0.3\textwidth}
        \centering
        \includegraphics[width=\textwidth]{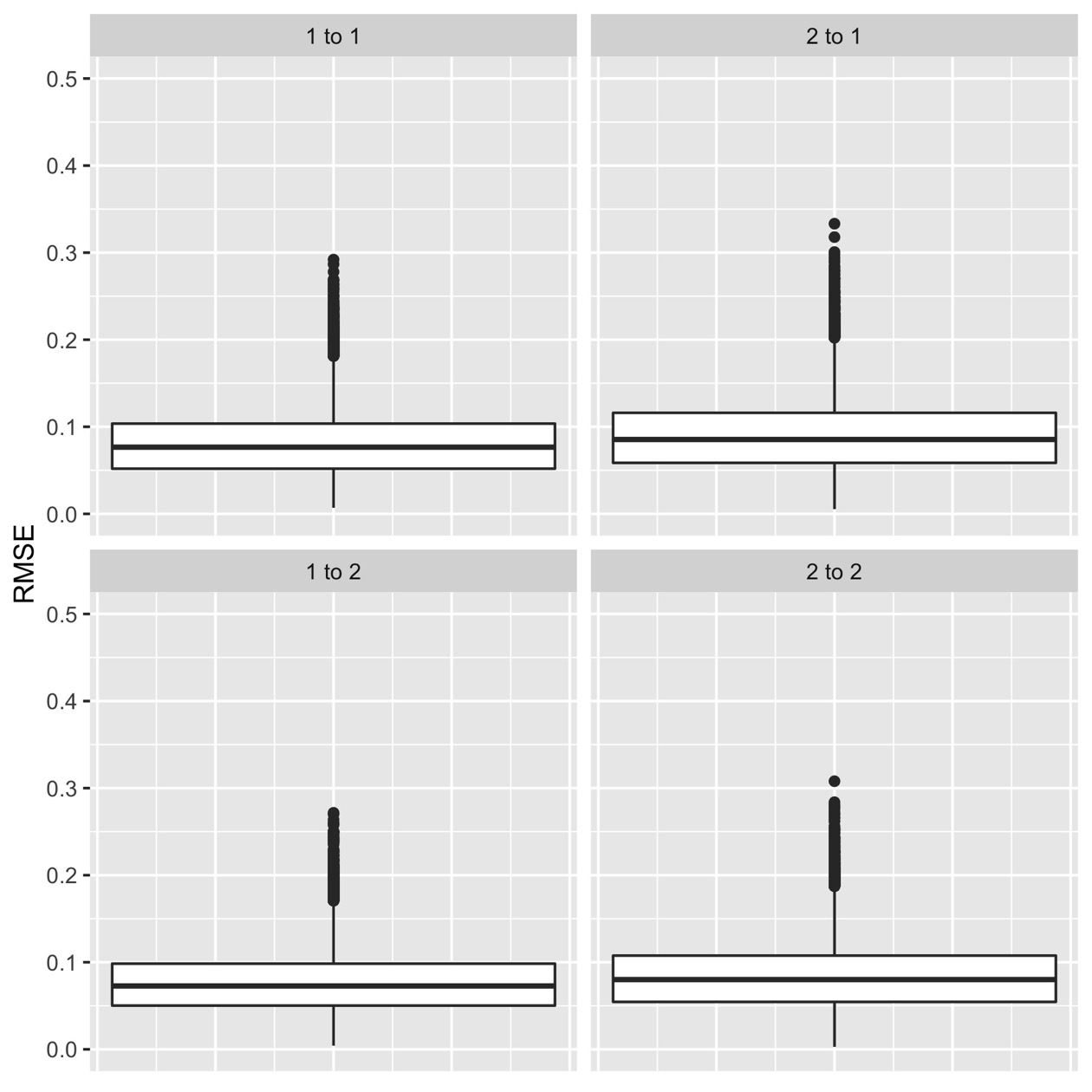}
        \caption{500 days}
    \end{subfigure}     
           \hfill
    \begin{subfigure}[b]{0.3\textwidth}
        \centering
        \includegraphics[width=\textwidth]{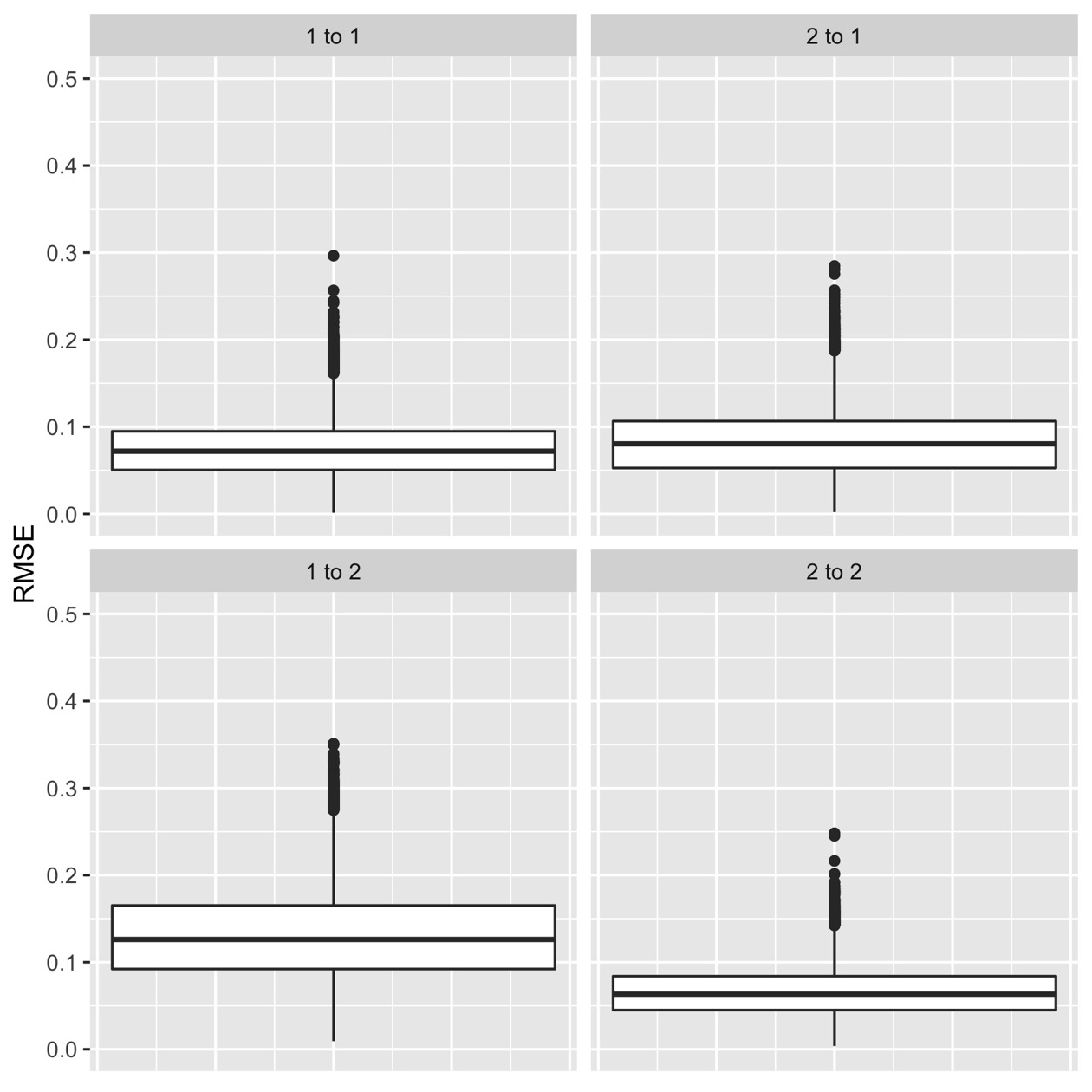}
        \caption{1000 days}
    \end{subfigure}
    \hfill
    \begin{subfigure}[b]{0.3\textwidth}
        \centering
        \includegraphics[width=\textwidth]{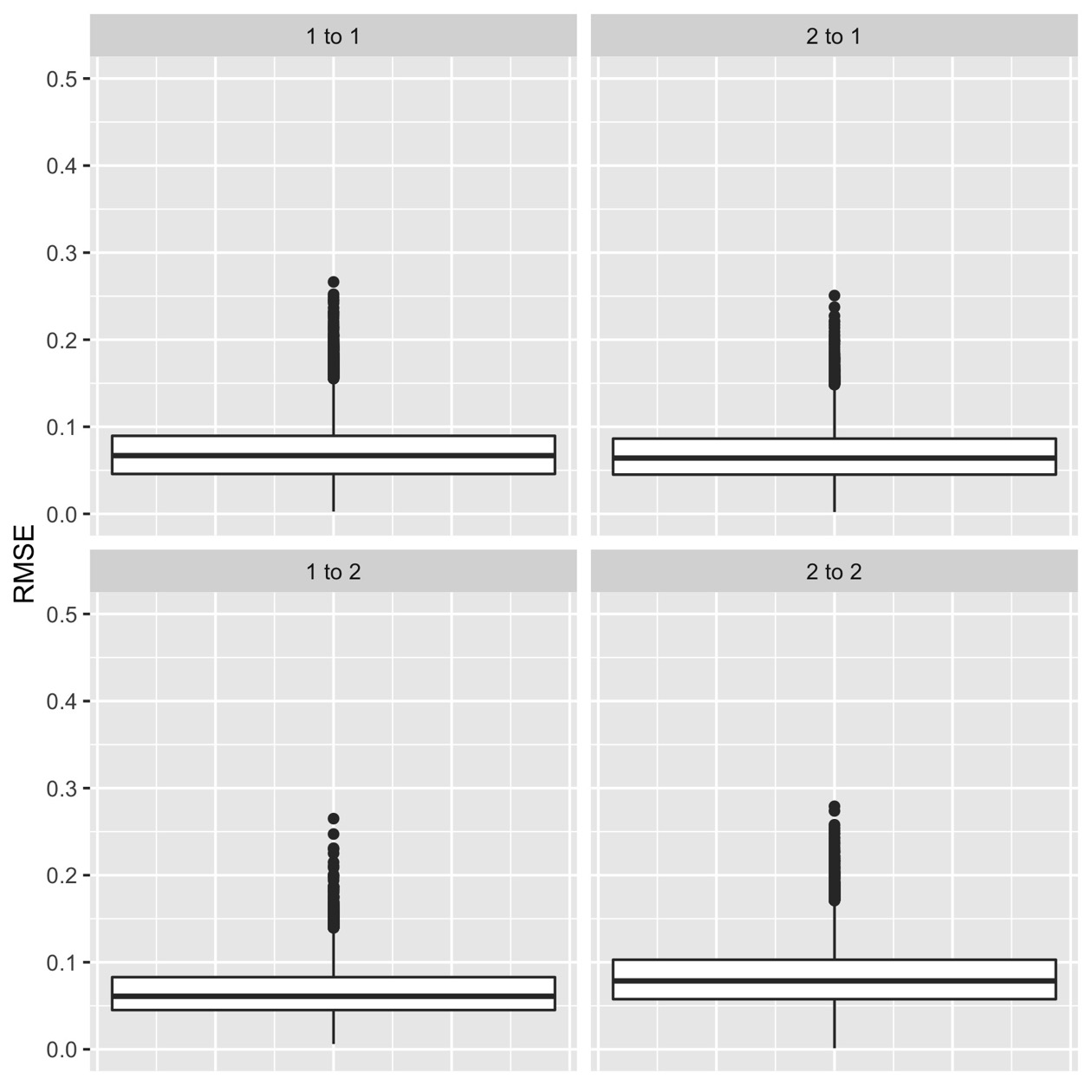}
        \caption{2000 days}
    \end{subfigure}
	\end{subfigure}    
	\caption[] 
	{\small Boxplots of the root mean squared error (comparing the estimated triggering kernel for each posterior sample to the true histogram) for the relatively informative prior setting.}
	\label{fig:rmse_2d_relinf}
\end{figure}

\begin{figure}[H]
    \centering
    \begin{subfigure}{0.8\textwidth}
    \begin{subfigure}[b]{0.3\textwidth}
        \centering
        \includegraphics[width=\textwidth]{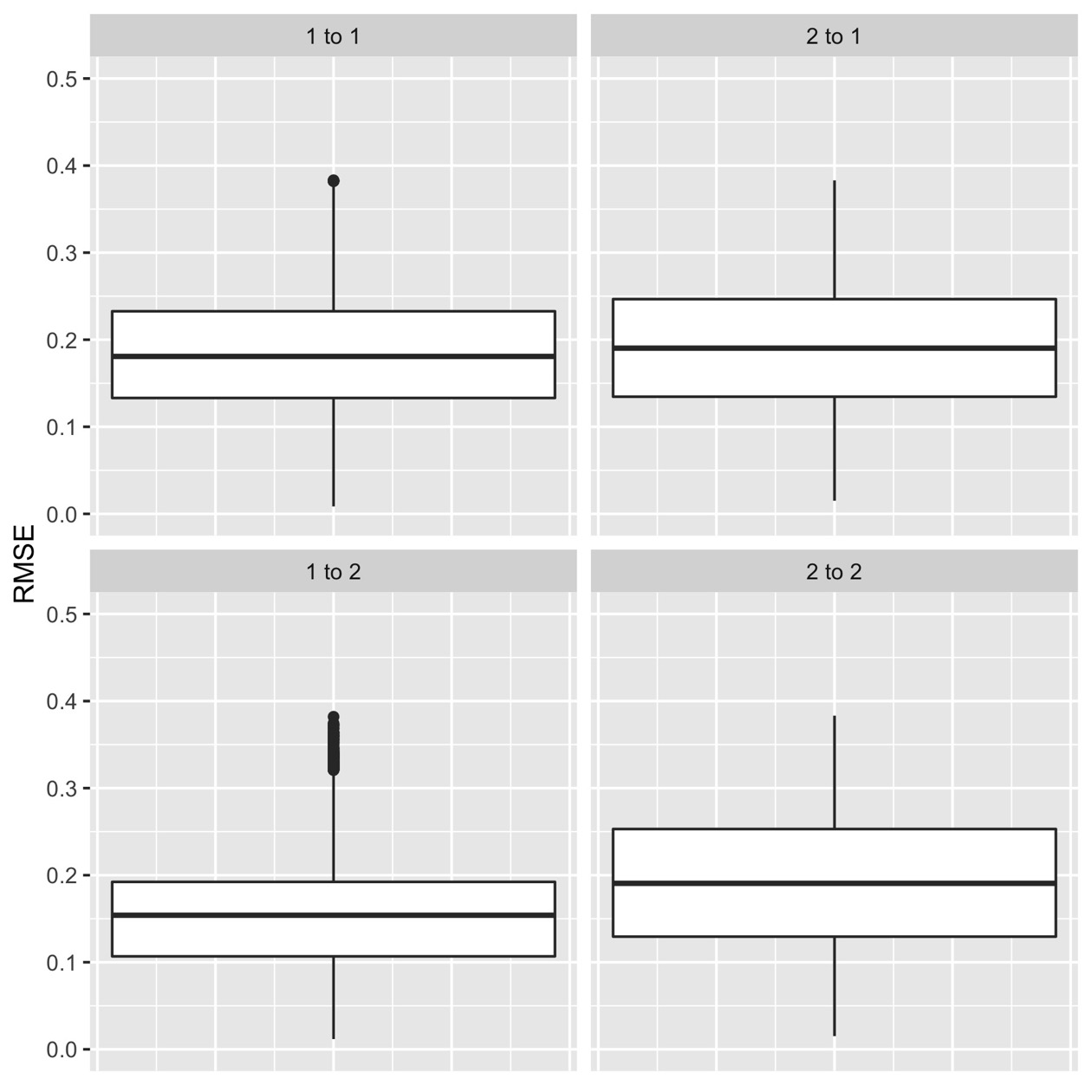}
        \caption{50 days}
    \end{subfigure}
    \hfill
    \begin{subfigure}[b]{0.3\textwidth}
        \centering
        \includegraphics[width=\textwidth]{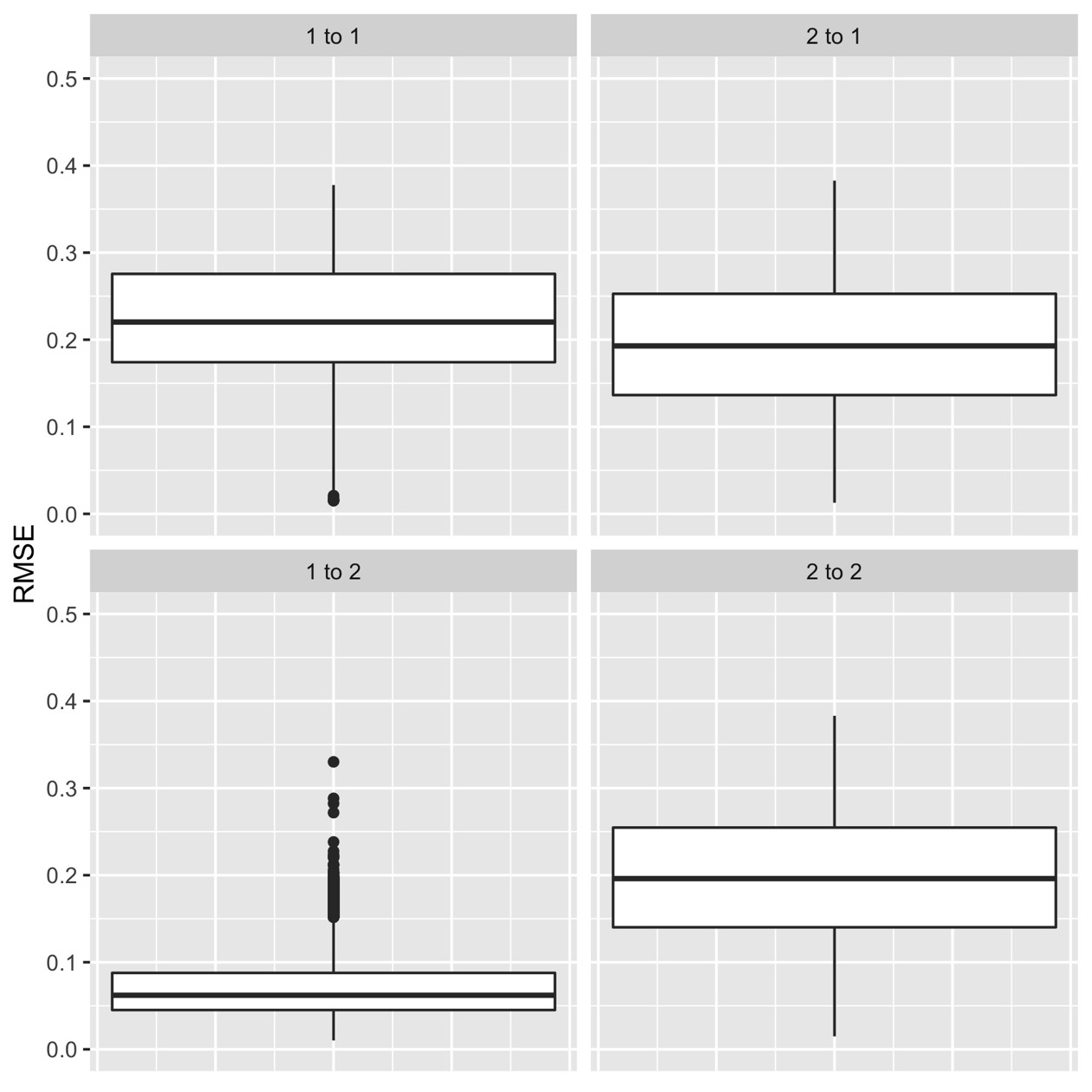}
        \caption{100 days}
    \end{subfigure}
    \hfill
    \begin{subfigure}[b]{0.3\textwidth}
        \centering
        \includegraphics[width=\textwidth]{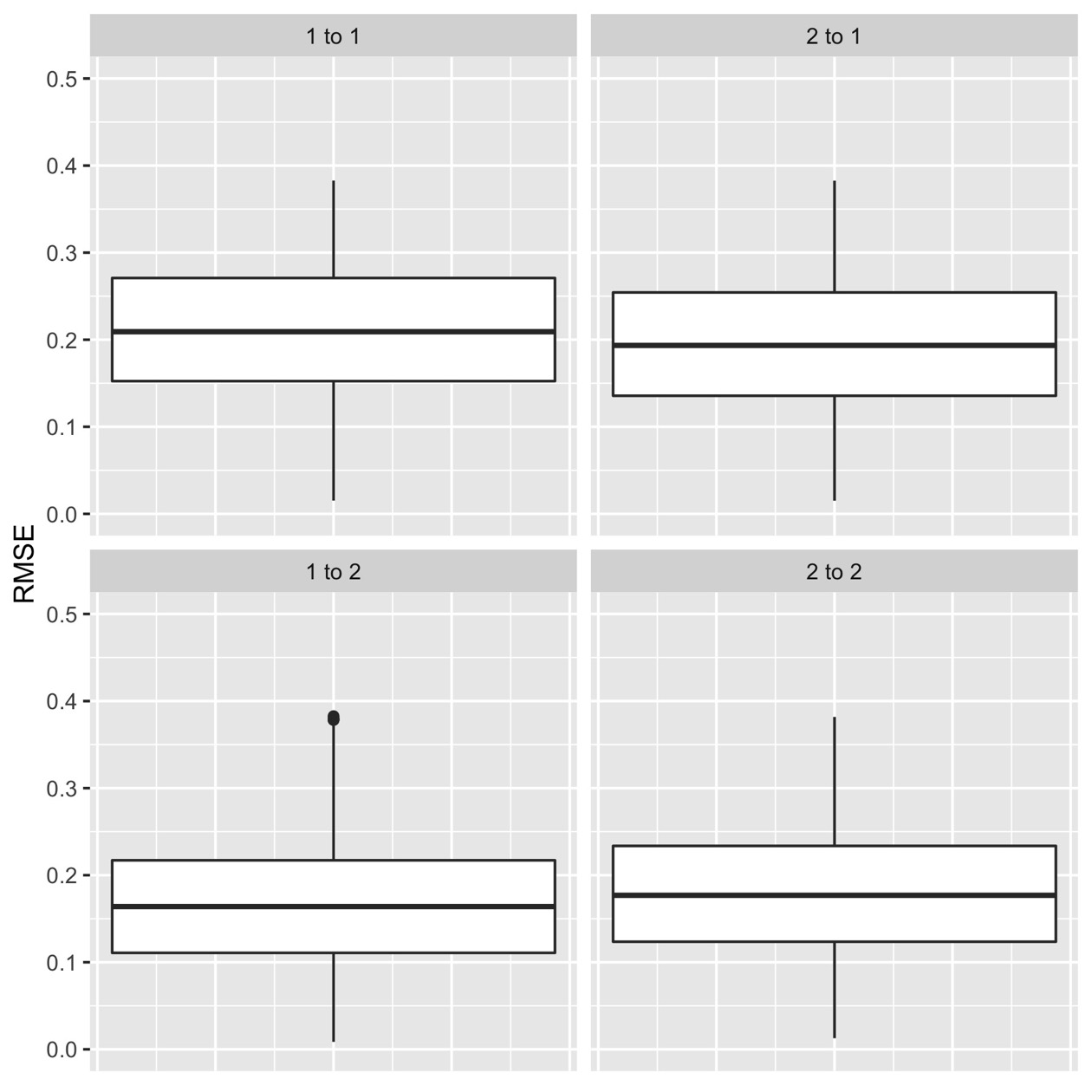}
        \caption{200 days}
    \end{subfigure}
    \hfill
    \begin{subfigure}[b]{0.3\textwidth}
        \centering
        \includegraphics[width=\textwidth]{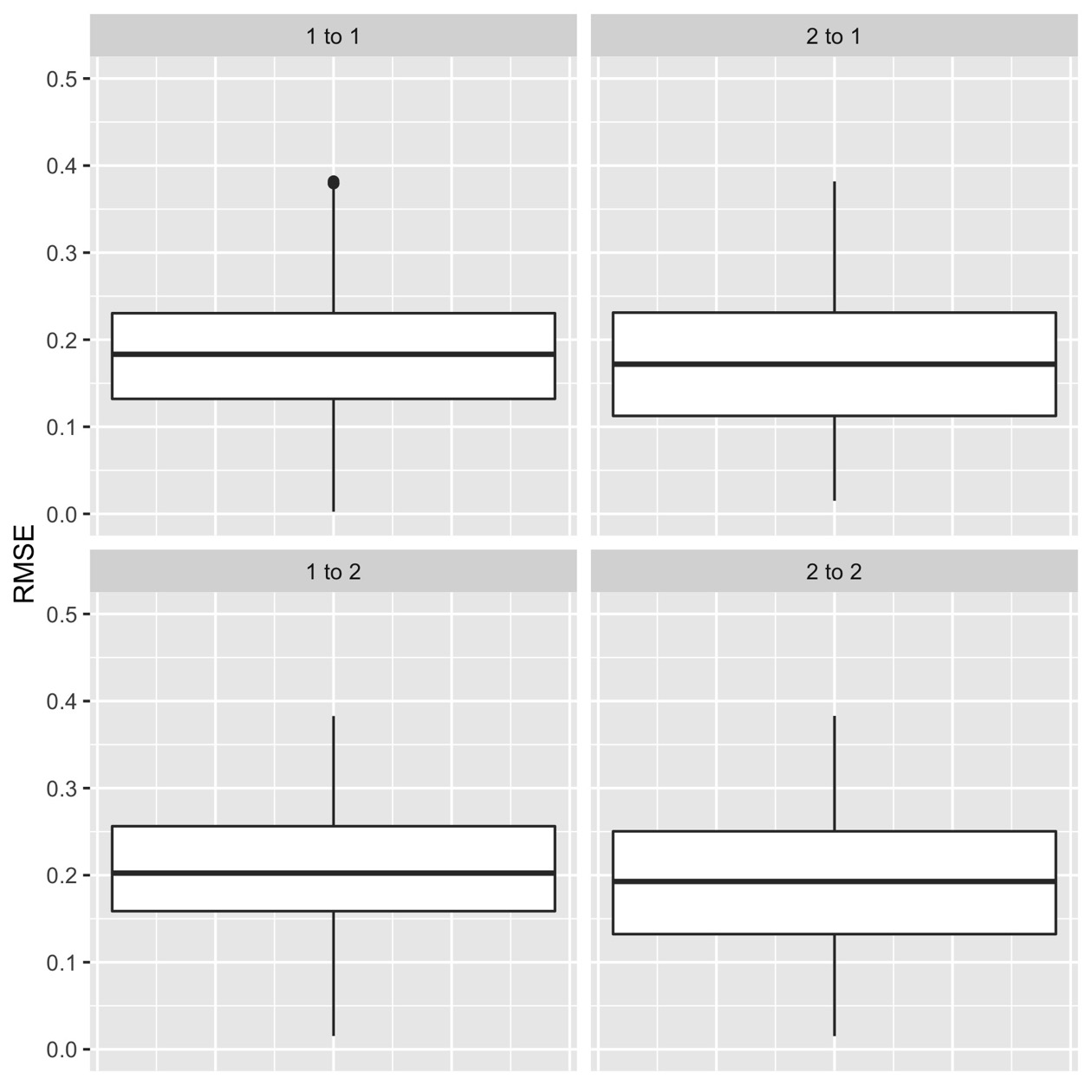}
        \caption{500 days}
    \end{subfigure}
                \hfill
    \begin{subfigure}[b]{0.3\textwidth}
        \centering
        \includegraphics[width=\textwidth]{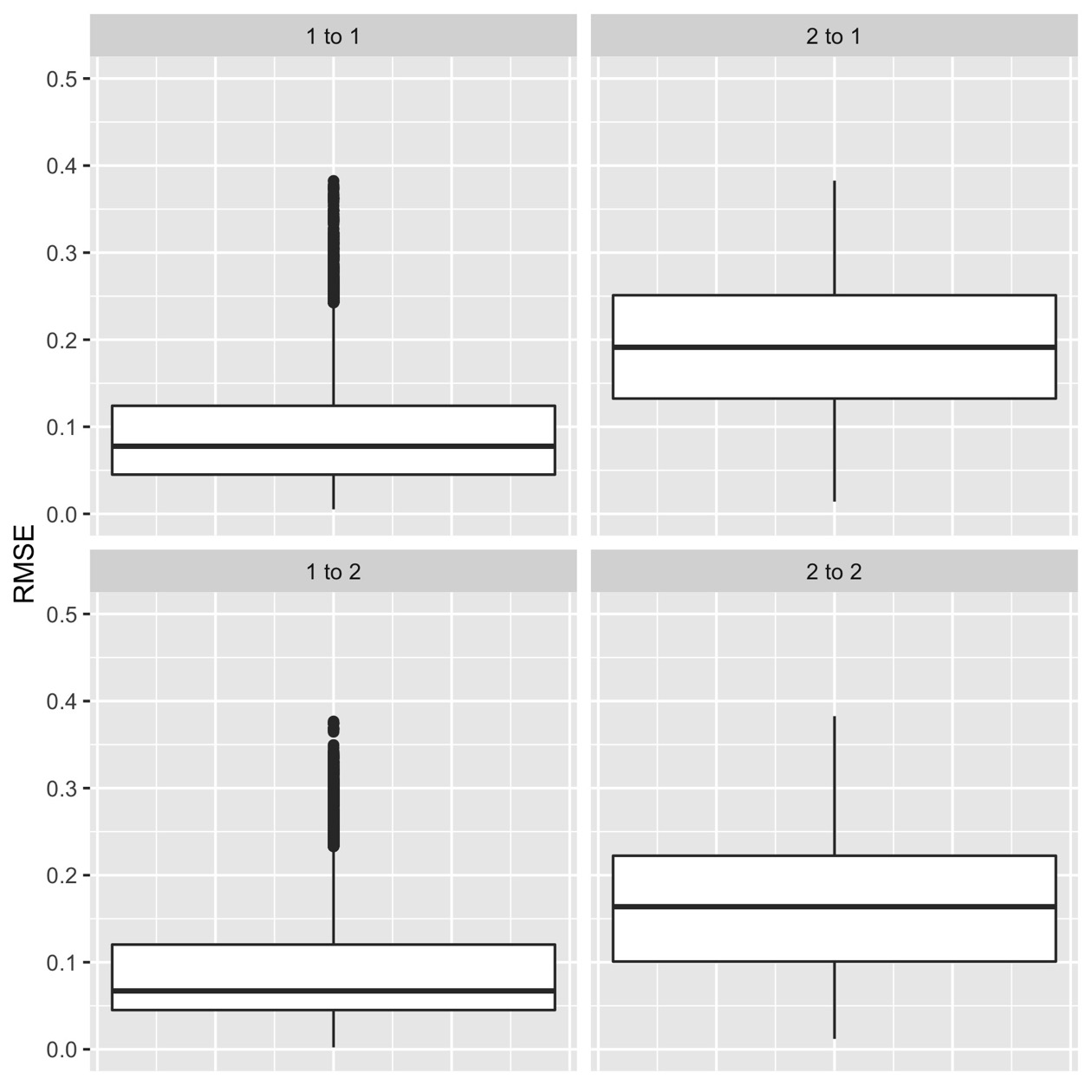}
        \caption{1000 days}
    \end{subfigure}
    \hfill
    \begin{subfigure}[b]{0.3\textwidth}
        \centering
        \includegraphics[width=\textwidth]{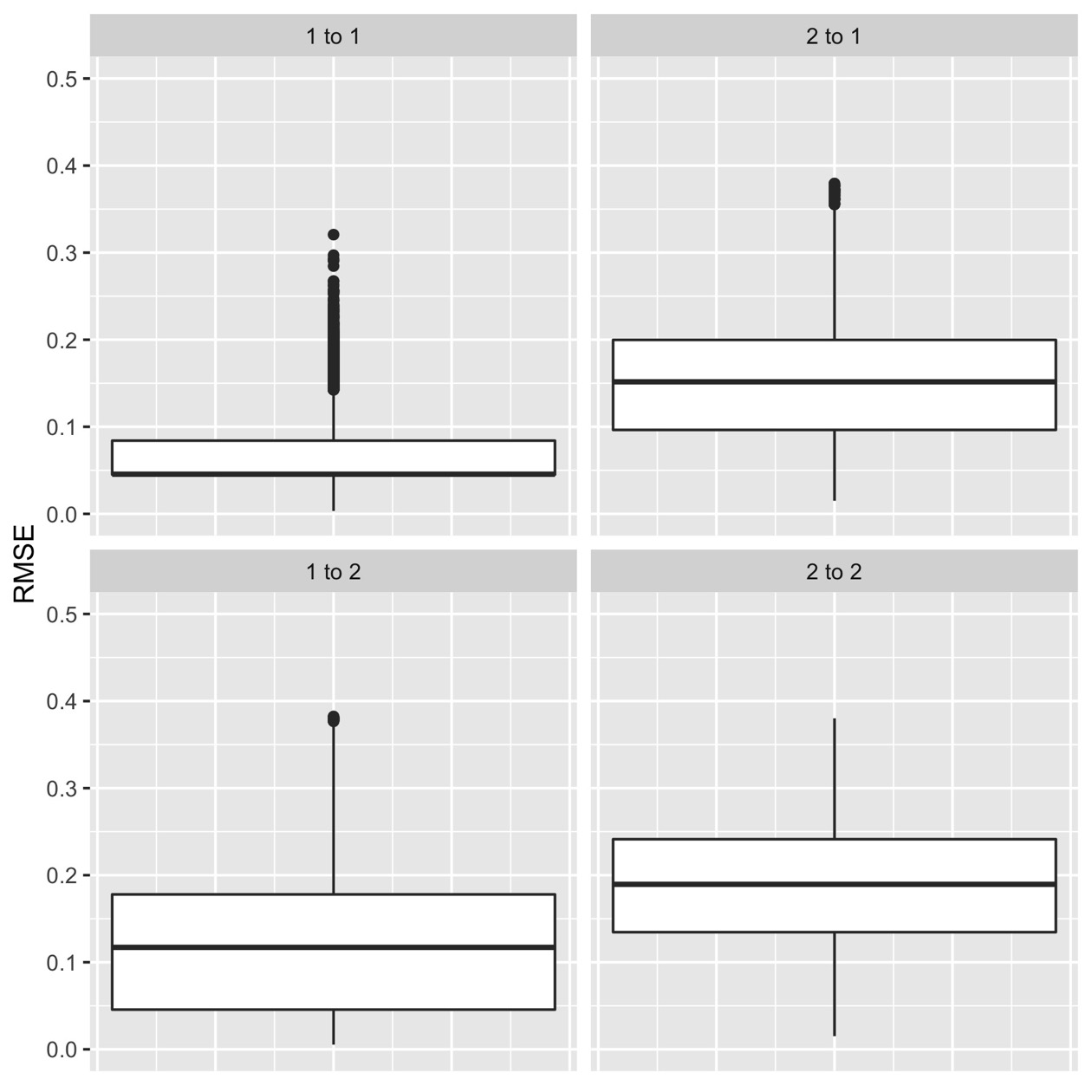}
        \caption{2000 days}
    \end{subfigure}
	\end{subfigure}    
	\caption[] 
	{\small Boxplots of the root mean squared error (comparing the estimated triggering kernel for each posterior sample to the true histogram) for the quite uninformative prior setting.}
	\label{fig:rmse_2d_uninf}
\end{figure}

Furthermore, Figures \ref{fig:rmse_2d_inf} - \ref{fig:rmse_2d_uninf} present box plots of the root mean squared error between the estimated triggering kernel for each posterior sample and the true kernel. The RMSE under the informative prior setting remains low and does not significantly differ for different time series lengths. As the prior is relaxed, increasing the length of the time series results in slight improvements to posterior accuracy. However these improvements are not as stark as seen in the univariate study in Section \ref{sec:uv_sim} due to the increased dimensionality of the parameter space in this example.

In terms of static parameters, we also observe the same phenomena as the univariate simulation study. Convergence of the MCMC algorithm and mixing were reasonable for the informative and relatively uninformative scenarios, whereas the quite uninformative scenarios often showed signs of non-convergence. This supports the observations made previously that the uninformative prior setting can be unreliable, particularly for smaller sample sizes. Further details on MCMC convergence diagnostics for the static parameters are provided in the supplementary material.

\subsubsection{Alternative approaches}\label{sec:mv_alt}

Other scenarios considered in the development of the proposed model are described here for completeness. A split-merge proposal was developed in addition to the birth-death proposal. The split-merge move was found to provide the most benefit when acceptance rates of the birth-death proposals were low. In our simulations, this only occurred when using the quite uninformative prior setting, which was shown to give unreliable inference. However for the other prior settings, the additional computational burden did not substantively improve posterior uncertainty or accuracy. Thus the final model we propose only considers a birth-death move. Further discussion on the split-merge move and corresponding results can be found in Section \ref{sec:discussion} and the supplementary material respectively. 

Alternative triggering kernels were also considered. The shape of the excitation function was varied and the length of the excitation ranged from 7 to 30 days of memory, aligning with the case studies considered in the following section. Simulations were also performed for a 3-dimensional process, once again to align with the case studies provided in this article. Each of these scenarios were repeated with various simulated datasets for a given set of parameters. The same outcomes as our main analysis were observed whereby the true parameters are generally within the range of the posterior interval given suitable prior conditions. We refer the reader to the supplementary material for further details on these.

\subsection{COVID-19 case study}

We fit the proposed multivariate DTHP model to describe the spread of the novel coronavirus disease COVID-19 between two countries, taking France and Italy as an example. These countries are selected as Italy was one of the first countries to experience widespread infection. A natural concern is how this affected not only the country itself, but also neighbouring countries. 

In the multivariate setting the triggering kernels describe the self-exciting and mutually-exciting activity between the countries with respect to spread of the disease. The maximum memory of the triggering kernel was 14 days, as transmission of known variants reportedly occurs within this timeframe in most cases \citep{He:2020, Rai:2021, Alene:2021}. In terms of prior settings for the continuous parameters, standard normal priors are used. This is equivalent to the relatively informative prior setting introduced in Section \ref{sec:sim_study}. It was selected as the true parameter values in this case study are unknown and this setting demonstrated good performance in our simulation studies. 

The infection patterns observed in the coronavirus pandemic are volatile and alternate between explosive and stable trajectories due to the interacting effects of factors such as non-pharmaceutical interventions, subsequent waves of infection and the availability of vaccinations. HPs cannot capture these alternating dynamics in a single model. Thus the time series is partitioned into phases based on the trajectory of the observed data, as suggested in \cite{Browning:2021}. 

We consider France as a reference country and create the partitions based on the trajectory of its mortality curve. Mortality data are used because, particularly in the early stages of the pandemic, infection data were found to be unreliable. To further justify this choice, \cite{Browning:2021} show that as a first-order approximation, the death dynamics are helpful to understand the infection dynamics and the parameters of the model can be interpreted with respect to infections. 
  
\begin{figure}[H]
	\centering	
	\includegraphics[width=0.5\textwidth]{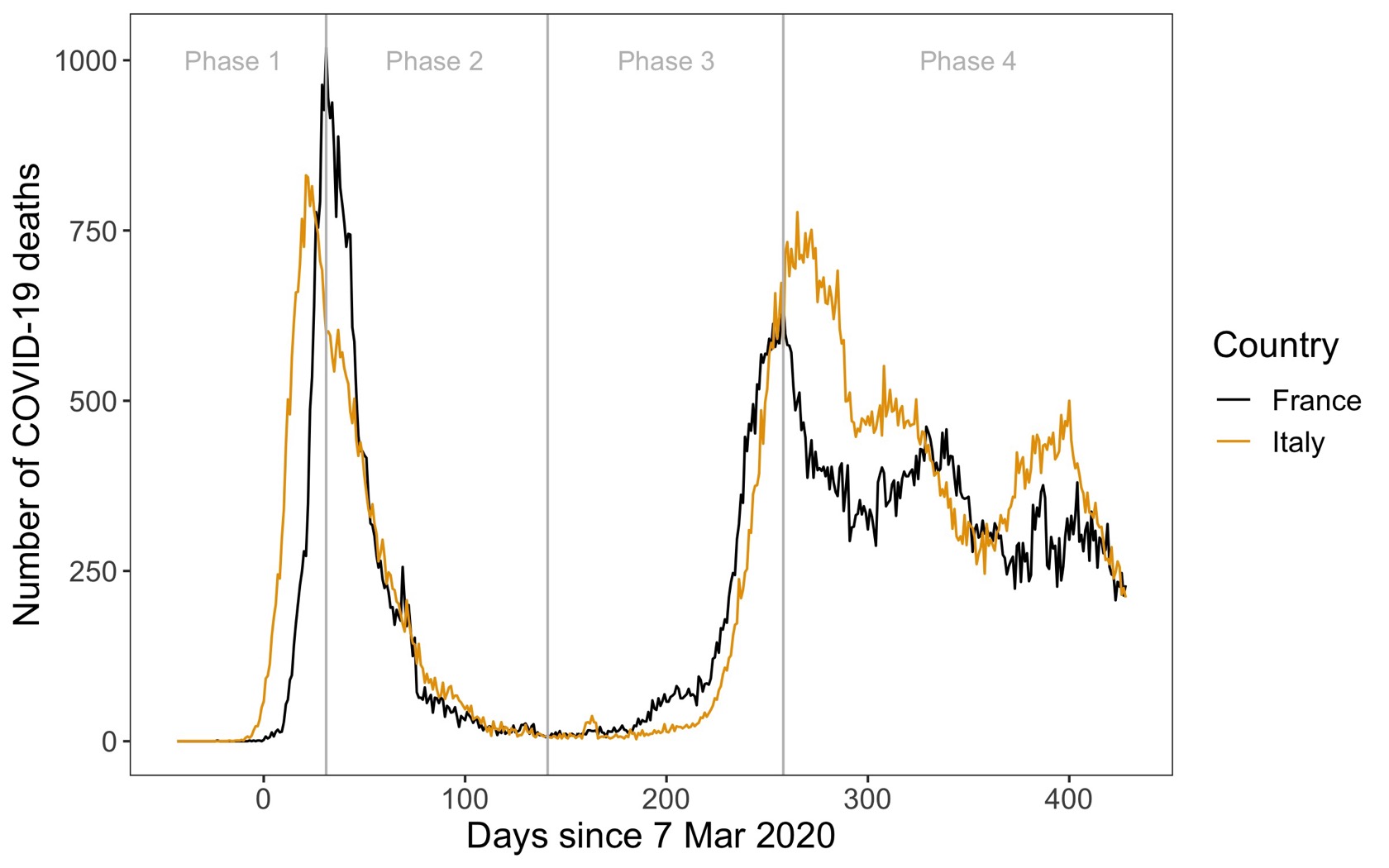}	
	\caption{Daily volume of deaths due to COVID-19 for France (black line) and Italy (orange line). Vertical lines represent partitions of the time series into phases.}
	\label{fig:covid_data}
\end{figure}

We use data gathered by the Johns Hopkins University \citep{JohnHopkinsCOVID}, containing daily counts by country of deaths due to COVID-19. Data collected from 7th March 2020 through to 8th May 2021 are used in this analysis. This period of the pandemic exhibits a wide range of patterns in the data, and thus was selected to demonstrate the flexibility of the proposed model. The data are smoothed over a rolling window of seven days. 
Fig \ref{fig:covid_data} shows the smoothed volume of daily deaths. The time series is partitioned into four phases, as depicted by the vertical lines in Figure \ref{fig:covid_data}. While the start of a new phase does not necessarily indicate a change in trajectory for Italy, in this case their peaks are roughly aligned with those in France.

\begin{figure}[H]
    \centering
    \begin{subfigure}{0.6\textwidth}
    \begin{subfigure}[b]{0.49\textwidth}
        \centering
        \includegraphics[width=\textwidth]{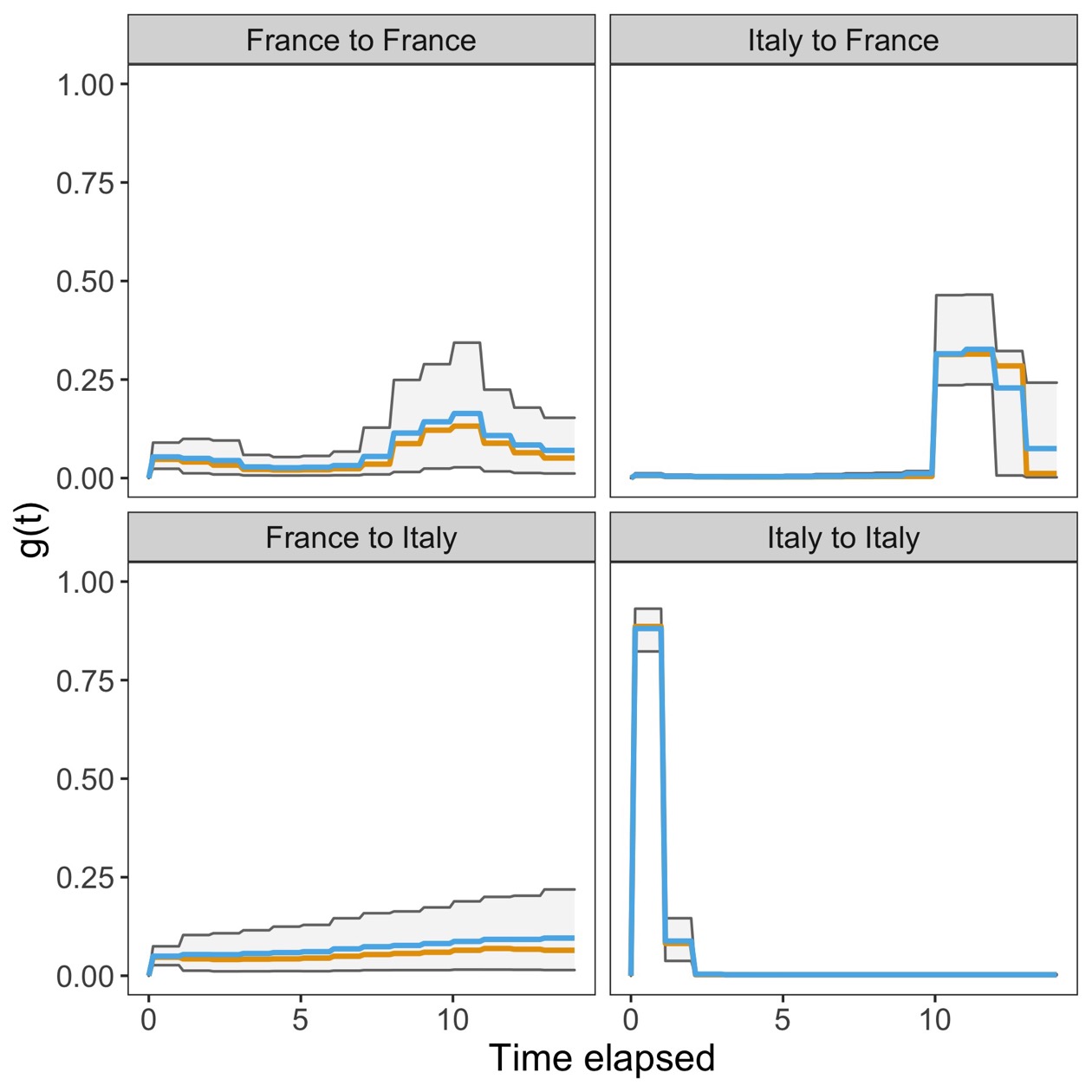}     
    \caption[]
    {\small Phase 1}
    \end{subfigure} 
    \hfill
    \begin{subfigure}[b]{0.49\textwidth}
        \centering
        \includegraphics[width=\textwidth]{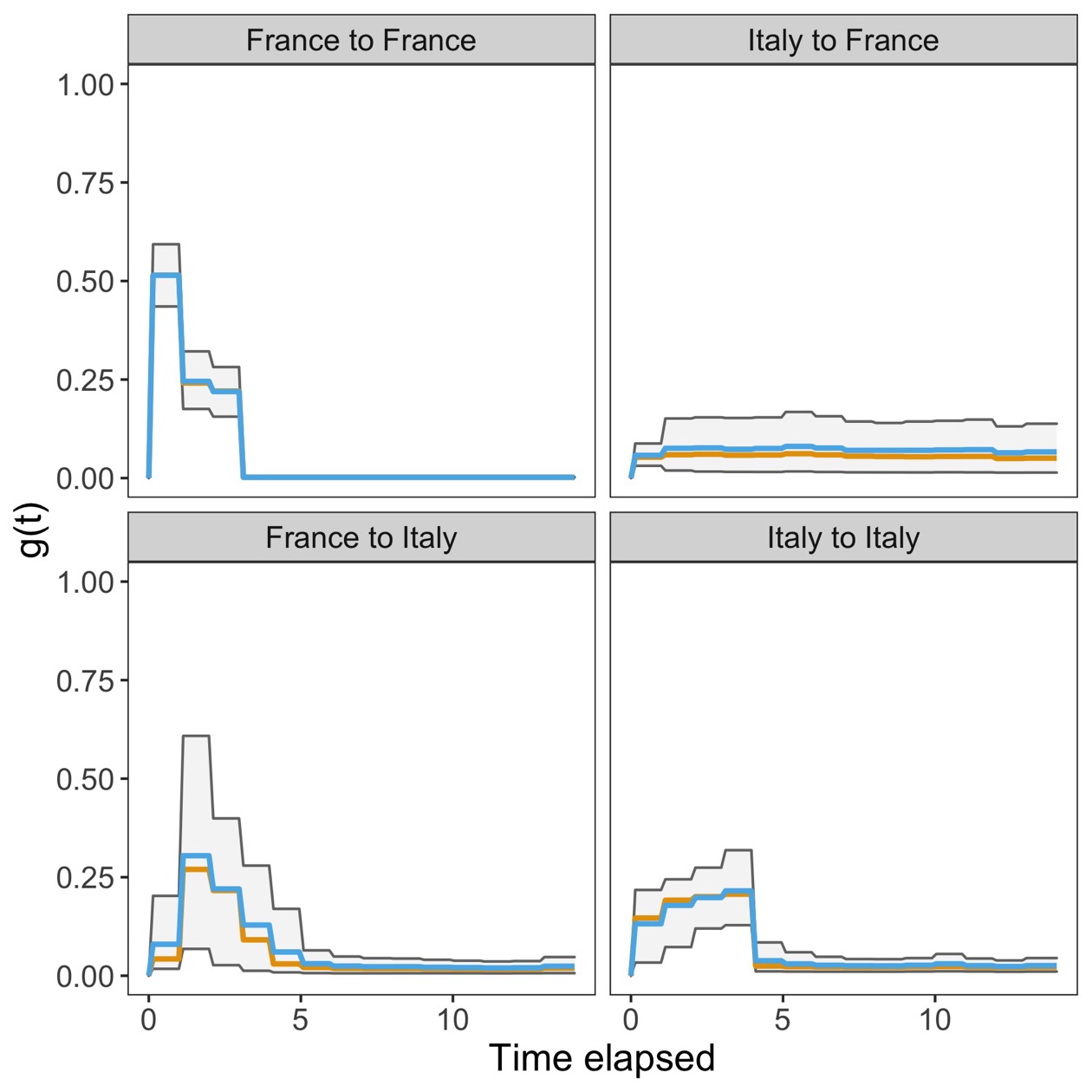}     
    \caption[]
    {\small Phase 2}
    \end{subfigure} 
    \hfill
    \begin{subfigure}[b]{0.49\textwidth}
        \centering
        \includegraphics[width=\textwidth]{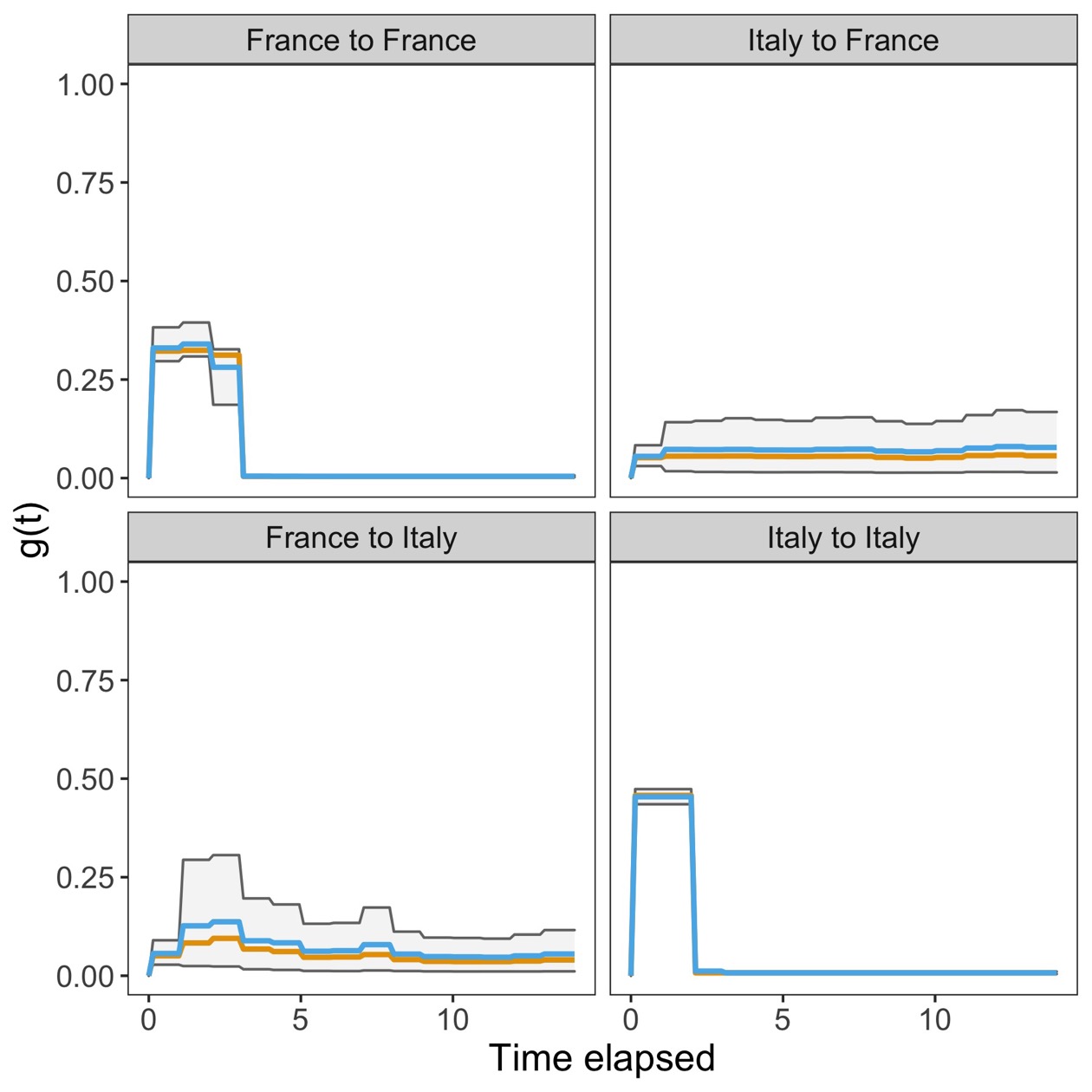}     
    \caption[]
    {\small Phase 3}
    \end{subfigure} 
    \hfill
    \begin{subfigure}[b]{0.49\textwidth}
        \centering
        \includegraphics[width=\textwidth]{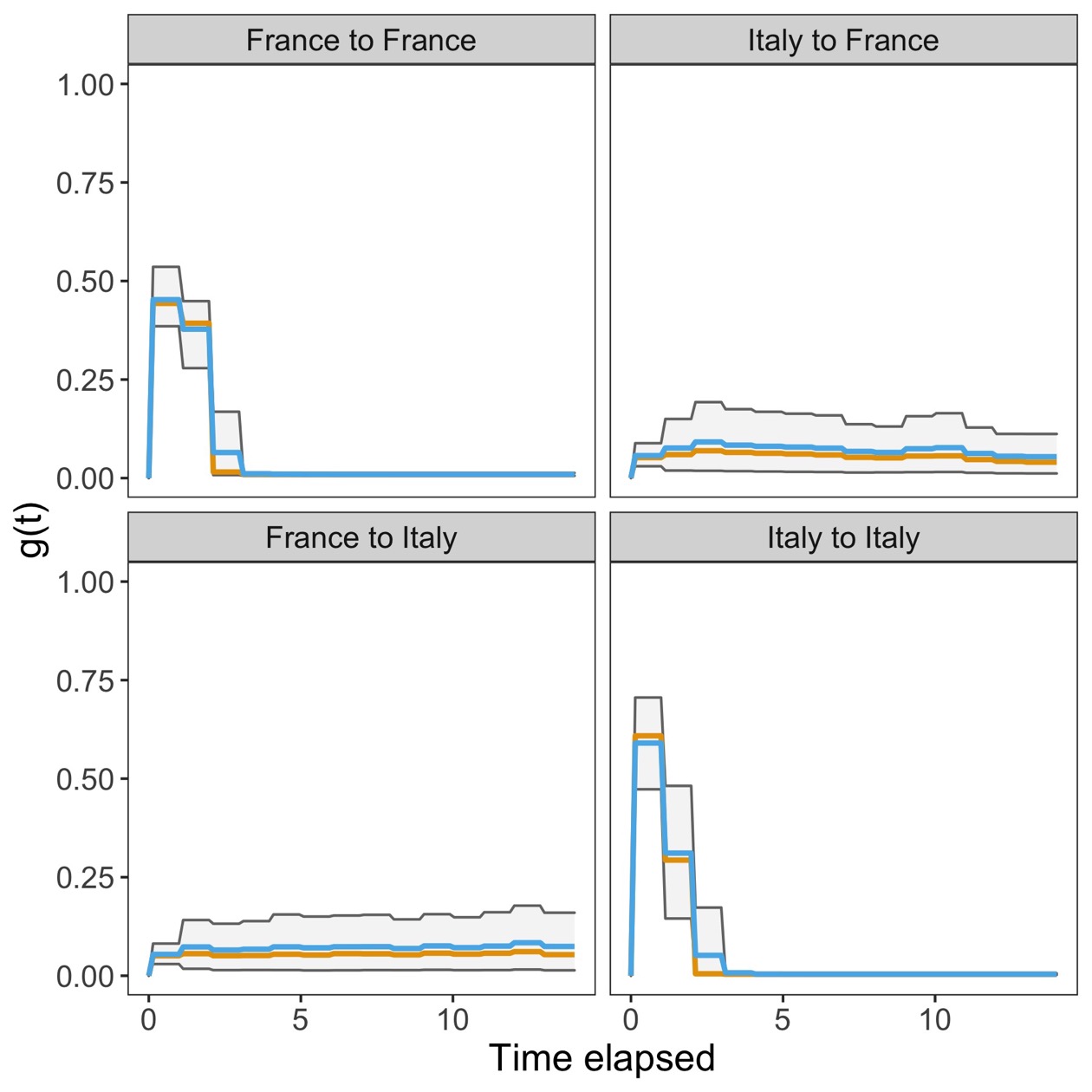}     
    \caption[]
    {\small Phase 4}
    \end{subfigure} 
	\end{subfigure}    
	\caption[]
    {\small Estimated excitation functions for COVID-19 case study. \\ 
    \textbf{Orange line:} posterior mean. \textbf{Blue line:} posterior median.\\ \textbf{Grey ribbon:} 80\% posterior interval.}
    \label{fig:covid_hist}
\end{figure}

Figure \ref{fig:covid_hist} presents the estimated triggering kernels both within and between France and Italy for each phase. In the first phase, the spread of COVID-19 in Italy is predominantly self-exciting, arising from within the country itself. This is in contrast to the excitation from Italy to France, where there is a sustained period of no excitation, followed by a spike at approximately 10 days post infection. This is indicative of the spread of the virus in France being excited in part due to international travel from Italy in the early stages of the pandemic, with a delay as infected individuals from Italy travel to France. This reflects what occurred historically since the virus was widely spread in Italy shortly before France. In subsequent phases, transmission of the virus appears to be largely contained within each country, with the self-excitation functions showing that the majority of excitation occurs within the first 4 days. The cross-excitation functions, on the other hand, are mostly flat, indicating that there is little excitatory behaviour between the countries. There is some indication of cross-excitation also from France to Italy in the second phase and less-so in the third phase. While this could indicate that some international transmission is still occurring, the uncertainty around these estimates is high so these particular interaction functions should be interpreted with caution. 

Table \ref{tab:covid_par_ests} presents the posterior medians for the static parameters in the model. In the context of the COVID-19 pandemic, these model parameters provide useful insights. The magnitude parameter $\alpha$ is particularly important, because if it is much less than one the shape of the triggering kernel is less influential to the overall intensity. Slow mixing was observed for some of these parameters, but multiple repetitions of this analysis suggest that the posteriors are consistent between samples. Further details can be found in the supplementary material.

The baseline parameter $\mu$ suggests that in Phase 1 there was a high level of imported cases in Italy driving the spread of the virus. This supports observations made previously as Italy was one of the first countries to experience widespread outbreaks. Subsequent phases show reduced levels in the baseline rate, indicative of reduced international travel. In Phase 4, France, and to some extent Italy, exhibit an increase to the baseline rate, indicating an increase in mobility as the pandemic progresses and interventions are relaxed. However it should be noted there is a large amount of uncertainty surrounding the estimates for larger values of $\mu$ due to slow mixing. The posterior intervals for each of the static model parameters are provided in the supplementary materials.

The estimates for the magnitude parameter $\alpha$ are reflective of the current trajectories in each of the phases, since $\alpha < 1 $ indicates a stationary process and $\alpha >1 $ results in exponential growth. In Phases 1 and 3, the magnitude parameter for self-excitation within France is greater than 1. For Italy it is less than 1 in Phase 1 due to the partition point being placed partway into the downward trajectory, resulting in an averaging of the parameter estimates. We discuss the matter of aligning the change points for this models further in Section \ref{sec:discussion}. 

Reduced interaction between countries throughout time, observed from the excitation functions, is also apparent in the cross-exciting magnitude parameters for Italy, with high levels in Phase 1 and a substantial reduction in subsequent phases. These low values for the cross-exciting magnitudes in later phases also dampens the impact of the higher uncertainty for the corresponding cross-excitation functions from Figure \ref{fig:covid_hist}.

\begin{table}[H]
\centering
\begin{tabular}{r|llll}
  \hline
Parameter & Phase 1 & Phase 2 & Phase 3 & Phase 4 \\ 
  \hline 
  $\mu^{\text{France}}$ & 0.716 & 0.478  & 0.503 & 6.047 \\   
  $\mu^{\text{Italy}}$ & 36.267  & 0.43  & 0.228 & 0.748  \\    
  $\alpha^{\text{France to France}}$ & 1.039 & 0.898  & 1.068  & 0.901  \\      
  $\alpha^{\text{Italy to France}}$ & 0.795 & 0.024   & 0.027   & 0.049 \\      
  $\alpha^{\text{France to Italy}}$ & 0.021 & 0.166 & 0.076  & 0.036 \\      
  $\alpha^{\text{Italy to Italy}}$ & 0.971 & 0.669  & 1.02  & 0.959 \\  \hline
  \end{tabular}
\caption{Median posterior estimates for the static model parameters. } 
\label{tab:covid_par_ests}
\end{table}

\begin{figure}[H]
	\centering	
	\includegraphics[width=0.6\textwidth]{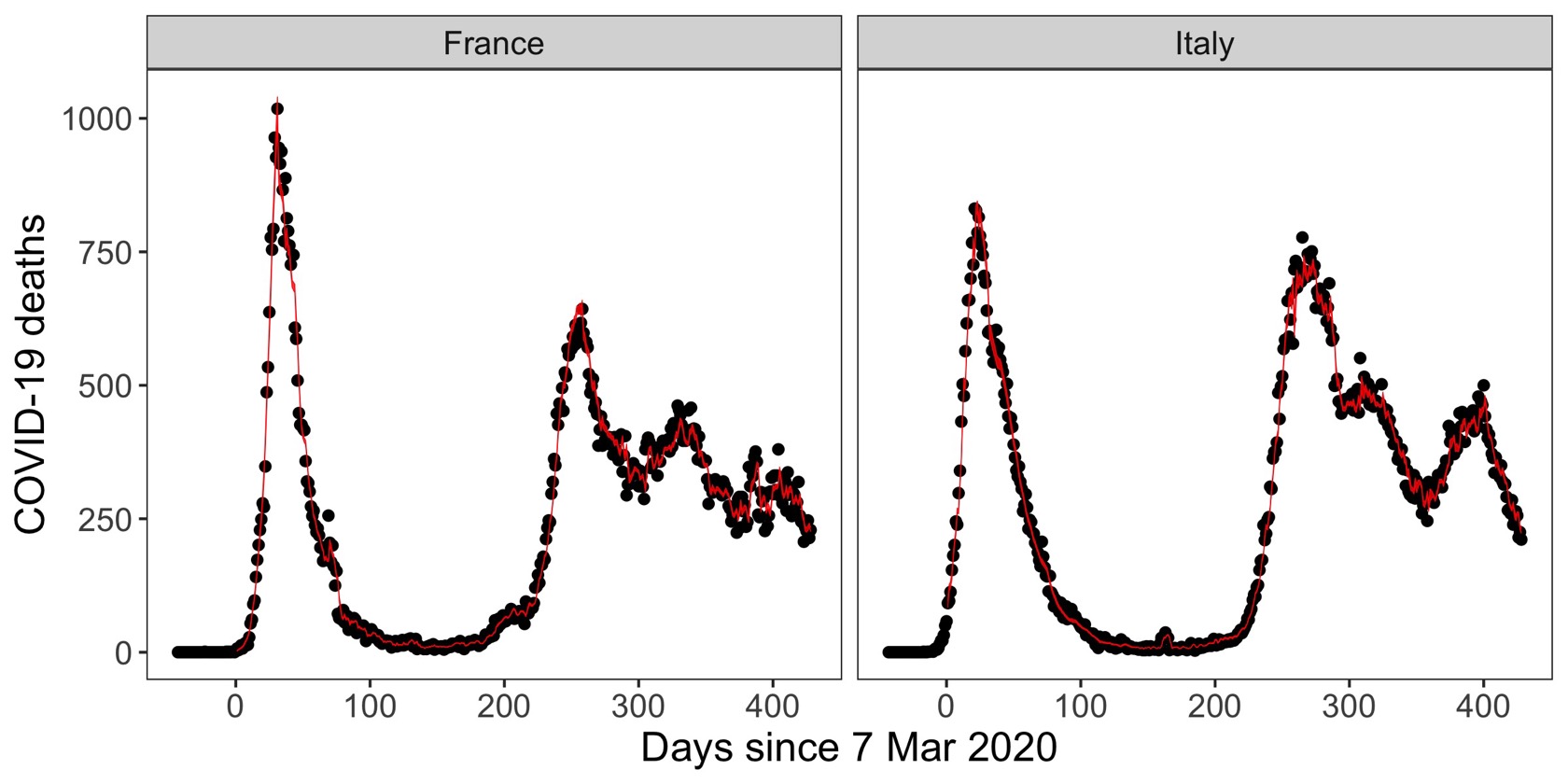}
	\caption{The observed number of deaths (black dots) compared to the 80\% posterior interval for the estimated expected number of events , i.e. $\lambda(t)$ (solid red ribbon).}
\label{fig:est_cif_vs_obs}
\end{figure}

Figure \ref{fig:est_cif_vs_obs} shows the observed data plotted against the estimated intensity function, $\lambda(t)$. As noted in Section \ref{sec:methods}, in this model $\lambda(t)$ represents the expected number of events on day $t$. We note that the estimated number of events closely follows the observed time series, and is able to react quickly to volatility in the observed data.

\subsection{Terrorism case study}

Next we apply the proposed model to a bivariate scenario describing the occurrence of terrorist activity between two countries of close spatial proximity, Indonesia and the Philippines, and a 3-dimensional model within the Philippines. For the latter, the Philippines was classified into the following three regions: Luzon, Mindanaos and Visayas. These countries were selected for this analysis because, over the selected time period, they experienced a high level of terrorist activity. The more granular perspective, considering regions within the Philippines, was considered to determine whether an increased level of interaction was observed within a country, as opposed to between two countries. The relatively informative prior setting was also selected for this case study. 

\begin{figure}[H]
    \centering
    \includegraphics[width=0.5\textwidth]{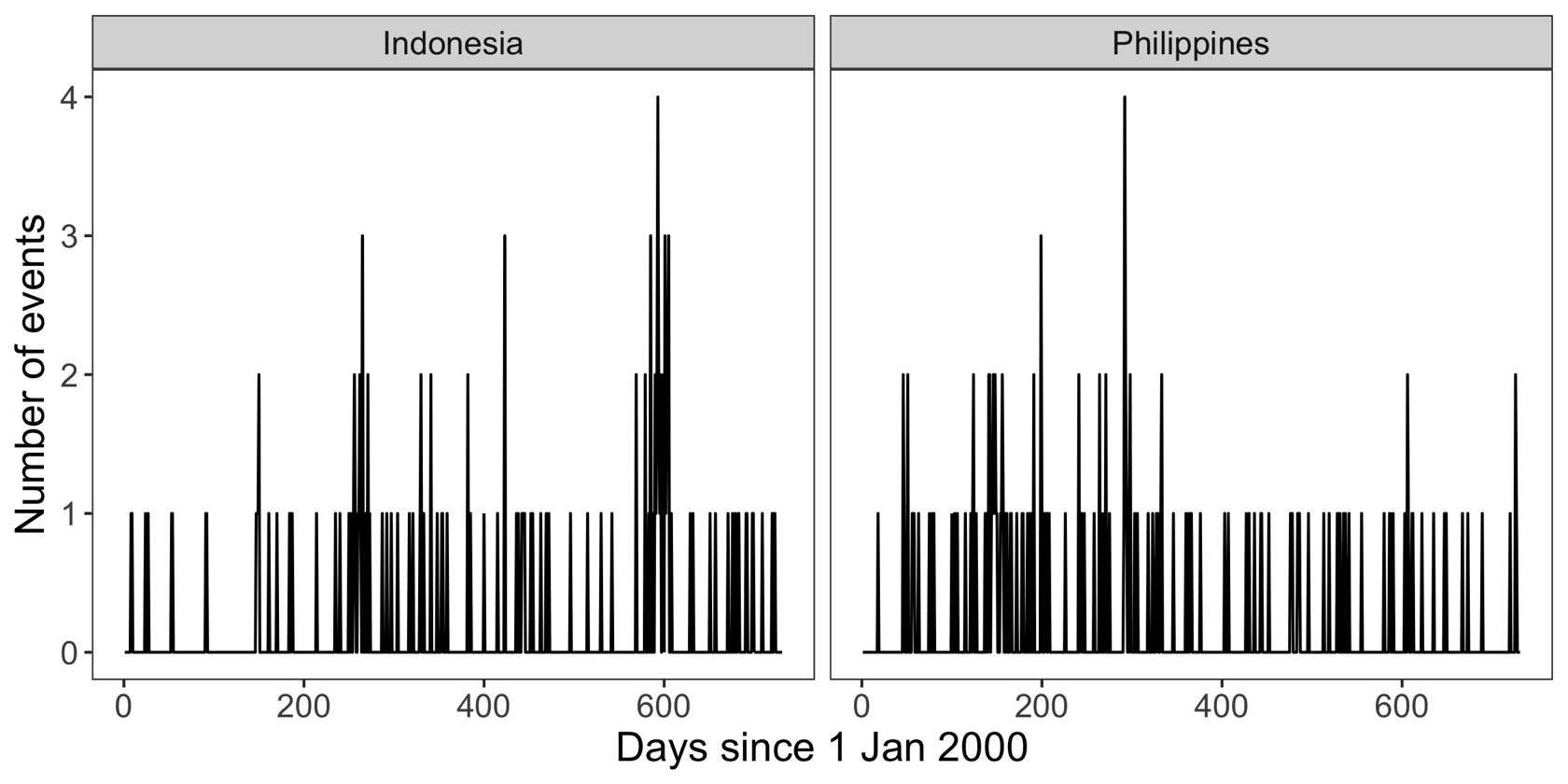}
    \caption{Observed number of terrorist events per day by country. \\ Left panel: event counts in Indonesia. Right panel: event counts in the Philippines.}    
    \label{fig:terr_data}
\end{figure}

\begin{figure}[H]
    \centering
    \includegraphics[width=0.5\textwidth]{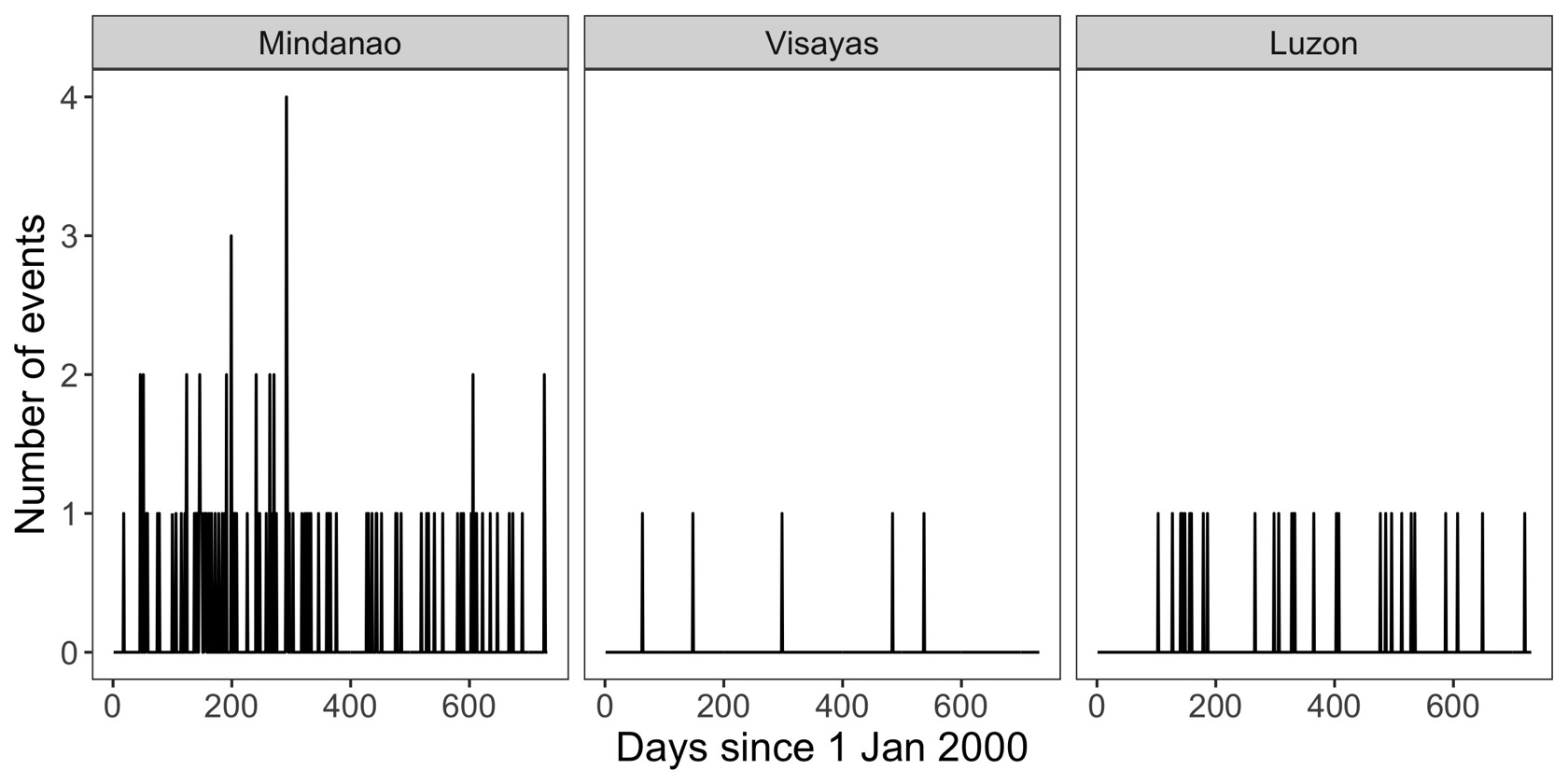}  
    \caption{Observed number of terrorist events per day by region in the Philippines. \\ Left panel: event counts in Mindanao. Middle panel: event counts in Visayas. \\ Right panel: event counts in Luzon.
    }    
    \label{fig:terr_data_withinPhil}
\end{figure}

The Global Terrorism Database (GTD) \citep{GTD} is an open source database containing information about terrorist events around the world from 1970 onwards. The database consists of 135 columns, containing details such as the dates and locations of terrorist attacks, the group responsible, weapons used and the number of fatalities. The daily count data are presented in Figures \ref{fig:terr_data} and \ref{fig:terr_data_withinPhil}. A 2 year period spanning from 1st January 2000 to 31st December 2001 is considered in this analysis. The maximum memory of the excitation kernel is 30 days, as the time for the effect of an event to subside is reportedly within this timeframe \citep{White:2012hy}.

\begin{figure}[H]
    \centering
    \begin{subfigure}{0.8\textwidth}
    \begin{subfigure}[b]{0.48\textwidth}
        \centering
        \includegraphics[width=\textwidth]{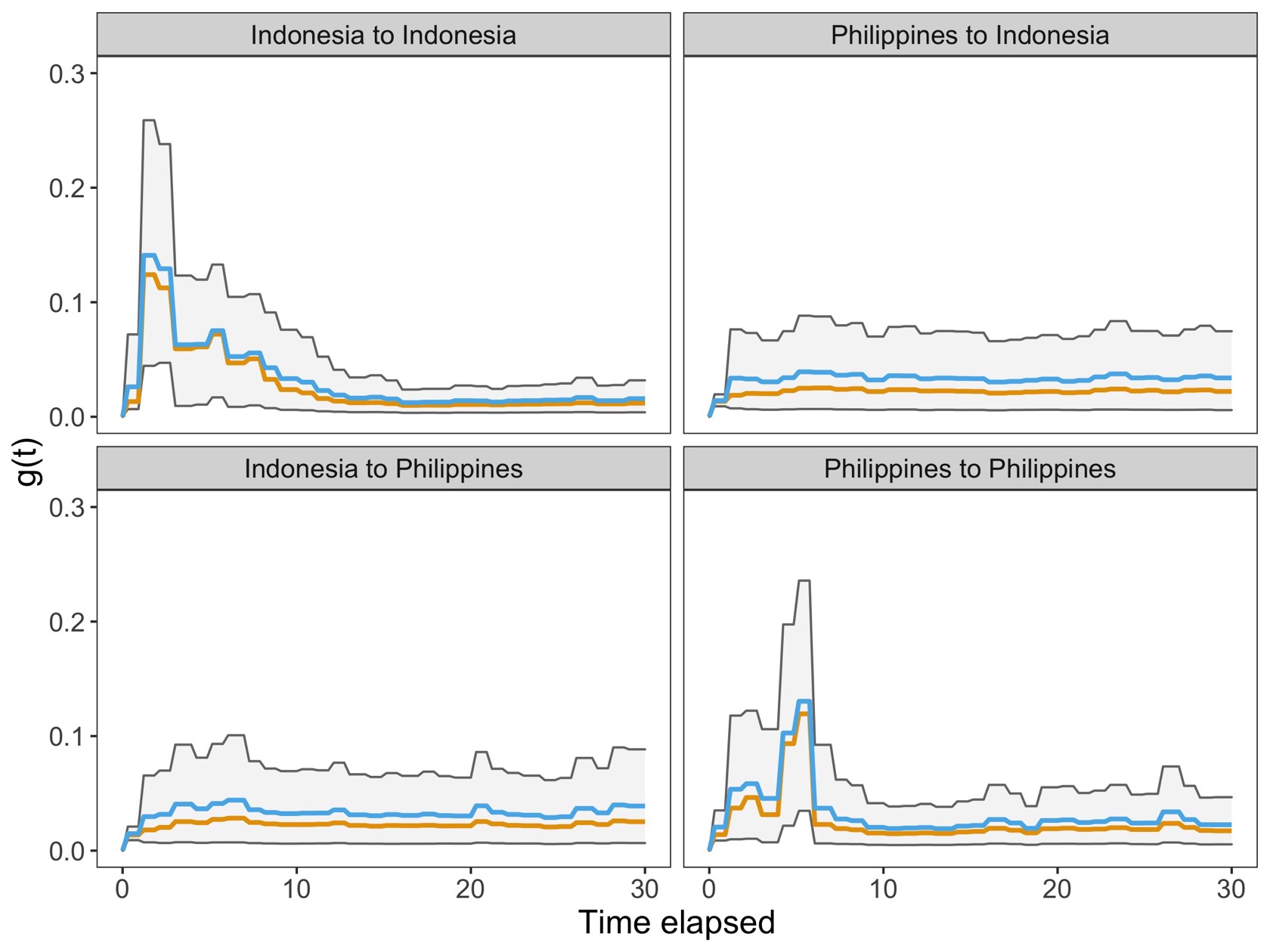}
        \caption{Indonesia and the Philippines}
    	\label{fig:terr_IndonPhil}
    \end{subfigure}
    \hfill
    \begin{subfigure}[b]{0.48\textwidth}
        \centering
        \includegraphics[width=\textwidth]{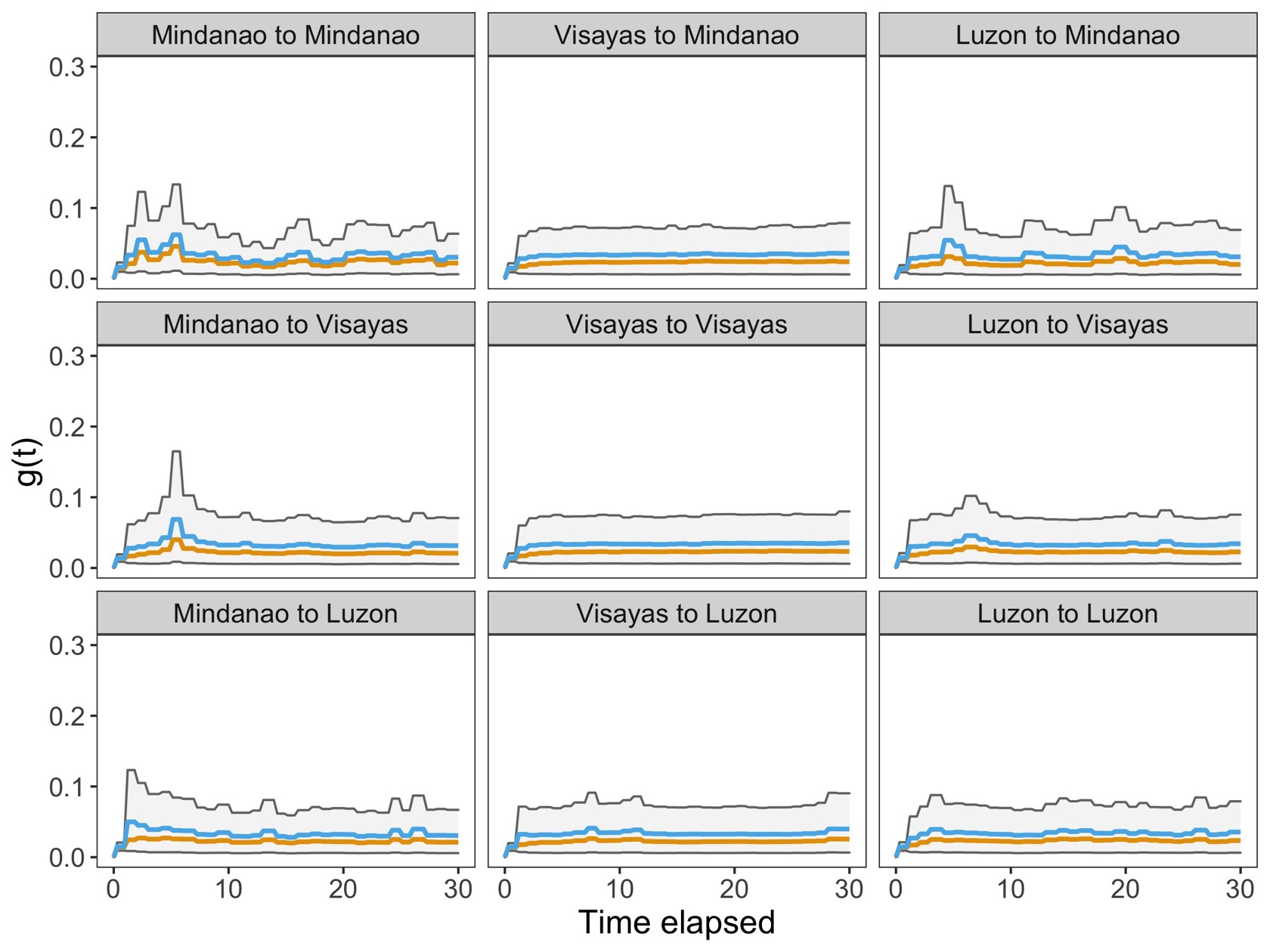}
        \caption{Luzon, Mindanaos and Visayas.}
    	\label{fig:terr_withinPhil}
    \end{subfigure}
    \end{subfigure}
	\caption[]
    {\small Estimated excitation functions for terrorism case study. \\
    \textbf{Orange line:} posterior mean. \textbf{Blue line:} posterior median.\\ \textbf{Grey ribbon:} 80\% posterior interval.}
\end{figure}

Figure \ref{fig:terr_IndonPhil} shows that there is minimal cross-excitation between Indonesia and the Philippines, despite their spatial proximity, as the level of excitation is relatively constant over the entire memory of the triggering kernel. The excitation is largely contained to within each country with a peak in the self-excitation functions for both countries occurring within the first week. 

Considering the model for regions within the Philippines, Figure \ref{fig:terr_withinPhil} suggests that Mindanao, the region with the highest occurrence of activity, is a predominant source of excitation both within Mindanao itself and also to the remaining regions. Visayas and to a lesser extent Luzon appear to exhibit very little self-excitation. However, Luzon is a moderate source of excitation for the remaining regions. We note that causal conclusions should be derived with caution since there could be other confounding factors, for example the influences of neighbouring countries or other aspects of the true generating process that we have not accounted for in our model. 
We note that good mixing and MCMC convergence diagnostics were observed throughout for this analysis. Further details can be found in the supplementary material.

\section{Discussion} \label{sec:discussion}

In this article we present a flexible approach for modelling discrete-time Hawkes processes. By incorporating trans-dimensional MCMC inference techniques, the triggering kernel of the DTHP can take the form of any random histogram function with integer-valued change points. This is in contrast to most standard models of Hawkes processes that rely on a parametric form for the triggering kernel. Particularly for complex data, this formulation is likely to be insufficient to capture the true excitation pattern. We illustrate through a comprehensive simulation study the situations in which the proposed model is likely to perform well or otherwise. The proposed model is also applied to real world data for several case studies: to model the interaction between France and Italy in the coronavirus pandemic, and to capture the excitation patterns of terrorist activity in Indonesia and the Philippines. In both our simulations and case studies, multivariate DTHPs of up to 3-dimensions are considered.

A comparison between our proposed model and a simple DTHP with a geometric triggering kernel was explored. The results suggest that if there is reason to believe the true triggering kernel follows a parametric form, then this may be preferable due to reduced complexity and depending on the kernel, reduced computational expense. However, this is rarely the case in practice. Our model, while it can be more computationally intensive compared to a parametric kernel with comparable compact support, provides a flexible function that is natural for a discrete time setting. Moreover, if parsimony is the priority, one could use our proposed method to determine whether a parametric form is appropriate for a given application, which should be determined on a case-by-case basis.

It was anticipated that this high-dimensional approach may not be universally effective. Numerous time series lengths and prior settings were considered to determine when the proposed model is most effective. The model performs well in situations in which the prior is sufficiently informative to account for lack of data, or if there is sufficient signal in the data to compensate for lack of prior information. In the absence of both of these, the proposed model will likely perform poorly. This is an expected results, and further not surprising due to the high-dimensional parameter space of this model, particularly for higher dimensional DTHPs. In determining whether a prior choice is sufficiently informative, methods such as history matching \citep{Drovandi:2021, Wang:2018} could be considered. It could also be useful to perform a prior sensitivity analysis, particularly when implementing this model on real data where the true parameters are unknown.


Although the results are not provided here, in our simulations we find that incorporating a split-merge move into the reversible-jump MCMC could improve model performance when acceptance rates for the birth-death move alone are low, resulting in slow mixing. This is likely due to the split-merge step proposing from regions of the parameter space that already have high posterior mass. However, in most of our examples the acceptance rate from the birth-death move alone was reasonable, and thus the benefit of the split-merge did not outweigh the additional computational cost.

We find that applying the model to real world data provides meaningful results. In the COVID-19 case study, the empirical results provide unique insights into the transmission between these countries and are supported historically. As such, this model could be applied more widely to study interactions between other countries of interest regarding the COVID-19 pandemic, and higher-dimensional models are also a possibility. In the case study of terrorist activity the results largely indicate that cross-excitation between the countries and regions is relatively low for the observation period considered. These results have potential implications to better understanding and managing terrorist activity in these regions, and this analysis could also be performed on other regions of interest. Both case studies showed reasonable model performance and convergence of the MCMC algorithm. Where slow mixing was observed, repeating the procedure numerous times indicated that the posteriors for each of the samples were consistent, suggesting that the chains converged to the same posterior.

Although our approach was shown to perform well under many scenarios, there are some limitations and possible extensions. A higher dimensional parameter space means that we require more from our data for reliable parameter inference, particularly if informative prior information is unavailable. The proposed model will not perform well if there is insufficient signal in the data. 

An area of future research is in developing an alternative algorithm, since reversible-jump MCMC is computationally resource-hungry and evaluation of these models often takes days, if not weeks depending on the volume of data and memory of the excitation function. Increasing the dimensionality of the problem beyond 3-dimensions is likely to be computationally intractable due to the volume of data that would be required to infer the parameters in conjunction with an exponentially increasing parameter space. 

One alternative extension of this model is in estimating the maximum excitation length, $s_\text{max}$. In this work it is specified apriori; however, estimating it would allow for further flexibility in varying the length of excitation possible and allow for a potentially Bayesian nonparametric specification of the histogram function. For example, in cases where the maximum length of excitation is unknown $T-1$, where $T$ is the length of the time series, is a natural maximum excitation time. A latent switching model on the knot points of the histogram could then be incorporated by assigning each latent integer-length segment of the histogram, from 1 to $T-1$, to a cluster with a common height parameter.

Another variation is introducing sparsity in the excitation relationships, either through the introduction of an indicator function to turn on and off each excitation function \citep{Donnet:2020} or by placing sparsity-inducing priors on the magnitude parameters \citep{Deutsch:2022}. This is useful for models in which the connectivity structure between dimensions is unknown and it is of interest to infer this structure. Whereas in our model we assume there is always some, possibly low, level of excitation present, this alternative variation would distinguish between a constant but low level of excitation, or a nonexistent excitation effect. 

Another area of future work, specifically for the multivariate COVID-19 case study, is regarding the partitioning of the time series into distinct phases. In this work we base the locations of the change points on the peaks and troughs of a single reference country. This could be improved by considering an approach such as dynamic time warping \citep{Sakoe:1978} to align the stationary points for each of the countries or regions of interest.

Notwithstanding these concerns and avenues for future research, to the authors' knowledge this study shows for the first time a flexible model for the triggering kernel of discrete-time Hawkes processes and the challenges and considerations necessary for these models with high-dimensional parameter spaces. We demonstrate that a parametric function is often not a sufficient for the triggering kernel, particularly for complex data. Although a similarly flexible model for continuous-time HPs is presented in \cite{Donnet:2020}, no such model exists for DTHPs. This paves the way for the detailed exploration of the interaction between complex multivariate self-exciting phenomena where event count data are collected at discrete-time intervals and parametric forms of the triggering kernel are too simplistic.

\newpage

\section*{Acknowledgements}

The project leading to this work has received funding from the Australian Research Council (ARC) Laureate Fellowship Program under the project ‘Bayesian Learning for Decision Making in the Big Data Era' (grant no.: FL150100150) and the European Research Council (ERC) under the European Union’s Horizon 2020 research and innovation programme (grant no.: 834175).

\bibliography{Bibliography,Bibliography_JHU}

\begin{thebibliography}{23}
\providecommand{\natexlab}[1]{#1}
\providecommand{\url}[1]{\texttt{#1}}
\expandafter\ifx\csname urlstyle\endcsname\relax
  \providecommand{\doi}[1]{doi: #1}\else
  \providecommand{\doi}{doi: \begingroup \urlstyle{rm}\Url}\fi

\bibitem[Alene et~al.(2021)Alene, Yismaw, Assemie, Ketema, Gietaneh, and
  Birhan]{Alene:2021}
M.~Alene, L.~Yismaw, M.~A. Assemie, D.~B. Ketema, W.~Gietaneh, and T.~Y.
  Birhan.
\newblock {Serial interval and incubation period of COVID-19: a systematic
  review and meta-analysis}.
\newblock \emph{BMC Infectious Diseases}, 21\penalty0 (1):\penalty0 257, 2021.
\newblock \doi{10.1186/s12879-021-05950-x}.

\bibitem[Bodin et~al.(2012)Bodin, Sambridge, Tkalčić, Arroucau, Gallagher,
  and Rawlinson]{Bodin:2012}
T.~Bodin, M.~Sambridge, H.~Tkalčić, P.~Arroucau, K.~Gallagher, and
  N.~Rawlinson.
\newblock {Transdimensional inversion of receiver functions and surface wave
  dispersion}.
\newblock \emph{Journal of Geophysical Research: Solid Earth}, 117\penalty0
  (B2):\penalty0 n/a -- n/a, 02 2012.
\newblock \doi{10.1029/2011jb008560}.
\newblock URL
  \url{https://agupubs.onlinelibrary.wiley.com/doi/full/10.1029/2011JB008560}.

\bibitem[Browning et~al.(2021)Browning, Sulem, Mengersen, Rivoirard, and
  Rousseau]{Browning:2021}
R.~Browning, D.~Sulem, K.~Mengersen, V.~Rivoirard, and J.~Rousseau.
\newblock {Simple discrete-time self-exciting models can describe complex
  dynamic processes: A case study of COVID-19.}
\newblock \emph{PloS one}, 16\penalty0 (4):\penalty0 e0250015, 2021.
\newblock \doi{10.1371/journal.pone.0250015}.

\bibitem[{Center for Systems Science and Engineering (CSSE) at Johns Hopkins
  University}(2021)]{JohnHopkinsCOVID}
{Center for Systems Science and Engineering (CSSE) at Johns Hopkins
  University}.
\newblock {COVID-19 data repository}.
\newblock \url{https://github.com/CSSEGISandData/COVID-19}, 2021.
\newblock Accessed: 8 May 2021.

\bibitem[Deutsch and Ross(2022)]{Deutsch:2022}
I.~Deutsch and G.~J. Ross.
\newblock {Bayesian Estimation of Multivariate Hawkes Processes with Inhibition
  and Sparsity}.
\newblock \emph{arXiv}, 2022.

\bibitem[Donnet et~al.(2020)Donnet, Rivoirard, and Rousseau]{Donnet:2020}
S.~Donnet, V.~Rivoirard, and J.~Rousseau.
\newblock {Nonparametric Bayesian estimation for multivariate Hawkes
  processes}.
\newblock \emph{The Annals of Statistics}, 48\penalty0 (5):\penalty0 2698 --
  2727, 00 2020.
\newblock \doi{10.1214/19-aos1903}.
\newblock URL \url{https://doi.org/10.1214/19-AOS1903}.

\bibitem[Drovandi et~al.(2021)Drovandi, Nott, and Pagendam]{Drovandi:2021}
C.~Drovandi, D.~J. Nott, and D.~E. Pagendam.
\newblock {A Semiautomatic Method for History Matching Using Sequential Monte
  Carlo}.
\newblock \emph{SIAM/ASA Journal on Uncertainty Quantification}, 9\penalty0
  (3):\penalty0 1034--1063, 2021.
\newblock \doi{10.1137/19m1286694}.

\bibitem[Green(1995)]{Green:1995}
P.~J. Green.
\newblock {Reversible jump Markov chain Monte Carlo computation and Bayesian
  model determination}.
\newblock \emph{Biometrika}, 82:\penalty0 711 -- 732, 00 1995.
\newblock URL
  \url{https://academic.oup.com/biomet/article-abstract/82/4/711/252058}.

\bibitem[Hawkes(1971)]{Hawkes:1971vr}
A.~G. Hawkes.
\newblock {Spectra of some self-exciting and mutually exciting point
  processes}.
\newblock \emph{Biometrika}, 58\penalty0 (1):\penalty0 83 -- 90, 04 1971.
\newblock \doi{https://doi.org/10.2307/2334319}.
\newblock URL \url{https://www.jstor.org/stable/2334319}.

\bibitem[He et~al.(2020)He, Lau, Wu, Deng, Wang, Hao, Lau, Wong, Guan, Tan, Mo,
  Chen, Liao, Chen, Hu, Zhang, Zhong, Wu, Zhao, Zhang, Cowling, Li, and
  Leung]{He:2020}
X.~He, E.~H.~Y. Lau, P.~Wu, X.~Deng, J.~Wang, X.~Hao, Y.~C. Lau, J.~Y. Wong,
  Y.~Guan, X.~Tan, X.~Mo, Y.~Chen, B.~Liao, W.~Chen, F.~Hu, Q.~Zhang, M.~Zhong,
  Y.~Wu, L.~Zhao, F.~Zhang, B.~J. Cowling, F.~Li, and G.~M. Leung.
\newblock {Temporal dynamics in viral shedding and transmissibility of
  COVID-19}.
\newblock \emph{Nature Medicine}, 26\penalty0 (5):\penalty0 672--675, 2020.
\newblock ISSN 1078-8956.
\newblock \doi{10.1038/s41591-020-0869-5}.

\bibitem[Kirchner(2016)]{Kirchner:2017es}
M.~Kirchner.
\newblock {An estimation procedure for the Hawkes process}.
\newblock \emph{Quantitative Finance}, 17\penalty0 (4):\penalty0 571 -- 595, 09
  2016.
\newblock ISSN 1469-7688.
\newblock \doi{10.1080/14697688.2016.1211312}.
\newblock URL
  \url{https://www.tandfonline.com/doi/full/10.1080/14697688.2016.1211312}.

\bibitem[Linderman and Adams(2015)]{Linderman:2015tk}
S.~W. Linderman and R.~P. Adams.
\newblock {Scalable Bayesian Inference for Excitatory Point Process Networks}.
\newblock \emph{arXiv}, 07 2015.
\newblock URL \url{https://arxiv.org/abs/1507.03228}.

\bibitem[Markwick(2020)]{Marwick:2020}
D.~Markwick.
\newblock \emph{{Bayesian Nonparametric Hawkes Processes with Applications}}.
\newblock 00 2020.
\newblock ISBN 1230856039.
\newblock URL \url{https://discovery.ucl.ac.uk/id/eprint/10109374/}.

\bibitem[Mohler(2013)]{Mohler:2013hy}
G.~Mohler.
\newblock {Modeling and estimation of multi-source clustering in crime and
  security data}.
\newblock \emph{Annals of Applied Statistics}, 7\penalty0 (3):\penalty0 1525 --
  1539, 09 2013.
\newblock ISSN 1932-6157.
\newblock \doi{https://doi.org/10.1214/13-AOAS647}.
\newblock URL \url{http://projecteuclid.org/euclid.aoas/1380804805}.

\bibitem[Rai et~al.(2021)Rai, Shukla, and Dwivedi]{Rai:2021}
B.~Rai, A.~Shukla, and L.~K. Dwivedi.
\newblock {Estimates of serial interval for COVID-19: A systematic review and
  meta-analysis}.
\newblock \emph{Clinical Epidemiology and Global Health}, 9:\penalty0 157--161,
  2021.
\newblock ISSN 2213-3984.
\newblock \doi{10.1016/j.cegh.2020.08.007}.

\bibitem[Sakoe and Chiba(1978)]{Sakoe:1978}
H.~Sakoe and S.~Chiba.
\newblock {Dynamic programming algorithm optimization for spoken word
  recognition}.
\newblock \emph{IEEE Transactions on Acoustics, Speech, and Signal Processing},
  26\penalty0 (1):\penalty0 43--49, 1978.
\newblock ISSN 0096-3518.
\newblock \doi{10.1109/tassp.1978.1163055}.

\bibitem[{START (National Consortium for the Study of Terrorism and Responses
  to Terrorism)}(2021)]{GTD}
{START (National Consortium for the Study of Terrorism and Responses to
  Terrorism)}.
\newblock {Global Terrorism Database 1970 - 2020}.
\newblock \url{https://www.start.umd.edu/gtd/}, 2021.
\newblock Accessed: 23 Feb 2021.

\bibitem[Sulem et~al.(2021)Sulem, Rivoirard, and Rousseau]{Sulem:2021}
D.~Sulem, V.~Rivoirard, and J.~Rousseau.
\newblock {Bayesian estimation of nonlinear Hawkes process}.
\newblock \emph{arXiv}, 2021.

\bibitem[Wang et~al.(2018)Wang, Nott, Drovandi, Mengersen, and
  Evans]{Wang:2018}
X.~Wang, D.~J. Nott, C.~C. Drovandi, K.~Mengersen, and M.~Evans.
\newblock {Using History Matching for Prior Choice}.
\newblock \emph{Technometrics}, 60\penalty0 (4):\penalty0 0--0, 2018.
\newblock ISSN 0040-1706.
\newblock \doi{10.1080/00401706.2017.1421587}.

\bibitem[White et~al.(2012)White, Porter, and Mazerolle]{White:2012hy}
G.~White, M.~D. Porter, and L.~Mazerolle.
\newblock {Terrorism Risk, Resilience and Volatility: A Comparison of Terrorism
  Patterns in Three Southeast Asian Countries}.
\newblock \emph{Journal of Quantitative Criminology}, 29\penalty0 (2):\penalty0
  295 -- 320, 10 2012.
\newblock ISSN 0748-4518.
\newblock \doi{https://doi.org/10.1007/s10940-012-9181-y}.
\newblock URL \url{http://link.springer.com/10.1007/s10940-012-9181-y}.

\bibitem[Zhang et~al.(2020)Zhang, Walder, and Rizoiu]{Zhang:2020vi}
R.~Zhang, C.~Walder, and M.-A. Rizoiu.
\newblock {Variational Inference for Sparse Gaussian Process Modulated Hawkes
  Process}.
\newblock \emph{Proceedings of the AAAI Conference on Artificial Intelligence},
  34\penalty0 (04):\penalty0 6803--6810, 2020.
\newblock ISSN 2159-5399.
\newblock \doi{10.1609/aaai.v34i04.6160}.

\bibitem[Zhou et~al.(2020)Zhou, Li, Fan, Wang, Sowmya, and Chen]{Zhou:2020em}
F.~Zhou, Z.~Li, X.~Fan, Y.~Wang, A.~Sowmya, and F.~Chen.
\newblock {Efficient Inference for Nonparametric Hawkes Processes Using
  Auxiliary Latent Variables}.
\newblock \emph{Journal of Machine Learning Research}, 21\penalty0
  (241):\penalty0 1 -- 31, 00 2020.
\newblock URL \url{https://www.jmlr.org/beta/papers/v21/19-930.html}.

\bibitem[Zhou et~al.(2021)Zhou, Luo, Li, Fan, Wang, Sowmya, and
  Chen]{Zhou:2021em}
F.~Zhou, S.~Luo, Z.~Li, X.~Fan, Y.~Wang, A.~Sowmya, and F.~Chen.
\newblock {Efficient EM-variational inference for nonparametric Hawkes
  process}.
\newblock \emph{Statistics and Computing}, 31\penalty0 (4):\penalty0 1 -- 11,
  07 2021.
\newblock \doi{10.1007/s11222-021-10021-x}.
\newblock URL
  \url{https://link.springer.com/article/10.1007/s11222-021-10021-x}.

\end{thebibliography}

\end{document}